\documentclass[12pt]{article}
\usepackage{amsmath}
\usepackage{graphicx,psfrag,epsf}
\usepackage{enumerate}
\usepackage{url} 
\usepackage{geometry} 						
\usepackage{amsmath,amsfonts,bm}			
\usepackage{enumitem} 						
\usepackage{natbib} 						
\usepackage[hidelinks, colorlinks=true, 
linkcolor=black, citecolor=blue]{hyperref} 	
\usepackage{fancyhdr, calc, lastpage} 		
\usepackage{booktabs} 						
\usepackage{lipsum} 						
\usepackage{endnotes}						
\usepackage{authblk}
\usepackage[utf8]{inputenc}
\usepackage[english]{babel}
\usepackage{amsthm}
\usepackage{tikz}
\usepackage{tikzscale}
\usetikzlibrary{bayesnet}
\usepackage{algorithm2e}
\usepackage{subcaption}
\usepackage{xcolor}
\usepackage{colortbl}
\usepackage{mathrsfs}
\usepackage{bbm}

\newcommand{\blind}{1}

\DeclareMathOperator{\Tr}{tr}

\newcommand{\xv}{\texttt{x}}

\newcommand{\zv}{\texttt{z}}
\newcommand{\ev}{\texttt{e}}
\newcommand{\vv}{\texttt{v}}
\newcommand{\bv}{\texttt{b}}

\newcommand{\R}{\ensuremath{\mathbb{R}}}

\newcommand{\Prob}{\ensuremath{\mathsf{p}}}
\newcommand{\Ex}{\ensuremath{\mathbb{E}}}
\newcommand{\sign}{\ensuremath{\text{sign}}}
\newcommand{\Ez}{\ensuremath{\mathbb{E}_{z\mid\hat{\Delta},X}}}
\newcommand{\Eg}{\ensuremath{\mathbb{E}_{\gamma \mid\hat{\Delta}}}}
\newcommand{\Ezg}{\ensuremath{\mathbb{E}_{z,\gamma\mid\hat{\Delta},X}}}
\newcommand{\Sb}{\ensuremath{\Tau_{\bv_i}}}

\newcommand{\I}{\mathbf{I}}

\newcommand{\Tau}{\mathcal{T}}
\newcommand{\loN}{\ensuremath{\lambda_{0}}}
\newcommand{\llN}{\ensuremath{\lambda_{1}}}
\newcommand{\loL}{\ensuremath{{\lambda}_{0}}}
\newcommand{\llL}{\ensuremath{{\lambda}_{1}}}
\newcommand{\loM}{\ensuremath{\tilde{\lambda}_{0}}}
\newcommand{\llM}{\ensuremath{\tilde{\lambda}_{1}}}
\newcommand{\loLM}{\ensuremath{\tilde{\lambda}_{0}}}
\newcommand{\llLM}{\ensuremath{\tilde{\lambda}_{1}}}

\theoremstyle{definition}
\newtheorem{definition}{Definition}[section]
\newtheorem{lemma}{Lemma}

\definecolor{mygray}{gray}{0.85}

\usepackage{xr}
\begin{document}

\def\spacingset#1{\renewcommand{\baselinestretch}%
{#1}\small\normalsize} \spacingset{1}


\if1\blind
{
  \title{\bf Heterogeneous large datasets integration using Bayesian factor regression}
  \author{
  Alejandra Avalos-Pacheco\vspace{-.35cm}\\
    Dept.\ of Statistics, University of Warwick, Coventry, United Kingdom\\
   Harvard Medical School, Boston, United States of America\\
    David Rossell \\
    Dept.\ of Business and Economics, Universitat Pompeu Fabra,  Barcelona, Spain\\
    and \\
    Richard S. Savage \\
    Dept.\ of Statistics, University of Warwick, Coventry, United Kingdom\\}
  \maketitle
} \fi

\if0\blind
{
  \bigskip
  \bigskip
  \bigskip
  \begin{center}
    {\LARGE\bf Heterogeneous large datasets integration using Bayesian factor regression}
\end{center}
  \medskip
} \fi

\bigskip
\begin{abstract}
Two key challenges in modern statistical applications are the large amount of information recorded per individual, and that such data are often not collected all at once but in batches.  
These batch effects can be complex, causing distortions in both mean and variance.  
We propose a novel sparse latent factor regression model to integrate such heterogeneous data.  
The model provides a tool for data exploration via dimensionality reduction while correcting for a range of batch effects.  
We study the use of several sparse priors (local and non-local) to learn the dimension of the latent factors. 
Our model is fitted in a deterministic fashion by means of an EM algorithm for which we derive closed-form updates, contributing a novel scalable algorithm for non-local priors of interest beyond the immediate scope of this paper. 
We present several examples, with a focus on bioinformatics applications. 
Our results show an increase in the accuracy of the dimensionality reduction, with non-local priors substantially improving the reconstruction of factor cardinality, as well as the need to account for batch effects to obtain reliable results. 
Our model provides a novel approach to latent factor regression that balances sparsity with sensitivity and is highly computationally efficient.
\end{abstract}

\noindent%
\vfill

\newpage
\spacingset{1.3}
\section{Introduction}
A first important task when dealing with large datasets is to conduct an exploratory analysis. 
Dimensionality reduction techniques have proven a highly popular tool for this purpose. 
Those techniques provide a lower-dimensional representation that can give insights into the underlying structure to visualise, denoise or extract meaningful features from the data.
See  \citet[chap. 3]{Johnson1988} or \citet[chap. 14]{statisticallearning} for a gentle introduction and 
 \citet{Burges10, cunningham2015ldr} for more recent reviews.\par
Large datasets are common in modern statistical applications. 
For instance, technological advances in bioinformatics such as high-throughput sequencing, microarrays, mass spectrometry and single cell genomics allow the gathering of a vast amount of biological data, enabling researchers to create models to explain the complex processes and interactions of biological systems (see \citet{Bersanelli2016} for a recent review). 
Cancer is a prominent example. 
Large-scale projects such as The Cancer Genome Atlas (TCGA), Cancer Genome Project (CGP) and the International Cancer Genome Consortium (ICGC), as well as many individual laboratories are generating extensive amounts of biological data (e.g.\ gene expression, mutation annotation, DNA methylation profiles, copy number changes) in addition to recording other covariates (e.g. gender, tumour stage, medical treatment and patient history). 
These projects aim to give a better understanding of the disease and improve prognosis, prevention and treatment. 
However, the large and heterogeneous nature of the data make the analyses and interpretations challenging.
Furthermore, such data are often generated under different experimental conditions, when new samples are incrementally added to existing samples, or in analyses coming from different projects, laboratories, or platforms; collecting data in this matter often produces batch effects \citep{Rhodes2004}.
These, unless properly adjusted for, may lead to incorrect conclusions \citep{Leek2010, BatchEffectsMatter2017}.
In the context of bioinformatics, several approaches have been developed for removing batch effects (see \citet{scherer2009} for a review and examples).
These include data ``normalization" methods using control metrics or regression methods \citep{Schadt2001, Yang2002}, matrix factorisation \citep{Alter2000, Benito2004} and location-scale methods \citep{Leek2007, BEcombat2009, Parker2014, Hornung2016}.
Strategies for batch effect correction include data pre-processing, for example via the so-called ComBat empirical Bayes approach \citep{COMBAT2007} or via singular value decomposition (SVD) \citep{Leek2007}.
As shown in our examples applying standard dimension reduction methods on such normalized data can produce unreliable results.
Intuitively this is due to using a two-step rather than a joint inference procedure on batch effects and dimension reduction.
Our examples focus on cancer-related gene expression; nonetheless, batch effects are also present in many other settings, e.g. structural magnetic resonance imaging (MRI) data from Alzheimer's  disease \citep{BrainImage2014, BrainImage2016}, multiple sclerosis \citep{SHAH2011267}, attention deficit hyperactivity disorder \citep{BrainImage2012} or even different tissues of marine mussels \citep{AVIO2015211}.\par
We address dimensionality reduction via a model-based framework relying on Bayesian factor analysis and latent factor regression. 
Our model builds on the approaches introduced by \citet{Lopes2004, Lucas2006, Carvalho2008} and \citet{Rockova2016}.
An important practical extension of these works is to increase the flexibility to account for systematic biases or sources of variation that do not reflect any underlying patterns of interest, i.e.\ batch effects. 
Our main contribution is to provide a model-based approach for tackling dimensionality reduction and batch effect correction simultaneously, avoiding the use of two-step procedures.
Another important contribution is to develop a scalable non-local prior based formulation to induce sparsity and learn the underlying number of factors; for this we provide a prior parameter elicitation, of practical importance to increase power to detect non-zero loadings.
A strategy related to ours is to use factor models to learn, on the one hand, the biological patterns via common factors shared across the different data sources and, on the other hand, the non-common sources of variation via data-specific factors \citep{RobertaMFA, RobertaBMFA}.
However, such a strategy is not designed for batch effects and requires MCMC estimation, making the inference slower.
Another related approach is to regress the covariance on batches and other explanatory variables, either parametrically or non-parametrically \citep{Hoff2012, JMLR:v16:fox15a}.
While useful, this method is not focused on dimension reduction and does not lead to sparse factor loadings that facilitate interpretation and, as shown in our examples, can improve inference.

We model observations with a regression on latent factors with sparse loadings, observed covariates, and batch effects that can alter the mean and intrinsic variance structures. 
Model fitting is done via a novel Expectation-Maximisation (EM) algorithm to obtain maximum posterior mode parameter estimates in a computationally efficient manner.
We focus on three different continuous prior formulations for the loadings: flat, Normal-spike-and-slab \citep{George93} and a novel Normal-spike-and-MOM-slab, based on a continuous relaxation of the non-local prior configuration by \cite{Rossell2010, Rossell2012}. 
We also discuss non-local Laplace-tailed extensions, along the lines of \citet{Rockova2016}. 
Spike-and-slab priors provide sparse loadings, effectively performing model selection on the number of required factors and non-zero loadings.
We obtain closed-form EM updates, a novel contribution to the non-local prior literature. 
As we will discuss later, the main advantage of non-local priors in this setting is to help achieve a better balance between sparsity and sensitivity in inferring non-zero loadings. 
To our knowledge, this is the first adaptation of non-local priors to factor models.
See also~\citet{Bar2018} who argued for improved sensitivity via 3-component mixture priors that resemble non-local priors in generalised linear models,
and \citet{Shi2019} for an application to linear regression via Gibbs sampling.\par
Our work is meant to contribute to applied aspects in dimension reduction that we show via examples to be of practical relevance,
as well as computational aspects related to high-dimensional sparse models that facilitate deploying non-local priors to applications.
As a motivating example, Figure~\ref{Cancer} (top row) displays systematic differences in mean and variance, thus showing the problem of not accounting for batch effects. 
After two-step procedures most of these differences are corrected, but distinct covariances are still present across batches (see rows 2 and 3).

The outline of this paper is as follows. 
Section \ref{sec:basicModel} reviews latent factor regression and introduces our extension, which includes a variance batch effect adjustment.
Section \ref{sec:Prior} proposes prior formulations including non-local priors on the loadings and important aspects related to prior parameter elicitation.
Section \ref{sec:EstimationNS} describes several EM algorithms for model fitting, parameter initialisation and post-processing steps required for effective model selection and dimension reduction.
Section \ref{sec:Results} presents applications on simulations and on cancer datasets under unsupervised and supervised settings. 
Section \ref{sec:Conclusion} concludes.
The supplementary material contains the derivation of the EM algorithm and additional results. 
Software implementing our methodology is available at \url{https://github.com/AleAviP/BFR.BE}. 
\section{Latent factor regression with batch effects}
\label{sec:basicModel}
Consider vectors $\xv_i = (x_{i1},x_{i2},\dots,x_{ip}) \in \R^{p}$, observed for $i=1,\dots,n$ individuals. 
The factor regression model defines $\xv_i$ as a regression on $p_v$ observed covariates denoted by $\vv_i \in \R^{p_v}$, and $q$ low-dimensional latent variables denoted $\zv_i \in \R^q$, also known as latent coordinates or factors. 
Let $X$ be the $n \times p$ matrix with the $i^{\text{th}}$ row equal to $\xv_i^\top$, $V$ the $n \times p_v$ matrix of known covariates with the $i^{\text{th}}$ row equal to $\vv_i^\top$ and $Z$ the $n \times q$ matrix of latent coordinates, containing $\zv_i^\top$ in the $i^{\text{th}}$ row.
The standard factor regression model is
\begin{align}
\label{LM}
\xv_i=\theta \vv_i + M \zv_i + \ev_{i},
\end{align}
where $\theta \in \R^{p \times p_v}$ is the matrix of regression coefficients, $M \in \R^{p \times q}$ is the matrix of factor loadings and $\ev_i \in \R^p$ is the error, distributed as $\ev_i \sim N(0, \Sigma)$ independently across $i=1,\dots,n$, where $\Sigma$ is a diagonal matrix. 
Factors are assumed to be standard normal, $\zv_i \sim N(0,\I)$, independent across $i=1,\dots,n$ and also independent of $\ev_i$.

Equation \eqref{LM} regresses the observed data $X$ on known covariates and on a latent factor structure. In particular, it allows additive batch effects to be accounted for by incorporating the variables recording the batches into $\vv_i$. 
However, 
in practice one often observes more complex batch effects; specifically in bioinformatics it is common to observe multiplicative effects on the variance \citep{COMBAT2007}.
We will later describe an example of this, shown in Figure~\ref{Cancer}.
Such artefacts cannot be captured by \eqref{LM} given that $\Sigma$ is assumed constant across all individuals.

To address this issue we extend \eqref{LM} by allowing $\Sigma$ to depend on $i$. 
Suppose the data were obtained in $p_b$ batches, e.g.\ from different days,  laboratories or instrumental calibrations, with $n_{l}$ individuals in batch $l$, for $l=1,\dots,p_b$, such that $n_{1} + n_{2} + \dots + n_{p_b} = n$. 
Let $\bv_i$ be the indicator vector of length ${p_b}$  defined as $b_{il} :=1$  if individual $i$ is in batch $l$, $b_{il} :=0$ otherwise.

We incorporate batch effects by adding a mean and variance adjustment. We let
\begin{align}
\label{BLM}
\xv_i=\theta \vv_i + M \zv_i + \beta \bv_i + \ev_i,
\end{align}
where $\theta$, $\vv_i$, $M$ and $\zv_i$ are as (\ref{LM}), $\beta \in \R^{p \times p_b}$ captures additive batch effects and the variance of $\ev_i$ captures multiplicative batch effects.
We denote by $\tau_{jl}$, $j=1,\dots,p$ and $l=1,\dots,p_b$ as the $j^{th}$ idiosyncratic precision element in batch $l$. 
Then, given $b_{il}=1$, the errors are independently distributed as $\ev_{ij} \sim N(0, \tau_{jl}^{-1})$. 
Further, denote by $\Tau$ the $p \times p_b$ matrix that has $\tau_{jl}$ as its $(j,l)$ element.

To help interpret the practical implications of the model, suppose that one has orthonormal factor loadings $M^\top M = \I$. Then \eqref{BLM} implies
\begin{align}
\zv_i=&M^{\top}\left(\xv_i-(\theta \vv_i+ \beta \bv_i + \ev_i)\right)
\end{align}
and thus, $\Ex(\zv_i \mid \xv_i, \vv_i, \bv_i, M, \theta, \beta)=M^{\top}\xv_i-M^{\top}\theta \vv_i-M^{\top}\beta \bv_i$. That is, the mean of the latent coordinates is the projection $M^{\top}\xv_i$ plus a translation given by the batch effect adjustment and (potentially) the observed covariates. 
An interesting observation is that their covariance Cov$(\zv_i \mid \xv_i, \vv_i, \bv_i, M, \theta, \beta, \Tau)=M^\top \Sb^{-1} M$ depends on the multiplicative batch-dependent noise.
As an example, the middle-left panel in Figure~\ref{Cancer} show the two first factors of an ovarian dataset pre-processed by ComBat. Relative to the unadjusted upper-left panel, ComBat removes systematic differences in mean and variance across the 2 batches, however the latent coordinates exhibit distinct covariances. To obtain suitably-adjusted low-dimension coordinates one should estimate $\Tau$ jointly with $(M, \theta,\beta)$.\par
Model (\ref{BLM}) can be represented in matrix notation as
\begin{align}
\label{ModelBatch}
X=& V \theta^\top + Z M^\top + B \beta^\top + E,
\end{align}
where $E \in \R^{n \times p}$  is the matrix of errors.\par
The latent factor model is non-identifiable up to orthogonal transformations, of the form $M^{*\top} = A^\top M^\top$ and $Z^{*} = ZA$, where $A$ is any orthogonal $q \times q$ matrix. 
Thus, the factor model in (\ref{ModelBatch}) can equivalently be rewritten as $X=V \theta^\top + Z^{*} M^{*\top} + B \beta^\top + E$. 
To obtain unique point estimates of $M$ and $Z$, several alternative prior specifications have been developed. 
One option is restricting the parameter space. 
\citet{Seber84} constrained $M$ such that $M^\top\Omega M$ is diagonal. \citet{Lopes2004} restricted $M$ to be lower-triangular with a strictly positive diagonal, $m_{jj}>0$, and assumed $M$ to be full-rank. 
More recently, \citet{Lopes2018} suggested a factor reordering via a Generalized Lower Triangular loading matrix.
However, under this approach the interpretation of $M$ depends on the arbitrary ordering of the columns in $X$, and it gives special roles to the first factors. 
Another option is to encourage sparsity in $M$, e.g.\ the classical varimax solution \citep{Kaiser1958} maximises the variance in the squared rotated loadings. 
A more modern strategy is to favour sparse solutions containing exact zero loadings, e.g. \citet{Rockova2016} proposed an EM algorithm that seeks rotations based on a so-called Parameter Expansion (PX) that aims to avoid local suboptimal regions. 
We adopt a similar strategy where sparse solutions are prefer by the introduced non-local penalties.
\section{Prior formulation}
\label{sec:Prior}
To complete Model~\eqref{BLM} we set priors for the loadings $M$, precisions $\tau_{jl}$, and regression parameters $(\theta,\beta)$. 
Through our proposed default prior formulation we assume that the columns in $X$ have been centred to zero mean and unit variance.
For the idiosyncratic precisions $\tau_{jl}$ we set
\begin{align}
\label{PriorSigma}
\tau_{jl} \mid \eta, \xi \sim & \text{ Gamma}(\eta/2,\eta \xi/2) 
\end{align}
independently across $j=1,\dots,p$ and $l=1,\dots,p_b$. By default in our examples we set the fairly informative values $\eta = \xi = 1$, leading to diffuse though proper priors.

For the regression parameters we set
\begin{align}
\label{PriorTheta}
(\theta_j,\beta_j) \sim & N(0,\psi \I),  \: \: \: \:j=1,\dots,p 
\end{align}
where $\psi$ is a user-defined prior dispersion that in our examples by default we set to $\psi=1$. 
The choice of $\psi = 1$ assigns the same marginal prior variances to elements in $(\theta_j,\beta_j)$ as the unit information prior often adopted as a default for linear regression \citep{schwarz1978}.

We remark that this prior does not encourage sparsity in the regression parameters $(\theta, \beta)$ or factor loadings, which we view as reasonable provided the number of variables $p_v$ and batches $p_b$ are moderate.
For large $p_v$ or $p_b$, a direct extension of our prior on the loadings $M$ could be adopted.\par
The loadings matrix $M$ plays an important role in improving shrinkage and simplifying interpretation. 
Some recent strategies include a LASSO-based method \citep{HastieT2009}, horseshoe priors \citep{Carvalho2009}, an Indian buffet process \citep{knowles2011nonparametric}, an infinite factor model  \citep{dunson11} among others. 
In this paper, we consider three priors on the loadings: an improper flat prior $\Prob(M) \propto 1$, a Normal spike-and-slab and a novel non-local pMoM spike-and-slab. 
The local and non-local spike-and-slab prior formulations are detailed bellow, along with Laplace-based extensions. 
These build on the approach by \citet{GeorgeEMVS, Rockova2016}, our main contribution being the introduction of non-local-based variations. 
\subsection{Local spike-and-slab prior}
\label{sec:SSL}
A traditional Bayesian approach to variable selection is the spike-and-slab prior, a two-component mixture prior \citep{Mitchell88, George93}. 
This prior aims to discriminate those loadings that warrant inclusion, modelled by the slab component,  from those that should be excluded, modelled by the spike component.

Specifically, a spike-and-slab prior density for the loadings $M$ has the form
\begin{equation}
\label{eq:localSS}
\Prob(M \mid \gamma, \lambda_0, \lambda_1 ) = \prod_{j=1}^{p}\prod_{k=1}^{q} (1-\gamma_{jk})\Prob(m_{jk} \mid \loN,\gamma_{jk}=0) + \gamma_{jk} \Prob(m_{jk} \mid \llN,\gamma_{jk}=1),
\end{equation}
where $\Prob(m_{jk} \mid \loN, \gamma_{jk} = 0)$ is a continuous density, $\loN$ is a given dispersion parameter of the spike component and $\llN> \loN$ is that of the slab component. 
The indicators $\gamma_{jk} \in \{0,1\}$ signal which $m_{jk}$ were generated by each component, and serve as a proxy for which loadings are significantly non-zero. 
We take as a base formulation the Normal-spike-and-slab prior by \citet{George93} were the spike is a Normal density with a small variance $\loN$ and the slab a Normal distribution with large variance $\llN$. 
Although Laplace-Spike-and-Slab priors have been shown to possess better properties for sparse inference \citep{Rockova2018}, as discussed bellow the introduction of non-local penalties improves certain undesirable features of the Normal-based prior. 
The elicitation of $\loN$ and $\llN$ is an important aspect of the formulation and will be discussed in Section \ref{Elicitation}. 
Specifically, the Normal-spike-and-slab is
\begin{equation}
\Prob(m_{jk} \mid \gamma_{jk}=l,  \lambda_l) = N (m_{jk};0,\lambda_l),
\label{eq:SSLprior1}
\end{equation}
The continuity of the spike distribution gives closed form expressions for the EM algorithm, making it computationally appealing. 
We refer to \eqref{eq:SSLprior1} as Normal-SS. 

We complete the model specification with a hierarchical prior over the latent indicator $\gamma= \{\gamma_{jk},  j=1,\dots,p,  k=1,\dots,q \}$ as follows,
\begin{align}
\label{PriorGamma}
\gamma_{jk}\mid \zeta_k &\sim \text{Bernoulli} (\zeta_k), \nonumber \\
\zeta_k \mid a_\zeta,b_\zeta&\sim \text{Beta}\left(\frac{a_\zeta}{k},b_\zeta\right),
\end{align}
with independence across $(j,k)$ where $a_\zeta>0$ and $b_\zeta>0$ are given prior parameters. 
By default we set $a_\zeta = b_\zeta=1$, which leads to a uniform prior for the first factor ($k=1$), $\zeta_k \mid a_\zeta,b_\zeta \sim \text{U}(0,1)$. 
Furthermore, note that $\frac{a_\zeta}{k}$ encourages increasingly sparse solutions in subsequent factors.
That is, related to our earlier discussion of non-identifiability (Section \ref{sec:basicModel}), we encourage loadings where the first factors have larger importance, leading to solutions that are sparse both in the rank of $M$ and its non-zero entries.

A potential concern with Normal-SS is that the slab density assigns non-negligible probability to regions of the parameter space that are also consistent with the spike, namely when $m_{jk}$ lies close to zero. 
We will address this via non-local priors and show that these, by enforcing separation between two components, help increase sensitivity.
\subsection{Non-local spike-and-slab prior}
\label{sec:pMoM}
Non-local priors (NLPs) are a family of distributions that assign vanishing prior density to a neighbourhood of the null hypothesis \citep{Rossell2010}. Definition~\ref{def:NLP} is an adaptation of the definition in \citet{Rossell2010} to \eqref{eq:localSS}. \par
\begin{figure}[h!]
\label{fig:NLPvsLP}
\centering
	\includegraphics[width=17cm]{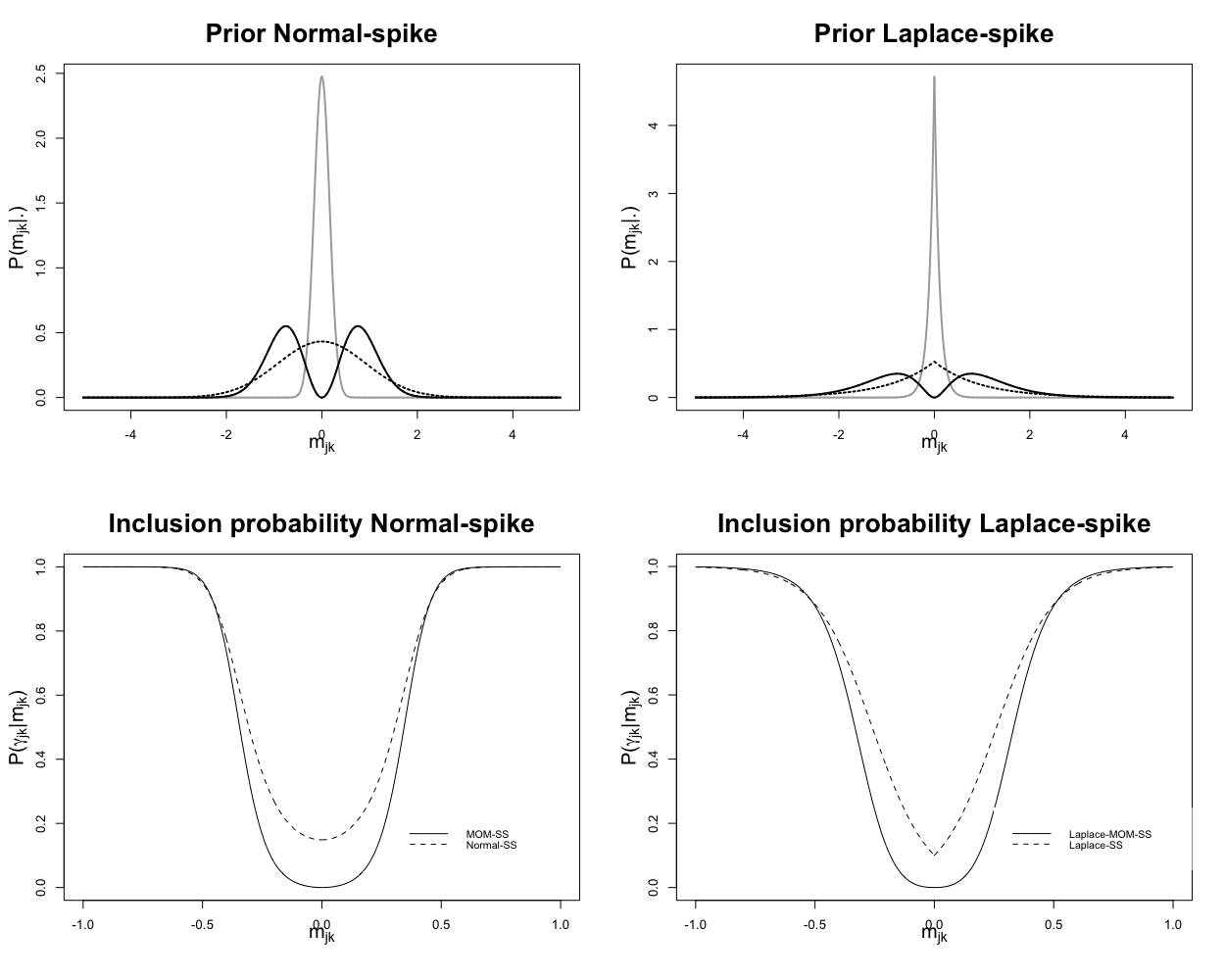}
	\caption{Prior comparison (top panels) for $m_{jk}$ under different prior specifications and its inclusion probabilities $\Prob(\gamma_{jk}\mid m_{jk})$ (bottom panels). Comparison between Normal-based (left) and Laplaced-based (right) priors.
Scales ($\loN,\llN$) are set to the defaults from Section~\ref{Elicitation}.}
	\label{fig:priors}
\end{figure}

\begin{definition}
\label{def:NLP}
An absolutely continuous measure with density $\Prob(m_{jk} | \gamma_{jk}=1)$ is a non-local prior if $\lim_{m_{jk} \to 0} \Prob(m_{jk} | \gamma_{jk}=1) = 0$.
\end{definition}
We call any prior not satisfying Definition~\ref{def:NLP} a local prior.
Non-local priors possess appealing properties for Bayesian model selection. 
They discard spurious parameters faster as the sample size $n$ grows, but preserve exponential rates to detect important coefficients \citep{Rossell2010, Jairo2016} and can lead to improved parameter estimation shrinkage \citep{Rossell2017}. 
To illustrate the motivation for NLPs in our setting consider Figure~\ref{fig:priors}. 
Normal-SS assigns positive probability to $m_{jk} = 0$.
Correspondingly, the conditional inclusion probability $\Prob(\gamma_{jk}=1 \mid m_{jk})$ remains non-negligible, even when $m_{jk}=0$ (lower left panel). 

As an alternative, we consider a product moment (pMOM) prior \citep{Rossell2012}.
\begin{equation}
\label{eq:priorpMOM}
\begin{split}
\Prob(m_{jk} \mid \gamma_{jk} = 0, \loM)&=\text{N}(m_{jk}; 0, \loM), \\
\Prob(m_{jk} \mid \gamma_{jk} = 1, \llM) &= \frac{m_{jk}^2}{\llM} \text{N}(m_{jk}; 0,\llM).
\end{split}
\end{equation}
We denote \eqref{eq:priorpMOM} as MOM-SS.
This prior assigns zero density to $m_{jk} = 0$ given $\gamma_{jk} = 1$, which implies $\Prob(\gamma_{jk}=1 \mid m_{jk}=0)=0$ (Figure \ref{fig:priors}). 
Prior elicitation for $\tilde{\lambda}_0$ and $\tilde{\lambda}_1$ is discussed in Section \ref{Elicitation}.
From a computational point of view, the EM algorithm can accommodate this extension by using a trivial extra gradient evaluation at negligible additional cost relative to the Normal-SS. 
Parameter estimation and algebraic details are described in Section \ref{sec:EstimationNS}. The prior on the inclusion indicators is set as in \eqref{PriorGamma}.

Beyond \eqref{eq:SSLprior1} and \eqref{eq:priorpMOM}, another natural extension is to use Laplace-based priors based on the Spike-and-Slab LASSO by \citet{Rockova2018}
\begin{align}
\Prob(m_{jk} \mid \gamma_{jk},  \loL, \llL) & = (1-\gamma_{jk})\text{Laplace} (m_{jk}; 0,\loL) + \gamma_{jk} \text{Laplace} (m_{jk}; 0,\llL),
\label{eq:SSLaplace}
\end{align}
with a slab component with variance $2 \loL^2$, and a spike component with $2 \llL^2$, where $\text{Laplace} (m_{jk}; 0,\lambda)=\frac{1}{2 \lambda} \exp \left(\frac{-\mid m_{jk} \mid}{\lambda} \right)$. 
We refer to \eqref{eq:SSLaplace} as Laplace-SS. 
As illustrated in Figure~\ref{fig:priors} (right panels) this prior can help encourage sparsity, setting $\Prob(\gamma_{jk}=1 \mid m_{jk}=0)$ to 
smaller values (though still non-zero) than the Normal-SS.

As an extension, akin to \eqref{eq:priorpMOM}, one could set a moment penalty on the Laplace density.
\begin{equation}
\label{eq:priorLMOM}
\begin{split}
\Prob(m_{jk} \mid \gamma_{jk} = 0, \loLM)&=\text{Laplace}(m_{jk}; 0,\loLM), \\
\Prob(m_{jk} \mid \gamma_{jk} = 1, \llLM) &= \frac{m_{jk}^2}{2\llLM^2} \text{Laplace}(m_{jk}; 0,\llLM).
\end{split}
\end{equation}
We denote \eqref{eq:priorLMOM} as Laplace-MOM-SS. 
Relative to \eqref{eq:priorpMOM}, as illustrated in Figure~\ref{fig:priors},  Laplace-MOM-SS leads to lower $\Prob(\gamma_{jk}=1 \mid m_{jk}=0)$ and higher $\Prob(\gamma_{jk}=1 \mid m_{jk})$ for moderately large $m_{jk}$.

We discuss prior elicitation for Laplace-MOM-SS in Section~\ref{Elicitation} and derive an EM algorithm in Section~\ref{sec:SSLalgorithm} but in our examples we focus on the MOM-SS for simplicity.
However, the Laplace-based \eqref{eq:priorLMOM} can also be shown to lead to closed-form EM updates. 

\subsection{Prior elicitation for the variance of the spike-and-slab priors}
\label{Elicitation}
A crucial aspect in a spike-and-slab prior is the choice of the prior scale parameters. 
It is common to fix the variance of the spike distribution $\loN$ to a value close to zero. 
Regarding $\llN$, one option is to set a hyper-prior or to try to estimate it from the data \citep{George93, George97, GeorgeEMVS, Rockova2018}. 
Setting a hyper-prior does not bypass prior elicitation, as one then needs to set the hyper-prior parameters, whereas estimating $\llN$ from the data increases the cost of computations. 
Instead, we capitalise on the fact that factor loadings have a natural interpretation in terms of the fraction of explained variance in $X$.
Thus, we propose default values that dictate which coefficients are considered as meaningfully different from zero. 
These defaults are guidelines in the absence of a priori knowledge.
A convenient feature of such an elicitation is that it can be easily extended to local priors and other non-Gaussian spike-and-slab priors.

Our goal is to find values $\loM$ and $\llM$ for the MOM-SS that distinguish practically relevant factors.
In the absence of covariates, the factor model decomposes the total variance in variable $j$ as Var$(\xv_{ij})=\sum_{k=1}^qm^2_{jk}+\tau^{-1}_{jj}$, hence $m^2_{jk}$ is the proportion of variance in variable $j$ explained by factor $k$. 
We take $m_{jk}^2>0.1$ as a threshold for practical relevance.
Specifically, we set $\loM$ such that $\Prob(|m_{jk}| \leq \sqrt{0.1} \mid \loM)=0.95$, that is $\loM =\frac{0.1}{(\Phi^{-1}(0.025))^2}\approx 0.026$, where $\Phi^{-1}$ denotes the standard normal quantile function. 
Likewise we set $\Prob(|m_{jk}| \geq \sqrt{0.1}  \mid \llM)=0.95$ under the MOM-SS, obtaining the default $\llM \approx 0.2842$.\par
Regarding the Normal-SS prior,we set $\loN=\loM$ and $\llN$ such that it is comparable to the MOM-SS in terms of informativeness, namely it matches the variance of the MOM-SS, obtaining that $\llN= 3 \llM \approx 0.8526$. \par
In Laplace-MOM-SS, we analogously set $\loLM = -\frac{\sqrt{0.1}}{\log(0.05)}\approx 0.1056$ so that $\Prob(|m_{jk}| \leq \sqrt{0.1} \mid \loLM)$ and $\llLM \approx 0.3867$ such that $\Prob(|m_{jk}| \geq \sqrt{0.1} \mid \llLM)=0.95$ for the Laplace-spike-and-MOM-slab prior. 
Finally for the Laplace-SS we set $\llL = \sqrt{6} \llLM \approx 0.9473$ and $\loL = \loLM$ for the spike and slab component, respectively, matching the variances of the non-local Laplace-based priors.\par

The resulting priors are in Figure~\ref{fig:priors}. 
We remark that a considerable difference can be observed between the local prior based and the non-local prior based formulations, particularly in the conditional inclusion probability around $m_{jk}=0$.
In our examples we will focus on the Normal MOM-SS.
Deeper analysis of Laplace-based non-local priors, whose thicker tails might help improve estimation accuracy, is left for interesting future work.

\section{Parameter estimation}
\label{sec:EstimationNS}
Parameter estimation in factor analysis is usually conducted using Expectation-Maximisation (EM, \citet{Dempster77}), MCMC algorithms \citep{Lopes2004} or approximated via variational inference \citep{Ghahramani00}. 
At the core of these algorithms is the fact that, conditional on the data and all other model parameters, we can set $\tilde{\xv}_i = \xv_i - \theta \vv_i -\beta \bv_i$ and express the model in (\ref{BLM}) as a linear regression $\tilde{\xv}_i=M\zv_i+\ev_i$, where $M$ and  $\Sigma$ are fixed at their current values of each MCMC iteration or maximisation step \citep{West2003, Carvalho2008}. 
We develop a deterministic optimisation along the lines of the EM algorithm of \citet{Rockova2016}. 
Section~\ref{sec:NSalgorithm} provides two EM algorithms to obtain posterior modes for our factor regression with batch effect correction with and without sparse formulation. 
Section~\ref{sec:SSLalgorithm} outlines an algorithm separately for Normal-SS, MOM-SS, Laplace-SS and Laplace-MOM-SS priors. 
Section~\ref{IniEMNS} discusses parameter initialisation and Section~\ref{Postprocessing} how to post-process the fitted model to obtain sparse solutions and variance-adjusted dimensionality reduction.

\subsection{EM algorithm under a uniform prior}
\label{sec:NSalgorithm}
We outline an EM algorithm to fit Model (\ref{BLM}) under a uniform prior $\Prob(M) \propto 1$ on the loadings via maximum a posteriori (MAP) estimation.
The algorithm maximises the log-posterior by treating the latent factors $Z$ as missing data and setting them to their expectation (conditional on all other parameters) in the E-step. 
Then, the remaining parameters $\Delta = (M,\theta,\beta,\Tau)$ are optimised in the M-step. 
In other words, the EM algorithm obtains a local mode of the log-posterior $\Prob (M, \theta,\beta,\Tau \mid X)$ by maximising the expected complete-data log-posterior $\Prob (M, \theta,\beta,\Tau \mid X,Z)$ iteratively. 
For convenience we denote by $\Tau_{\bv_i}$ the idiosyncratic precision matrix in batch $l$, i.e. if $b_{il}=1$ by $\tau_{jl}$, then the errors are distributed as $\ev_i \sim N(0, \Tau_{\bv_i}^{-1})$. 
We also denote with $\hat{\Delta}=(\hat{M},\hat{\theta},\hat{\beta},\hat{\Tau})$ the current value of the parameters
We briefly describe the algorithm; see Supplementary Section~\ref{appendix:EMV1} for its full derivation.

The E-step takes the expectation of $\log\Prob(M, \theta, \beta, \Tau \mid X,Z) $ with respect to $\Prob(Z\mid \hat{\Delta},X)$
Specifically, let

\begin{equation}
\label{EMFlat}
\begin{split}
Q(\Delta)=& \Ez\left[ \log \Prob (M, \theta,\beta,\Tau \mid X, Z)\right] \\
=& C -\frac{1}{2} \sum_{i=1}^n \left[(\xv_i-\theta \vv_i - \beta \bv_i)^\top \Tau_{\bv_i} (\xv_i-\theta \vv_i - \beta \bv_i) \right.  \\
&\left. -2 (\xv_i-\theta \vv_i - \beta \bv_i)^\top \Tau_{\bv_i} M\Ex[\zv_i \mid\hat{\Delta},X]   + \Tr\left(M^{\top} \Tau_{\bv_i} M \Ex[\zv_i \zv_i^\top \mid\hat{\Delta},X]\right) \right] \\
&+ \sum_{l=1}^{p_b} \frac{n_l+ \eta  -2 }{2} \log \mid\Tau_{l}\mid -\sum_{l=1}^{p_b} \frac{\eta \xi}{2} \Tr( \Tau_{l} ) -\frac{1}{2} \sum_{j=1}^p (\theta_{j}^\top ,\beta_{j}^\top )\frac{1}{\psi}\I (\theta_{j},\beta_{j}),
\end{split}
\end{equation}
where $C$ is a constant. 
Expression \eqref{EMFlat} only depends on $Z$ through the conditional posterior mean 
\begin{equation}
\label{EZx1}
\Ex[\zv_i| \hat{\Delta},X] = (\I_q + \hat{M}^\top \hat{\Sb}\hat{M})^{-1} \hat{M}^\top\hat{\Sb}(\xv_i -\hat{\theta} \vv_i - \hat{\beta} \bv_i)
\end{equation}
and the conditional second moments
\begin{equation}
\label{EZZx1}
\Ex[\zv_i \zv_i^\top\mid\hat{\Delta},X] = (\I_q + \hat{M}^\top \hat{\Sb}\hat{M})^{-1}  +\Ex[\zv_i\mid\hat{\Delta},X]\Ex[\zv_i\mid\hat{\Delta},X]^\top,
\end{equation}
where $(\I_q + \hat{M}^\top \hat{\Sb}\hat{M})^{-1}=\text{Cov}[\zv_i| \hat{\Delta},X]$ is the conditional covariance matrix of the latent factors.
We emphasise that \eqref{EZx1} and \eqref{EZZx1} depend on batch-specific precisions $\Sb$.\par
The M-step maximises $Q(\Delta)$ with respect to $M,\theta,\beta,\Tau$. 
Setting its partial derivatives to 0 gives the updates
\begin{equation}
\label{MSigmaVanilla}
\hat{m}_j=\left[  \sum_{i=1}^n \left( \hat{\tau}_j^\top\bv_i \tilde{x}_{ij} \Ex[\zv_i^\top \mid\hat{\Delta},X] \right)  \right] \left[ \sum_{i=1}^n \left( \hat{\tau}_j^\top \bv_i \Ex[\zv_i \zv_i^\top \mid\hat{\Delta},X] \right) \right]^{-1}
\end{equation}
\begin{equation}
\label{TauSigmaVanilla}
\hat{\Tau}_l^{-1}=\frac{1}{n_l+ \eta  -2}\text{diag}\left\{\sum_{i\colon 
	b_{il} = 1}\left( \tilde{\xv}_i \tilde{\xv}_i^\top  -2\tilde{\xv}_i \Ex[\zv_i \mid\hat{\Delta},X]^\top  \hat{M}^\top + \hat{M}  \Ex[\zv_i \zv_i^\top \mid\hat{\Delta},X] \hat{M}^\top \right) + \eta \xi \I_p \right\}
\end{equation}
where $\tilde{\xv}_i=\xv_i-\hat{\theta} \vv_i - \hat{\beta} \bv_i$ and $\tilde{x}_{ij}=x_{ij}-\hat{\theta} v_{ij} - \hat{\beta} b_{ij}$. 

The updates for $(\theta_j, \beta_j)$ are
\begin{equation}
\label{ThetaFlatT}
(\hat{\theta}^{\top}_j, \hat{\beta}^{\top}_j) = \sum_{i=1}^n \left[ \hat{\tau}_{j}^{\top} \bv_i (x_{ij} - \hat{m}_{j}^{\top}  \Ex[\zv_{i} \mid\hat{\Delta},X] ) (\vv_i,\bv_i)^\top \right] \left[ \sum_{i=1}^n \left[ \hat{\tau}_{j}^{\top} \bv_i (\vv_i,\bv_i)(\vv_i,\bv_i)^{\top} \right]  +\frac{1}{\psi} \I\right]^{-1}
\end{equation}
Equation \eqref{ThetaFlatT} has the form of a ridge regression estimator with penalty $\psi$.\par
Algorithm \ref{alg:EMvanilla} summarises the EM algorithm.
The stopping criteria is reaching a tolerance $\epsilon^*$ in the log-posterior change, a maximum number of iterations $T$ or a change $\epsilon^*_M$ on the loadings. 
By default we set $\epsilon^* = 0.001$, $T = 100$ and $\epsilon^*_M = 0.05$. 
Parameter initialisation is an important aspect that helps obtain better local modes and reduce computational time; its discussion is deferred to Section~\ref{IniEMNS}.\par
\RestyleAlgo{boxruled}
\begin{algorithm}[h!]
 \textbf{initialise} $\hat{M}=M^{(0)}$, $\hat{\theta}=\theta^{(0)}$, $\hat{\beta}=\beta^{(0)}$, $\hat{\Sb}=\Sb^{(0)}$\\
 \While{$\epsilon>\epsilon^*$, $\epsilon_M>\epsilon^*_M$ and $t<T$}{
  \textbf{E-step}:\\
  \begin{tabular}{rl}
Latent factors: & $\Ex[\zv_i| \hat{\Delta},X] = (\I_q + \hat{M}^\top \hat{\Sb}\hat{M})^{-1} \hat{M}^\top\hat{\Sb}(\xv_i -\hat{\theta} \vv_i - \hat{\beta} \bv_i)$
\end{tabular}\\

  \textbf{M-step}:\\ 
  \begin{tabular}{rl}
Loadings: & {\scriptsize $\hat{m}_j=\left[  \sum_{i=1}^n \left( \hat{\tau}_j^\top\bv_i \tilde{x}_{ij} \Ex[\zv_i^\top \mid\hat{\Delta},X] \right)  \right] \left[ \sum_{i=1}^n \left( \hat{\tau}_j^\top \bv_i \Ex[\zv_i \zv_i^\top \mid\hat{\Delta},X] \right) \right]^{-1}$}\\
Variances: &\scriptsize $\hat{\Tau}_l^{-1}=\frac{1}{n_l+ \eta  -2}\text{diag}\left\{\sum_{i\colon 
	b_{il} = 1}\left( \tilde{\xv}_i \tilde{\xv}_i^\top  -2\tilde{\xv}_i \Ex[\zv_i \mid\hat{\Delta},X]^\top  \hat{M}^\top + \hat{M}  \Ex[\zv_i \zv_i^\top \mid\hat{\Delta},X] \hat{M}^\top \right) + \eta \xi \I_p \right\}$ \\
Coefficients:& {\scriptsize $(\hat{\theta}^{\top}_j, \hat{\beta}^{\top}_j) = \sum_{i=1}^n \left[ \hat{\tau}_{j}^{\top} \bv_i (x_{ij} - \hat{m}_{j}^{\top}  \Ex[\zv_{i} \mid\hat{\Delta},X] ) (\vv_i,\bv_i)^\top \right] \left[ \sum_{i=1}^n \left[ \hat{\tau}_{j}^{\top} \bv_i (\vv_i,\bv_i)(\vv_i,\bv_i)^{\top} \right]  +\frac{1}{\psi} \I\right]^{-1}$}
\end{tabular}

\textbf{set} $\Delta^{(t+1)}= \hat{\Delta}$ and $M^{(t+1)}=\hat{M}$ 
 
\textbf{compute} $\epsilon=Q(\Delta^{t+1}) - Q(\Delta^t)$, $\epsilon_M =\max |m_{jk}^{(t+1)}-m_{jk}^{(t)}|$ and $t=t+1$ 
 }
 \caption{EM algorithm for factor regression model with uniform $\Prob(M)$}
 \label{alg:EMvanilla}
\end{algorithm}

\subsection{EM algorithm for spike-and-slab priors}
\label{sec:SSLalgorithm}
The algorithm is derived analogously to Section \ref{sec:NSalgorithm}. 
The expected complete-data log-posterior can be split into $Q(\Delta) =C+ Q_1(\theta, M, \beta, \Tau)+Q_2(\zeta)$,  where

\begin{align}
\label{eq:QSSL}
Q_1(\theta, M, \beta, \Tau)=
&-\frac{1}{2} \sum_{i=1}^n \left[(\xv_i-\theta \vv_i - \beta \bv_i)^\top \Tau_{\bv_i} (\xv_i-\theta \vv_i - \beta \bv_i) -2 (\xv_i-\theta \vv_i - \beta \bv_i)^\top \Tau_{\bv_i}M\Ex[\zv_i \mid\hat{\Delta},X]   \right. \nonumber \\
&\left.  + \Tr\left(M^{\top} \Tau_{\bv_i} M \Ex[\zv_i \zv_i^\top \mid\hat{\Delta},X]\right) \right]+ \sum_{l=1}^{p_b} \frac{n_l+ \eta  -2}{2} \log \mid\Tau_{l}\mid -\sum_{l=1}^{p_b} \frac{\eta \xi}{2} \Tr( \Tau_{l} ) \nonumber \\
& -\frac{1}{2} \sum_{j=1}^p (\theta_{j},\beta_{j})^\top \frac{1}{\psi}\I (\theta_{j},\beta_{j})  +\sum_{j=1}^p \sum_{k=1}^q \Eg \left[ \log \Prob (m_{jk} \mid \gamma_{jk} , \lambda_0, \lambda_1) \right],
\end{align}

\begin{equation}
Q_2(\zeta)=\sum_{j=1}^p\sum_{k=1}^q  \log\left( \frac{\zeta_k}{1-\zeta_k}\right)  \Ex [\gamma_{jk}\mid \hat{\Delta}] +\sum_{k=1}^q \left((\frac{a_\zeta}{k}-1) \log(\zeta_k)+(p+b_\zeta-1) \log(1-\zeta_k)\right) .
\end{equation}
with $C$ a constant and $\Ex[\zv_i\mid\hat{\Delta},X]$ and $\Ex[\zv_i\zv_i^\top\mid\hat{\Delta},X]$ as in \eqref{EZx1} and \eqref{EZZx1}.

$Q_1(\theta, M, \beta, \Tau)$ resembles the E-step for the flat prior in Section \ref{sec:NSalgorithm}, plus an extra conditional expectation $\Eg \left[ \log \Prob (m_{jk} \mid \gamma_{jk} , \lambda_0, \lambda_1) \right]$. 
$Q_2(\zeta)$ arises from the Beta-Binomial prior on $\gamma_{jk}$ and the $\Ex[\gamma_{jk} \mid \cdot]$ are straightforward to compute. 
In the M-step we maximise $Q_1$ w.r.t.\ $(\theta, M, \beta, \Tau)$, this can be
done in a completely independent fashion from optimising $Q_2$ w.r.t.\ $\zeta$.\par
Further the conditional expectation of $\Ex[\gamma_{jk} \mid \hat{\Delta}] =\hat{p}_{jk}$ is\par
\begin{equation}
\label{eq:Inclusion}
\hat{p}_{jk} =\frac{\Prob(\hat{m}_{jk} \mid \gamma_{jk}=1, \loN, \llN) \Prob(\gamma_{jk}=1)}{\Prob(\hat{m}_{jk} \mid \gamma_{jk}=0, \loN, \llN) \Prob(\gamma_{jk}=0) + \Prob(\hat{m}_{jk} \mid \gamma_{jk}=1, \loN, \llN) \Prob(\gamma_{jk}=1)}.
\end{equation}
For the Normal-SS prior, Equation~\eqref{eq:Inclusion} is
\begin{equation}
\label{eq:InclusionSSL}
\hat{p}_{jk}=\left[1+\sqrt{\frac{\llN}{\loN}} \exp\left(-\frac{1}{2} \hat{m}_{jk}^2 \left( \frac{1}{\loN} - \frac{1}{\llN} \right) \right) \frac{1-\Ex[\zeta_j]}{\Ex[\zeta_j]}\right]^{-1},
\end{equation}
for the MOM-SS 
\begin{align}
\label{eq:InclusionMOM}
\hat{p}_{jk} = \left[1+ \frac{\llM}{\hat{m}^2_{jk}}\sqrt{\frac{\llM}{\loM}} \exp\left(-\frac{1}{2} \hat{m}_{jk}^2 \left( \frac{1}{\loM} - \frac{1}{\llM} \right) \right) \frac{1-\Ex[\zeta_j]}{\Ex[\zeta_j]}\right]^{-1},
\end{align}
for the Laplace-SS 
\begin{equation}
\label{eq:InclusionSSLap}
\hat{p}_{jk} =\left[1+\frac{\llL}{\loL} \exp\left(- \mid \hat{m}_{jk} \mid \left( \frac{1}{\loL} - \frac{1}{\llL} \right) \right) \frac{1-\Ex[\zeta_j]}{\Ex[\zeta_j]}\right]^{-1},
\end{equation}
and for the Laplace-MOM-SS
\begin{equation}
\label{eq:InclusionSSM}
\hat{p}_{jk} =\left[1+\frac{2 \llLM^2}{\hat{m}^2_{jk}}\frac{\llLM}{\loLM} \exp\left(- \mid \hat{m}_{jk} \mid \left( \frac{1}{\loLM} - \frac{1}{\llLM} \right) \right) \frac{1-\Ex[\zeta_j]}{\Ex[\zeta_j]}\right]^{-1}.
\end{equation}
Equations~\eqref{eq:InclusionSSL} and \eqref{eq:InclusionSSLap} are analogous to the EM posterior update for $m_{jk}$ in a two-component Gaussian or Laplace mixture \citep{GeorgeEMVS}.
Equations~\eqref{eq:InclusionMOM} and \eqref{eq:InclusionSSM} are similar to their local counterparts, but incorporate a penalty for small $m^2_{jk}$.

The main difference between the local and non-local priors lies in updating the loadings and the idiosyncratic variances. 
We discuss these separately for each prior later in this section.

The updates for the precision $\Tau_l$ and the regression parameters $(\theta, \beta)$ are given in Equations \eqref{TauSigmaVanilla} and \eqref{ThetaFlatT} respectively.

Maximising $Q_2(\zeta)$ with respect to $\zeta_k$ gives
\begin{align}
\hat{\zeta}_k=\frac{\sum_{j=1}^p \hat{p}_{jk} + \frac{a_\zeta}{k} -1}{\frac{a_\zeta}{k} + b_\zeta + p -1}
\end{align}
for $k = 1, \dots , q$. \par
Algorithm \ref{alg:EMSSL}  summarises the algorithm. 
It is initialised with the two-stage least-squares method described in Section \ref{IniEMNS} and $\zeta_k = 0.5$ for $k = 1, \dots, q$. 
The stopping criteria are as in Algorithm \ref{alg:EMvanilla}. 
The different updates for $M$ are outlined below, separately for each prior specification.\par
\RestyleAlgo{boxruled}
\begin{algorithm}[h!]
 \textbf{initialise} $\hat{M}=M^{(0)}$, $\hat{\theta}=\theta^{(0)}$, $\hat{\beta}=\beta^{(0)}$, $\hat{\Sb}=\Sb^{(0)}$, $\hat{\zeta}=\zeta^{(0)}$\\
 \While{$\epsilon>\epsilon^*$, $\epsilon_M>\epsilon^*_M$ and $t<T$}{
  \textbf{E-step}:\\
  \begin{tabular}{rl}
Latent factors: & {\scriptsize $\Ex[\zv_i| \hat{\Delta},X] = (\I_q + \hat{M}^\top \hat{\Sb}\hat{M})^{-1} \hat{M}^\top\hat{\Sb}(\xv_i -\hat{\theta} \vv_i - \hat{\beta} \bv_i)$}\\
{\small Latent indicators$^+$}: & $\Ex[\gamma_{jk}\mid \hat{\Delta}]= \hat{p}_{jk}$
\end{tabular}

  \textbf{M-step}:\\
  \begin{tabular}{rl}
Loadings$^+$: & $\hat{m}_{jk}=$arg max$_{m_{jk}} Q_1(\hat{\Delta})$\\
Variances: &\scriptsize $\hat{\Tau}_l^{-1}=\frac{1}{n_l+ \eta  -2}\text{diag}\left\{\sum_{i\colon 
	b_{il} = 1}\left( \tilde{\xv}_i \tilde{\xv}_i^\top  -2\tilde{\xv}_i \Ex[\zv_i \mid\hat{\Delta},X]^\top  \hat{M}^\top + \hat{M}  \Ex[\zv_i \zv_i^\top \mid\hat{\Delta},X] \hat{M}^\top \right) + \eta \xi \I_p \right\}$\\
Coefficients:&{\scriptsize $(\hat{\theta}^{\top}_j, \hat{\beta}^{\top}_j) = \sum_{i=1}^n \left[ \hat{\tau}_{j}^{\top} \bv_i (x_{ij} - \hat{m}_{j}^{\top}  \Ex[\zv_{i} \mid\hat{\Delta},X] ) (\vv_i,\bv_i)^\top \right] \left[ \sum_{i=1}^n \left[ \hat{\tau}_{j}^{\top} \bv_i (\vv_i,\bv_i)(\vv_i,\bv_i)^{\top} \right]  +\frac{1}{\psi} \I\right]^{-1}$}\\
Weights: &$\hat{\zeta}_k=\frac{\sum_{j=1}^p \hat{p}_{jk} + \frac{a_\zeta}{k} -1}{\frac{a_\zeta}{k} + b_\zeta + p -1}$
\end{tabular}

\textbf{set} $\Delta^{(t+1)}= \hat{\Delta}$ and $M^{(t+1)}=\hat{M}$ 
 
\textbf{compute} $\epsilon=Q(\Delta^{t+1}) - Q(\Delta^t)$, $\epsilon_M =\max |m_{jk}^{(t+1)}-m_{jk}^{(t)}|$ and $t=t+1$ 
 }
 $^{+}$ see Section~\ref{sec:SSLalgorithm}, Supplementary Sections~\ref{app:EMSSL} and \ref{app:EMpMOM} for details.
 \caption{EM algorithm for factor regression model with spike-and-slab $\Prob(M)$}
 \label{alg:EMSSL}
\end{algorithm}

Let $d_{jk}=[(1-\gamma_{jk}) \lambda_{0} + \gamma_{jk}\lambda_{1}]^{-1}$.
In Expression \eqref{eq:QSSL}, under a Normal-SS prior
\begin{equation}
\label{eq:Q1SSL}
\Eg \left[ \log \Prob (m_{jk} \mid \gamma_{jk} , \lambda_0, \lambda_1) \right] \propto -\frac{1}{2} \hat{m}^2_{jk} \Ex \left[ d_{jk} \mid \hat{\Delta}  \right] = -\frac{1}{2} \hat{m}^2_{jk} \left[ \frac{1-\hat{p}_{jk} }{\loN} +\frac{\hat{p}_{jk}}{\llN} \right]
\end{equation}
where $\hat{p}_{jk}$ is as in \eqref{eq:InclusionSSL}.

Thus, the EM update for the $j^{th}$ row of matrix $M$ is,
\footnotesize
\begin{align}
\hat{m}_j=\left[\sum_{i=1}^n \left( \hat{\tau}_j^{\top}\bv_i\tilde{x}_{ij} \Ex[\zv_i^\top \mid\hat{\Delta},X] \right)\right]\left[  \text{diag}\{\Ex[d_{j1}\mid \hat{\Delta}], \dots ,\Ex[d_{jq}\mid \hat{\Delta}] \}+\sum_{i=1}^n \left( \hat{\tau}_j^{\top} \bv_i \Ex[\zv_i \zv_i^\top \mid\hat{\Delta},X] \right) \right]^{-1},
\end{align}
\normalsize
for $j=1,\dots,p$, where $\tilde{x}_{ij}=x_{ij}-\theta v_{ij} - \beta b_{ij}$. 
A full derivation is given in Supplementary Section~\ref{app:EMSSL}.\par
For the MOM-SS 
\begin{equation}
\label{EzxpMoM}
\Eg \left[ \log \Prob (m_{jk} \mid \gamma_{jk} , \loM, \llM) \right] \propto  -\frac{1}{2} m_{jk}^{2}  \left[ \frac{1-\hat{p}_{jk} }{\loM} +\frac{\hat{p}_{jk}}{\llM} \right]+ \hat{p}_{jk} \log(m_{jk}^2	).
\end{equation}
where $\hat{p}_{jk}$ is given in \eqref{eq:InclusionMOM}

For the M-step, we use a coordinate descent algorithm (CDA) that performs successive univariate optimisation on \eqref{eq:QSSL} with respect to each $m_{jk}$.
An advantage is that the updates have a closed-form that is computationally inexpensive. 
As a potential drawback it could require a larger number of iterations to converge relative to performing joint optimisation with respect to multiple elements in $M$. 
However, we have not found this to be a practical problem in our examples.

Viewed as a function of only $m_{jk}$, it is possible to express $Q_1(m_{jk})$ as
\begin{equation}
\label{mjkpMoM1}
Q_1(m_{jk})= a m_{jk}^2 + b m_{jk} + c \log(m_{jk}^2),
\end{equation}
where 
\begin{equation}
\begin{split}
a&=-\frac{1}{2} \left( \left[ \frac{1-\hat{p}_{jk}}{\loM }+\frac{\hat{p}_{jk}}{\llM} \right]+\sum_{i=1}^n \hat{\tau}_j^{\top} \bv_i \Ex[z_{ik} z_{ik}^\top \mid \hat{\Delta},X] \right)\\
b&= \sum_{i=1}^n \left[\hat{\tau}_j^{\top}\bv_i(x_{ij}-\hat{\theta} v_{ij} -\hat{\beta} b_{ij}) \Ex[z_{ik} \mid\hat{\Delta},X]-\sum_{r\neq k}^q \hat{m}_{jr}  \hat{\tau}_j^{\top} \bv_i \Ex[z_{ir} z_{ik}^\top \mid\hat{\Delta},X] \right]\\
c&=\hat{p}_{jk}
\end{split}
\end{equation}
See Supplementary Section~\ref{app:EMpMOM}. 
The global maximum of \eqref{mjkpMoM1} is summarised in Lemma~\ref{Lemma:maxMjk}.
\begin{lemma}
\label{Lemma:maxMjk} 
Let $f(m_{jk})= a m_{jk}^2 + b m_{jk} + c \log(m_{jk}^2)$, where $a<0$ and $c>0$. Define $\underbar{m}_{jk} = \frac{-b - \sqrt{b^2 - 16ac}}{4a}$ and $\bar{m}_{jk} = \frac{-b + \sqrt{b^2 - 16ac}}{4a}$.\\
If $b>0$, then $\underbar{m}_{jk}= \arg \max_{m_{jk}} f(m_{jk})$. If $b<0$, then $\bar{m}_{jk}= \arg \max_{m_{jk}} f(m_{jk})$. If $b=0$, then $\bar{m}_{jk}=\underbar{m}_{jk}= \arg \max_{m_{jk}} f(m_{jk})$ 
\end{lemma}

Akin to the MOM-SS, we can express $Q_1(m_{jk})$ as function of $m_{jk}$ for the Laplace-based priors as:
\[
Q_1(m_{jk})= a m_{jk}^2 + b m_{jk} + c |m_{jk}|+ d \log(m_{jk}^2) 
\]
\begin{equation}
\begin{split}
a=&-\frac{1}{2} \sum_{i=1}^n \hat{\tau}_j^{\top} \bv_i \Ex[z_{ik} z_{ik}^\top \mid\hat{\Delta},X] \\
b=& \sum_{i=1}^n \left[\hat{\tau}_j^{\top}\bv_i(x_{ij}-\hat{\theta} v_{ij} - \hat{\beta} b_{ij}) \Ex[z_{ik} \mid\hat{\Delta},X]-\sum_{r\neq k}^q \hat{m}_{jr}  \hat{\tau}_j^{\top} \bv_i \Ex[z_{ir} z_{ik}^\top \mid\hat{\Delta},X] \right]\\
c=&-\left[ \frac{1-\hat{p}_{jk} }{\loL} +\frac{\hat{p}_{jk}}{\llL} \right]\\
d=& 
\begin{cases}
    0 & \text{for Laplace-SS} \\ 
    \hat{p}_{jk} & \text{for Lapace-MOM-SS} 
\end{cases}
\end{split}
\end{equation}
for $j=1,\dots,p$ and where $\hat{p}_{jk}$ is as in 
\eqref{eq:InclusionSSLap} and \eqref{eq:InclusionSSM} for Laplace-SS and Laplace-MOM-SS respectively.

Lemma~\ref{Lemma:maxMjkLaplaceSS} summarises the global maximum for Laplace-SS

\begin{lemma}
\label{Lemma:maxMjkLaplaceSS}
Let $f(m_{jk})=  a m_{jk}^2 + b m_{jk} + c |m_{jk}|$, where $a<0$ and $c<0$. 
Define $m_{jk}^+ = \frac{-(b+c)}{2a}$ and $m^-_{jk} = \frac{-(b-c)}{2a}$.

If $b>-c$, then $m_{jk}^+= \arg \max_{m_{jk}} f(m_{jk})$. If $b<c$, then $m_{jk}^-= \arg \max_{m_{jk}} f(m_{jk})$. If $c \leq b \leq -c$, then $0= \arg \max_{m_{jk}} f(m_{jk})$.
\end{lemma}

Finally for the Laplace-MOM-SS, we emphasise that when $m_{jk}=0$, $Q_1(m_{jk}=0)=-\infty$. Thus the solution for $m_{jk}$ is given by setting $\frac{\partial Q_1}{\partial m_{jk}}=0$ as given in Lemma~\ref{Coro:maxMjk}.

\begin{lemma}
\label{Coro:maxMjk}
Let $f(m_{jk})=  a m_{jk}^2 + b m_{jk} + c |m_{jk}|+ d \log(m_{jk}^2)$, where $a<0$, $c<0$ and $d>0$. 
Define $m_{jk}^+ = \frac{-(b+c) - \sqrt{(b+c)^2 - 16ad}}{4a}$ and $m^-_{jk} = \frac{-(b-c) + \sqrt{(b-c)^2 - 16ad}}{4a}$.

If $b>0$, then $m_{jk}^+= \arg \max_{m_{jk}} f(m_{jk})$. If $b<0$, then $m_{jk}^-= \arg \max_{m_{jk}} f(m_{jk})$. If $b=0$, then $m_{jk}^+=m_{jk}^-= \arg \max_{m_{jk}} f(m_{jk})$.
\end{lemma}
We remark that if  either $\xv_i$ or $\vv_i$ are continuous, the event of $b=0$ has zero probability. If both $\xv_i$ and $\vv_i$ are discrete and in presence of the rare event of $b=0$, then the sign of the update for $m_{jk}$ is set to the previous one.

\subsection{Initialisation of parameters}
\label{IniEMNS}

The EM algorithm can be sensitive to parameter initialisation. 
We propose two different strategies: least-squares and least-squares with rotation.\par
The first option is a simple two-step least-squares that is computationally efficient and performs well in many of our examples.\par
Step 1: initialise $(\theta^{(0)},\beta^{(0)})=[(V,B)^{\top}(V,B)]^{-1}(V,B)^{\top}X$. \par
Step 2: Let $\hat{E}=X-(V\theta^{(0)\top}+B\beta^{(0)\top})$. 
Consider the eigendecomposition of $\frac{1}{n}\hat{E}^\top \hat{E}$ where $l_1\geq l_2 \geq\dots \geq l_q$ are the eigenvalues and $u_1,\dots,u_q$ the eigenvectors. 
Set $M^{(0)}=[\sqrt{l_1}u_1\mid \dots \mid \sqrt{l_q}u_q]$ and $\Tau_l^{(0)}=[$diag$\{\frac{1}{n}\hat{E}^\top \hat{E}-M^{(0)}M^{(0)\top}\}]^{-1}$ for $l=1,\dots,p_b$

The rotated least-squares adds an extra step.\par 
Step 3: varimax rotation for the loadings obtained in Step 2. 

The reason for this extra step is to help escape local modes.
The EM algorithm does not guarantee convergence to a global maximum, but it increases the log-posterior at each iteration. 
This local maxima issue is intensified by the non-identifiability of the factor model through the rotational ambiguity of the likelihood and the strong association between the updates of loadings and factors. 

\subsection{Post-processing for model selection and dimensionality reduction}
\label{Postprocessing}
The EM algorithm gives point estimates $(\hat{M},\hat{\theta},\hat{\Tau},\hat{\zeta})$. 
Under Laplace-SS one can obtain exact sparsity via $\hat{m}_{jk}=0$, however this is not the case for our other priors. 
To address this, we define $\hat{\gamma}$ as the solution of the following optimisation problem
\begin{equation}
\hat{\gamma}= \text{argmax}_{\gamma} \Prob(\gamma \mid X, \hat{M}, \hat{\theta}, \hat{\Tau}, \hat{\zeta}) = \text{argmax}_{\gamma} \prod_{jk} \Prob(\gamma_{jk} | \hat{m}_{jk}, \hat{\zeta}_k)
\end{equation}
where the right-hand side follows from the assumed conditional independence of $m_{jk}$. 
That is, we set $\hat{\gamma}_{jk}=1$ if $\Prob(\gamma_{jk}=1 | \hat{m}_{jk}, \hat{\zeta}_k) > 0.5$ and $\gamma_{jk}=0$ otherwise. 
When $\hat{\gamma}_{jk}=0$ we set $\hat{m}_{jk}=0$ effectively selecting the number of factors and the non-zero loadings within each factor. 

As an alternative post-processing step we consider that in some applications one may want to select only the number of factors. 
We then consider to setting $\tilde{\gamma}_{j k}=1$ if $\sum_{j=1}^p \hat{\gamma}_{jk} \neq 0$ and $\gamma_{jk}=0$ otherwise.\par
The combination of the two initialisation alternatives and two different post-processing options gives four possible solutions for $\hat{M}$.
To choose which is best in our examples, we use weighted 10-fold cross-validation, where the weights reflect that batches with higher variance should receive lower weight, selecting the model with smallest weighted  cross validation reconstruction error (See Supplementary Section~\ref{app:WCV} for details ).

Finally we re-order of the factors so that $\sum_{j=1}^p \gamma_{jk}$ is decreasing in $k$, which under our prior~\eqref{PriorGamma} is guaranteed to increase the log-posterior. 
This is the so-called left-ordered inclusion matrix of \citet{Griffiths2011}. 
This facilitates the interpretation of latent factors.

Latent factors are also post-processed for data visualisation purposes. 
The aim of this is to obtain new standardised factors $\tilde{\zv}_i=[\text{Cov}(\zv_i\mid\hat{\Delta},X)]^{-1}\Ex[\zv_i\mid\hat{\Delta},X]$, with $\text{Cov}(\zv_i\mid\hat{\Delta},X) = (\I_q + \hat{M}^\top \hat{\Sb}\hat{M})^{-1}$, whose covariance does not depend on their batch.
\section{Results}
\label{sec:Results}
We assess our approach on simulated and on experimental datasets. 
Section~\ref{Nobatch} assesses the accuracy of our prior in obtaining sparse factor loadings, estimating the covariance and low-dimensional representations, by comparing its performance to competing methods in a setting where there are no batch effects.
 Then Section~\ref{Sec:BE} studies the importance of accounting for batch effects in simulations and Section~\ref{Sec:Cancer} in two cancer datasets.
In the latter we also assess the ability of the obtained dimension reduction to predict survival outcomes.

Sections~\ref{Nobatch} and \ref{Sec:BE} study simulations under two different loading matrices $M$ (truly sparse and dense) and two different scenarios (without and with batch effects).
We compare our methods with the Fast Bayesian Factor Analysis via Automatic Rotations to Sparsity (FastBFA) of \citet{Rockova2016} and the Penalized Likelihood Factor Analysis with a LASSO penalty (LASSO-BIC) of \citet{Hirose2015}.
We also use the ComBat empirical Bayes batch effect correction of \citet{COMBAT2007} for scenarios with batch effects, doing an MLE estimation of the factor analysis model (ComBat-MLE). 
In Section~\ref{Sec:Cancer} we analyse a high-dimensional gene expression data under a supervised and an unsupervised framework. 
We use the clinically annotated data for the ovarian cancer transcriptome from \textbf{R} package \texttt{curatedOvarianData 1.16.0} \citep{curatedOvarianData} and the lung cancer data from The Cancer Genome Atlas (TCGA) from \textbf{R} package \texttt{TCGA2STAT 1.2} \citep{TCGA2STAT}.

The \textbf{R} code for our model is available at \url{https://github.com/AleAviP/BFR.BE}. 
We used \textbf{R} function \texttt{FACTOR\_ROTATE} of \citet{Rockova2016} for FastBFA, the \textbf{R} package \texttt{fanc 2.2} for LASSO-BIC \citep{fanc} and package \texttt{sva 3.26.0} for ComBat \citep{svaR}. 
Hyper-parameters for the Normal-SS and MOM-SS were set as in section~\ref{Elicitation}, the hyper-parameters for FastBFA were set via Dynamic Posterior Exploration as in \citet{Rockova2016} with $1/\lambda_0=0.001$ and $ 1/\lambda_1\in \{5, 10, 20, 30\}$ and using varimax robustifications.
For the LASSO-BIC we selected the model with smallest BIC to set the regularization parameter.
Finally, for scenarios with batch effects, we adjusted the data via a ComBat correction and performed a Factor Analysis via EM algorithm to maximise likelihood with the \texttt{fa.em} function in the \texttt{cate} package \citep{cateR}. 

\begin{figure}[h!]
\centering
\begin{subfigure}{.475\textwidth}
  \centering
  \includegraphics[width=1\linewidth]{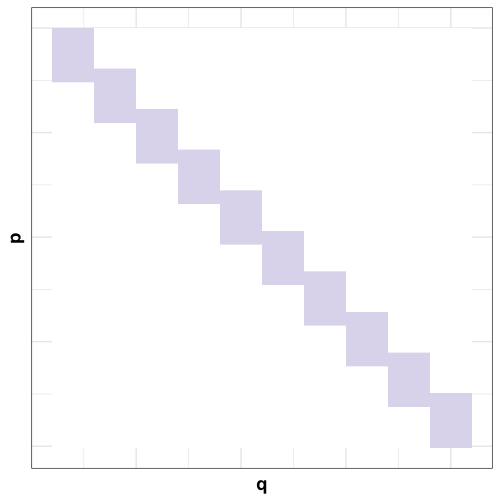}
  \caption{Loadings of truly sparse $M^*$}
  \label{fig:MSparse}
\end{subfigure}
\begin{subfigure}{.475\textwidth}
  \centering
  \includegraphics[width=1\linewidth]{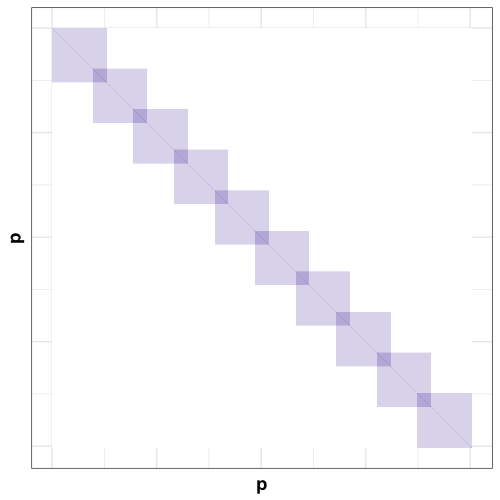}
  \caption{Covariance of truly sparse $M^*$}
  \label{fig:MMSparse}
\end{subfigure}\\
\begin{subfigure}{.475\textwidth}
  \centering
  \includegraphics[width=1\linewidth]{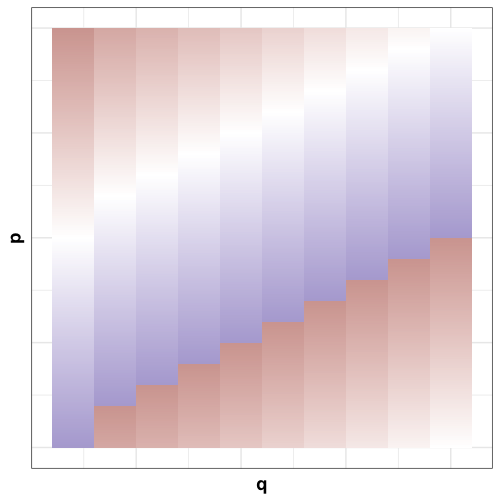}
  \caption{Loadings of dense $M^*$}
  \label{fig:MFix}
\end{subfigure}
\begin{subfigure}{.475\textwidth}
  \centering
  \includegraphics[width=1\linewidth]{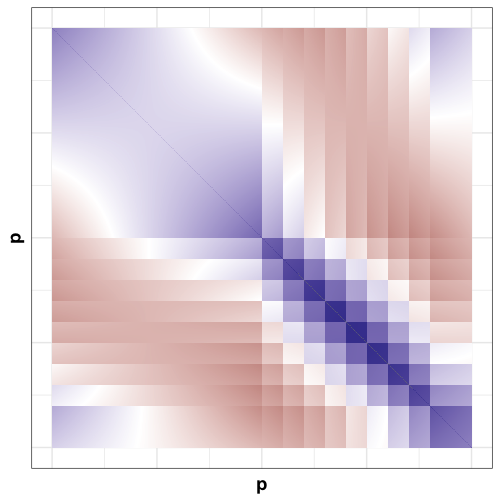}
  \caption{Covariance of dense $M^*$}
  \label{fig:MMFix}
\end{subfigure}
\caption{Synthetic data. Heatmaps of data-generating loadings and covariance with red highly negative, blue highly positive and white zero values.}
\label{HeatmapMtoy}
\end{figure}

\subsection{No batch effect}
\label{Nobatch}
To assess the precision of the parameter estimates returned by the EM algorithm, we simulated data from two different data-generating truths: truly sparse and dense for the loadings $M$. 
In both, the truth was set to $q^*=10$ factors. 
The dense loadings matrix has a grid of elements set uniformly between $(-1,1)$, whereas the truly sparse $M$ has a banded-diagonal structure with $m_{jk}=1$ for the non-zero elements, as shown in Figure~\ref{HeatmapMtoy}. 

Some visual representations of our findings are display in the Supplementary Figures~\ref{fig:ZMrecons}-\ref{fig:HeatGammapMOMvsSSLBEsup}.

\begin{table}[h!]
\centering
\caption{Synthetic data without batch effects for $n=100$, $q^*=10$, $p=1,000$ or $1,500$ parameters, truly sparse and dense loadings $M^*$.} 
\scalebox{0.75}{
\begin{tabular}{r|rrccr|rrccr}
	\hline
 &\multicolumn{5}{c|}{$p=1,000$}&\multicolumn{5}{c}{$p=1,500$} \\
	\hline
	Model&$\hat{q}$  &$|\hat{M}|_0$& \scriptsize{$|| \Ex[X] -\hat{ \Ex}[X]||_F$} & \scriptsize{$|| \text{Cov}[x_i] -\widehat{ \text{Cov}}[x_i]||_F$}    & it &$\hat{q}$ &$|\hat{M}|_0$& \scriptsize{$|| \Ex[X] -\hat{ \Ex}[X]||_F$}  &  \scriptsize{$|| \text{Cov}[x_i] -\widehat{ \text{Cov}}[x_i]||_F$}  & it \\ 
  \hline
  &\multicolumn{10}{c}{Dense $M$, $q=10$} \\
	\hline
Flat & 10.0 & 10000.0 & 104.8 & \textbf{1173.3} & 2.0 & 10.0 & 15000.0 & 126.5 & \textbf{1895.7} & 2.0 \\ 
  Normal-SS & 10.0 & 1859.7 & \textbf{92.4} & 1266.4 & 9.3 & 10.0 & 2461.0 & \textbf{112.5} & 1988.2 & 6.9 \\ 
  \rowcolor{lightgray}
  MOM-SS & 10.0 & 1468.6 & 93.5 & 1294.3 & 9.7 & 10.0 & 2059.1 & 114.3 & 1998.5 & 6.3 \\ 
  FastBFA & 9.6 & 976.9 & 137.9 & 1738.2 & 153.6 & 9.4 & 1400.4 & 163.2 & 5638.7 & 162.0 \\ 
  LASSO-BIC & 10.0 & 5331.3 & 110.6 & 1682.5 & NA & 10.0 & 8607.7 & 137.4 & 2524.8 & NA \\ 
\hline
  &\multicolumn{10}{c}{Dense $M$, $q=100$} \\
	\hline  
  Flat & 100.0 & 100000.0 & 313.2 & 1200.6 & 3.0 & 100.0 & 150000.0 & 376.2 & 1925.3 & 2.5 \\ 
  Normal-SS & 34.8 & 3418.5 & 190.5 & 1190.7 & 4.2 & 14.8 & 5083.8 & 154.4 & 1911.8 & 4.0 \\ 
  \rowcolor{lightgray}
  MOM-SS & 10.5 & 3215.9 & 108.9 & 1178.7 & 5.0 & 11.2 & 4232.8 & 135.6 & 1902.8 & 4.0 \\ 
  FastBFA & 96.6 & 3379.3 & 297.4 & \textbf{451.2} & 11.3 & 97.3 & 4558.4 & 362.1 & \textbf{670.5} & 10.5 \\ 
  LASSO-BIC & 11.0 & 4829.2 & \textbf{80.5} & 1682.7 & NA & 11.1 & 7839.6 & \textbf{99.5} & 2524.8 & NA \\ 
  \hline
  &\multicolumn{10}{c}{Sparse $M$, $q=10$} \\
	\hline  
  Flat & 10.0 & 10000.0 & 104.8 & 184.1 & 2.0 & 10.0 & 15000.0 & 126.4 & 301.4 & 2.0 \\ 
  Normal-SS & 10.0 & 1300.1 & 55.8 & 124.1 & 3.9 & 10.0 & 1942.1 & \textbf{68.0} & 248.5 & 3.0 \\ 
  \rowcolor{lightgray}
  MOM-SS & 10.0 & 1299.9 & \textbf{53.8} & \textbf{122.5} & 4.3 & 10.0 & 1943.0 & 69.4 & \textbf{235.2} & 2.3 \\ 
  FastBFA & 8.7 & 1076.3 & 74.8 & 176.8 & 93.1 & 7.1 & 1320.4 & 84.7 & 344.2 & 122.7 \\ 
  LASSO-BIC & 10.0 & 5304.3 & 77.4 & 424.0 & NA & 10.0 & 8397.0 & 93.3 & 636.3 & NA \\ 
 \hline
  &\multicolumn{10}{c}{Sparse $M$, $q=100$} \\
	 \hline  
  Flat & 100.0 & 100000.0 & 313.7 & 310.8 & 3.0 & 100.0 & 150000.0 & 375.4 & 446.7 & 2.5 \\ 
  Normal-SS & 22.0 & 2801.8 & 165.3 & 203.7 & 4.0 & 42.5 & 2795.9 & 230.0 & 335.1 & 4.3 \\ 
  \rowcolor{lightgray}
  MOM-SS & 10.5 & 2156.8 & 109.7 & \textbf{194.1} & 5.0 & 11.2 & 2430.5 & 136.4 & \textbf{324.8} & 4.0 \\ 
  FastBFA & 97.9 & 1508.9 & 283.0 & 215.2 & 9.9 & 97.6 & 2229.7 & 363.0 & 326.4 & 9.2 \\ 
  LASSO-BIC & 10.0 & 4815.5 & \textbf{75.0} & 425.1 & NA & 10.0 & 7980.8 & \textbf{91.2} & 637.1 & NA \\ 
   \hline
\end{tabular}}
\label{table:n100NB}
\end{table}

We simulated $n = 100$ observations from $\xv_i =M^* \zv_i, + \ev_i$, with growing $p = 1,000$ and $1,500$, where the factors $\zv_i \sim N(0,\I_q)$, the errors $\ev_i \sim N(0,\Tau^{-1})$ with $\Tau^{-1}=\I_p$, and the loadings $M^*$ are set as dense or sparse as in Figure~\ref{HeatmapMtoy}. 
For comparison, FastBFA was initialised as our models via two-step least-squares (Section~\ref{IniEMNS}). 

Table~\ref{table:n100NB} shows the selected number of factors $\hat{q}$, the number of estimated non-zero loadings $| \hat{M} |_0=\sum_{j,k}\mathbbm{1}(\hat{m}_{jk} \neq 0)$, the Frobenius norm (F.N.) between the true expected value and its reconstruction $|| E[X]-\hat{E}[X]||_F=|| ZM^\top-\Ex[Z\mid \hat{\Delta},X]\hat{M}^\top||_F$ and between the true and reconstructed covariances $|| \text{Cov}[x_i] -\widehat{ \text{Cov}}[x_i]||_F = || (MM^\top + \Tau^{-1}) - (\hat{M} \hat{M}^\top + \hat{\Tau}^{-1}) ||_F$, and the number of iterations until convergence.
The mean across 100 different simulations is displayed and the model with smallest mean Frobenius norm per scenario is indicated in bold.

We first considered the unrealistic scenario where $M$ is dense and one guessed correctly the true number of factors $q=q^*=10$. 
The aim of this setting was to investigate if MOM-SS shrinkage provided a poor estimation when the factors were not truly sparse.
MOM-SS and Normal-SS performed similarly as $p$ grew, and competitively relative to the flat prior.
To extend our example, we then set $q=100$ to illustrate the performance when there is sparsity in terms of the number of factors, but not within factors. 
LASSO-BIC had the best reconstruction for the mean but performed poorly on the covariance, whereas FastBFA outperformed all the models to estimate the covariance but performed poorly for the mean. 
However, MOM-SS had a good balance in terms of estimating the expected value and the covariance, being the second best in both cases.

We further illustrate our model under the arguably more interesting case of truly sparse loadings. 
First we set $q=10$ the true cardinality. 
In this scenario MOM-SS and Normal-SS presented the best results both for mean and covariance.
This example reflects the advantages of shrinkage and the varimax rotation for the initialisation in the loadings, leading to good sparse solutions. 
Finally we considered the same scenario with $q=100$. 
LASSO-BIC was best to estimate the mean at the cost of reduced precision in the covariance reconstruction. 
MOM-SS displayed the lowest error for the covariance and second smallest for the mean, showing a good balance between those metrics.

In general, MOM-SS achieved a good balance between estimating the mean, which is useful for dimensionality reduction, and sparse covariance estimation. 
Recall that we used a coordinate descent algorithm for the non-local prior, which as a potential drawback could require a larger number of iterations than performing jointly optimising multiple elements in $M$.  
However, Table~\ref{table:n100NB} showed that MOM-SS required roughly the same number of iterations to converge as the Normal-SS.
We can see that MOM-SS and LASSO-BIC estimated $\hat{q}$ accurately. 
Note that in general FastBFA had the highest estimated latent cardinality $\hat{q}$, due to the fat tails of the Laplace priors, which adds some columns of $M$ that contain very few non-zero loadings after the tenth factor, as shown in Supplementary Sections~\ref{Sup:PlotsS} and \ref{Sup:PlotsD}. 
Nonetheless, this model displayed a mean number of non-zero loadings closer to the ground truth (1,300 and 1,940 for the $p=1,000$ and $p=1,500$ respectively under sparse $M$). 

\begin{figure}[h!]
\centering
\begin{subfigure}{.19\textwidth}
  \centering
  \includegraphics[width=1\linewidth]{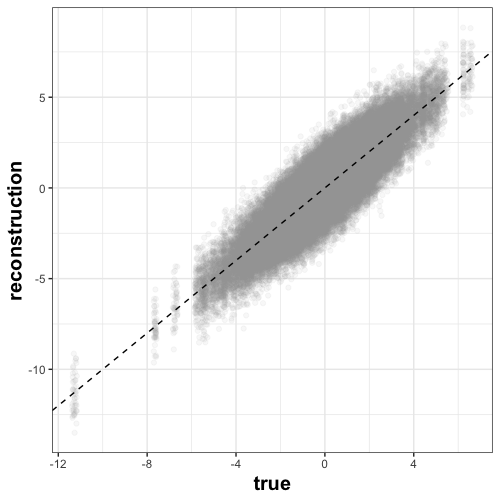}
  \caption*{\tiny Flat Dense $M$}
\end{subfigure}
\begin{subfigure}{.19\textwidth}
  \centering
  \includegraphics[width=1\linewidth]{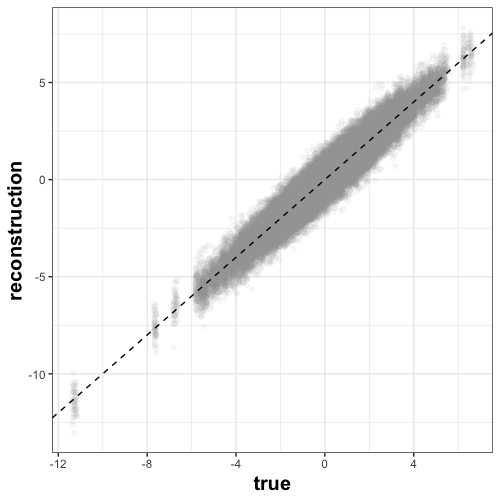}
  \caption*{\tiny Normal-SS Dense $M$}
\end{subfigure}
\begin{subfigure}{.19\textwidth}
  \centering
  \includegraphics[width=1\linewidth]{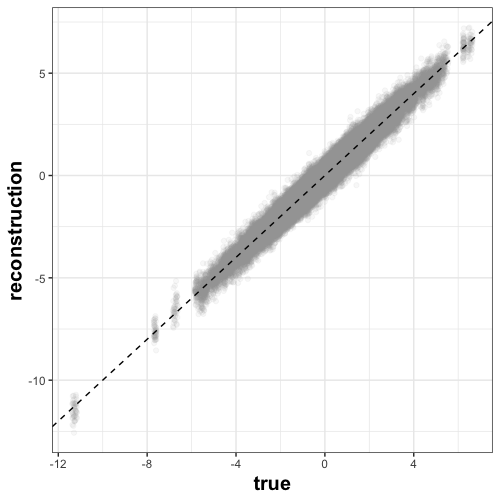}
  \caption*{\tiny MOM-SS Dense $M$}
\end{subfigure}
\begin{subfigure}{.19\textwidth}
  \centering
  \includegraphics[width=1\linewidth]{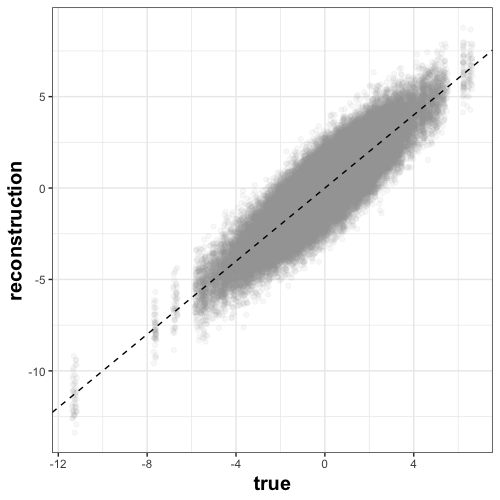}
  \caption*{\tiny FastBFA Dense $M$}
\end{subfigure}
\begin{subfigure}{.19\textwidth}
  \centering
  \includegraphics[width=1\linewidth]{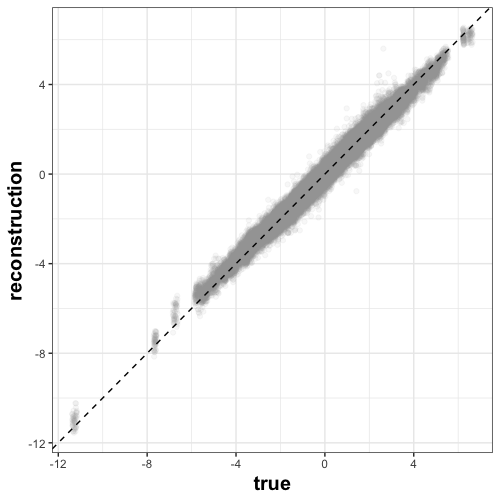}
  \caption*{\tiny LASSO Dense $M$}
\end{subfigure}\\
\begin{subfigure}{.19\textwidth}
  \centering
  \includegraphics[width=1\linewidth]{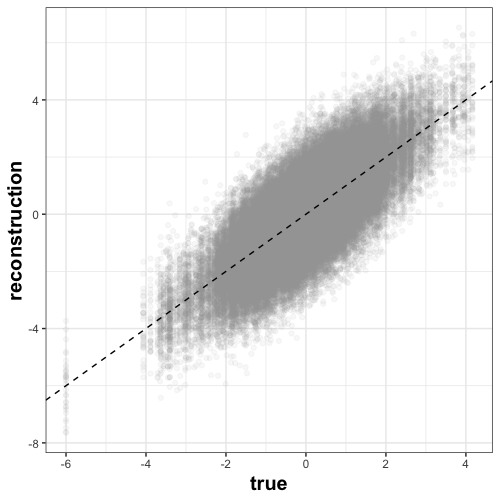}
  \caption*{\tiny Flat Sparse $M$}
\end{subfigure}
\begin{subfigure}{.19\textwidth}
  \centering
  \includegraphics[width=1\linewidth]{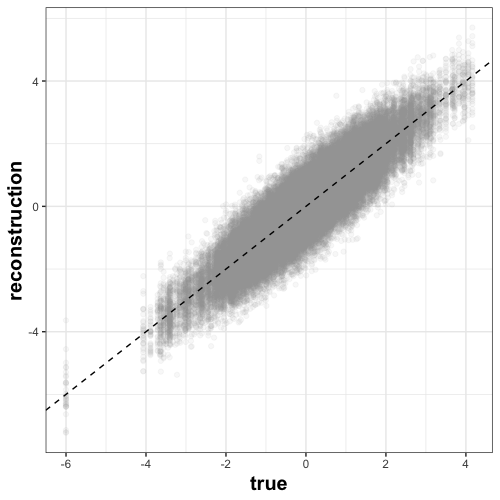}
  \caption*{\tiny Normal-SS Sparse $M$}
\end{subfigure}
\begin{subfigure}{.19\textwidth}
  \centering
  \includegraphics[width=1\linewidth]{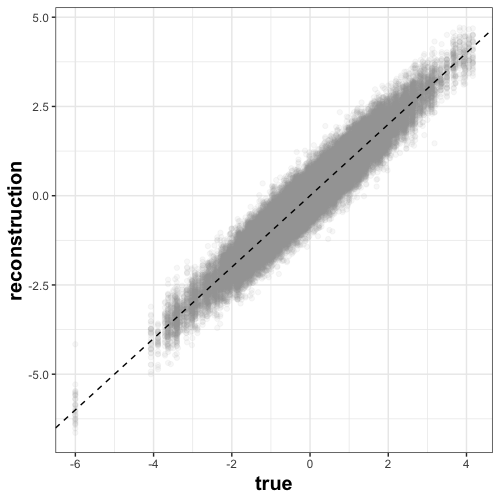}
  \caption*{\tiny MOM-SS Sparse $M$}
\end{subfigure}
\begin{subfigure}{.19\textwidth}
  \centering
  \includegraphics[width=1\linewidth]{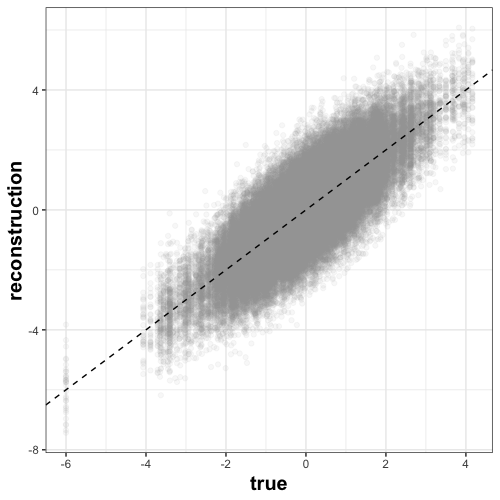}
  \caption*{\tiny FastBFA Sparse $M$}
\end{subfigure}
\begin{subfigure}{.19\textwidth}
  \centering
  \includegraphics[width=1\linewidth]{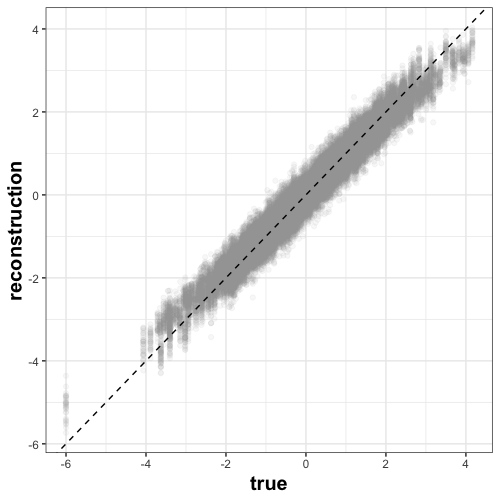}
  \caption*{\tiny LASSO Sparse $M$}
\end{subfigure}
\caption{Scatterplots comparing $ZM^\top$ vs. $\Ex[Z\mid\hat{\Delta},X]\hat{M}^\top$ between the different models  under dense (top) and truly sparse (bottom) loadings $M$ with $q=100$ in simulations without batch effect.}
\label{fig:ZMrecons}
\end{figure}

\subsection{Batch effects}
\label{Sec:BE}
\begin{table}[h!]
\centering
\caption{Synthetic data with batch effects for $n=200$, $q^*=10$, $p=250$ or $500$ parameters, truly sparse and dense loadings $M^*$.}
\scalebox{0.75}{
\begin{tabular}{r|rrccr|rrccr}
	\hline
 &\multicolumn{5}{c|}{$p=250$}&\multicolumn{5}{c}{$p=500$} \\
	\hline
	Model&$\hat{q}$  &$|\hat{M}|_0$& \scriptsize{$|| \Ex[X] -\hat{ \Ex}[X]||_F$} &\tiny{$|| ZM^\top-  \Ex[Z\mid\hat{\Delta},X]\hat{ M}^\top||_F$} & it &$\hat{q}$ &$|\hat{M}|_0$& \scriptsize{$|| \Ex[X] -\hat{ \Ex}[X]||_F$} &\tiny{$|| ZM^\top-  \Ex[Z\mid\hat{\Delta},X]\hat{ M}^\top||_F$}& it \\ 
	\hline
  &\multicolumn{10}{c}{Dense $M$, $q=10$} \\
	\hline
  Flat & 10.0 & 2500.0 & 56.5 & 88.2 & 4.4 & 10.0 & 5000.0 & 71.9 & 120.2 & 4.0 \\ 
  Normal-SS & 10.0 & 727.6 & \textbf{54.0} & \textbf{83.9} & 8.0 & 10.0 & 1398.7 & \textbf{68.6} & \textbf{116.5} & 4.5 \\ 
  \rowcolor{lightgray}
  MOM-SS & 10.0 & 1097.3 & 55.1 & 84.6 & 15.2 & 10.0 & 1257.5 & 70.1 & 127.4 & 81.1 \\ 
  ComBat-MLE & 10.0 & 2500.0 & 178.5 & 810.2 & 3.1 & 10.0 & 5000.0 & 249.2 & 1144.9 & 3.2 \\ 
  FastBFA & 10.0 & 1153.0 & 89.0 & 834.5 & 12.3 & 10.0 & 2343.1 & 106.6 & 1182.6 & 10.9 \\ 
  LASSO-BIC & 10.0 & 2109.9 & 99.2 & 833.1 & NA & 10.0 & 4377.1 & 118.1 & 1182.9 & NA \\ 
  \hline
  &\multicolumn{10}{c}{Dense $M$, $q=100$} \\
	\hline
  Flat & 100.0 & 25000.0 & 140.7 & 157.6 & 5.0 & 100.0 & 50000.0 & 208.8 & 231.2 & 10.7 \\ 
  Normal-SS & 29.7 & 983.5 & 87.4 & 111.2 & 6.3 & 10.0 & 2725.1 & \textbf{73.4} & \textbf{119.8} & 5.6 \\ 
  \rowcolor{lightgray}
  MOM-SS & 10.0 & 1216.7 & \textbf{57.4} & \textbf{87.7} & 7.2 & 10.0 & 2293.5 & 74.0 & 120.4 & 6.3 \\ 
  ComBat-MLE & 100.0 & 25000.0 & 70.6 & 822.6 & 33.8 & 100.0 & 50000.0 & 123.3 & 1161.0 & 14.8 \\ 
  FastBFA & 35.3 & 1285.5 & 79.3 & 826.7 & 19.6 & 59.9 & 2589.0 & 126.8 & 1181.6 & 12.5 \\ 
  LASSO-BIC & 12.9 & 1579.6 & 59.4 & 827.8 & NA & 11.1 & 2939.6 & 75.8 & 1171.2 & NA \\ 
  \hline
  &\multicolumn{10}{c}{Sparse $M$, $q=10$} \\
	\hline
  Flat & 10.0 & 2500.0 & 49.7 & 68.5 & 4.1 & 10.0 & 5000.0 & 60.8 & 90.7 & 4.1 \\ 
  Normal-SS & 10.0 & 330.0 & 45.7 & 58.7 & 4.9 & 10.0 & 650.0 & \textbf{55.9} & 77.0 & 4.1 \\ 
  \rowcolor{lightgray}
  MOM-SS & 10.0 & 330.0 & \textbf{45.5} & \textbf{57.8} & 5.4 & 10.0 & 650.0 & 56.0 & \textbf{76.6} & 4.1 \\ 
  ComBat-MLE & 10.0 & 2500.0 & 171.4 & 807.8 & 2.0 & 10.0 & 5000.0 & 244.5 & 1140.3 & 1.0 \\ 
  FastBFA & 10.0 & 817.1 & 78.1 & 832.1 & 9.8 & 10.0 & 1617.5 & 104.2 & 1178.1 & 9.9 \\ 
  LASSO-BIC & 10.0 & 2307.4 & 73.3 & 835.0 & NA & 10.0 & 4835.0 & 97.9 & 1181.1 & NA \\ 
  \hline
  &\multicolumn{10}{c}{Sparse $M$, $q=100$} \\
	\hline
  Flat & 100.0 & 25000.0 & 140.4 & 146.4 & 5.0 & 100.0 & 50000.0 & 207.9 & 216.0 & 10.4 \\ 
  Normal-SS & 93.2 & 372.9 & 139.9 & 143.7 & 7.2 & 10.0 & 2675.5 & 74.4 & \textbf{91.2} & 5.6 \\ 
  \rowcolor{lightgray}
  MOM-SS & 10.0 & 1286.2 & 59.1 & \textbf{70.2} & 7.1 & 10.0 & 2197.0 & 75.6 & 92.8 & 6.3 \\ 
  ComBat-MLE & 100.0 & 25000.0 & 70.8 & 821.1 & 42.6 & 100.0 & 50000.0 & 123.1 & 1157.3 & 14.3 \\ 
  FastBFA & 41.5 & 976.5 & 84.8 & 828.2 & 18.1 & 65.8 & 1956.8 & 130.9 & 1179.8 & 13.7 \\ 
  LASSO-BIC & 12.3 & 1663.3 & \textbf{56.0} & 824.7 & NA & 12.9 & 3794.4 & \textbf{70.2} & 1167.7 & NA \\
  \hline
  \end{tabular}}
\label{table:n200BE}
\end{table}

\begin{figure}[h!]
\begin{subfigure}{.16\textwidth}
  \centering
  \includegraphics[width=1\linewidth]{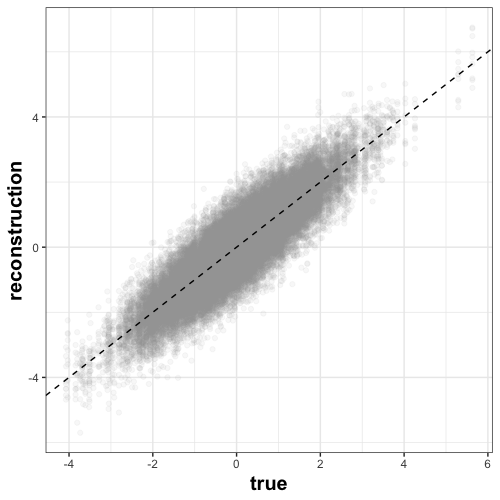}
  \caption*{\tiny Flat Dense $M$}
\end{subfigure}
\begin{subfigure}{.16\textwidth}
  \centering
  \includegraphics[width=1\linewidth]{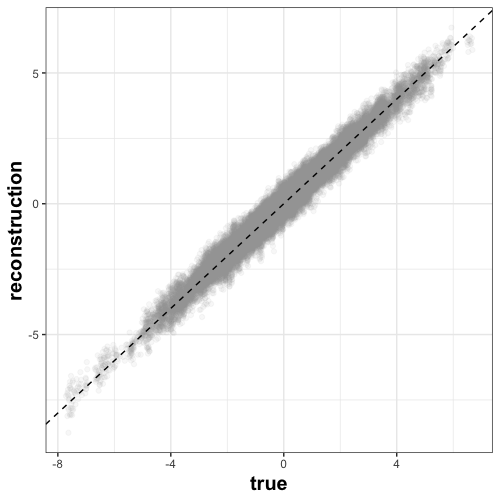}
  \caption*{\tiny Normal-SS Dense $M$}
\end{subfigure}
\begin{subfigure}{.16\textwidth}
  \centering
  \includegraphics[width=1\linewidth]{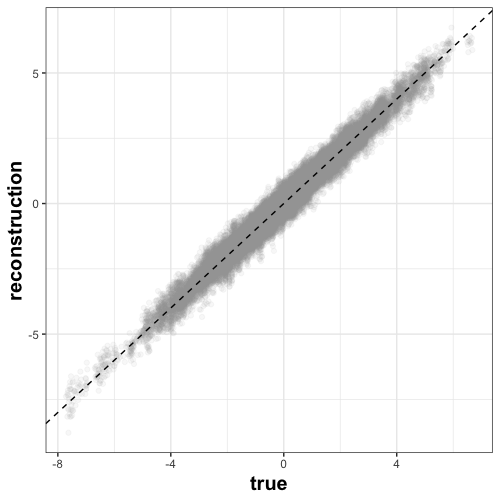}
  \caption*{\tiny MOM-SS Dense $M$}
\end{subfigure}
\begin{subfigure}{.16\textwidth}
  \centering
  \includegraphics[width=1\linewidth]{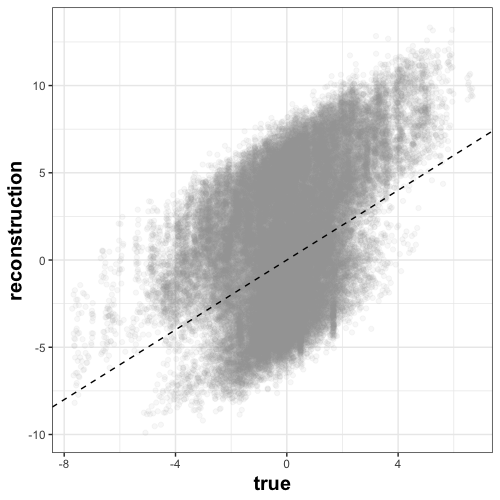}
  \caption*{\tiny ComBat-MLE Dense $M$}
\end{subfigure}
\begin{subfigure}{.16\textwidth}
  \centering
  \includegraphics[width=1\linewidth]{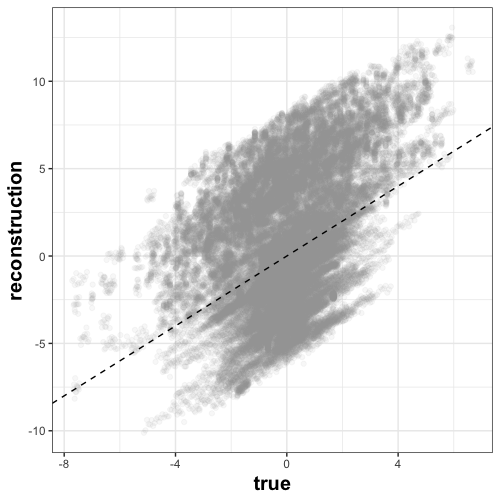}
  \caption*{\tiny FastBFA Dense $M$}
\end{subfigure}
\begin{subfigure}{.16\textwidth}
  \centering
  \includegraphics[width=1\linewidth]{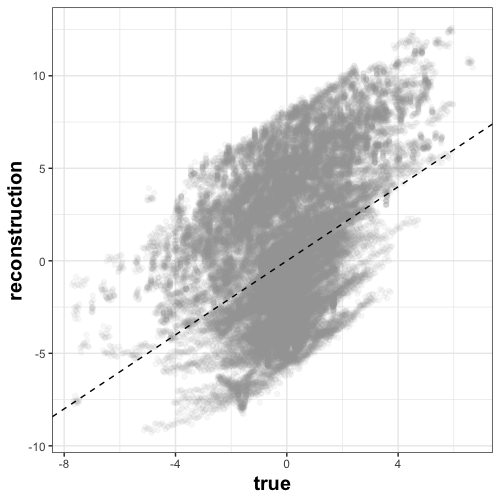}
  \caption*{\tiny LASSO Dense $M$}
\end{subfigure}\\
\begin{subfigure}{.16\textwidth}
  \centering
  \includegraphics[width=1\linewidth]{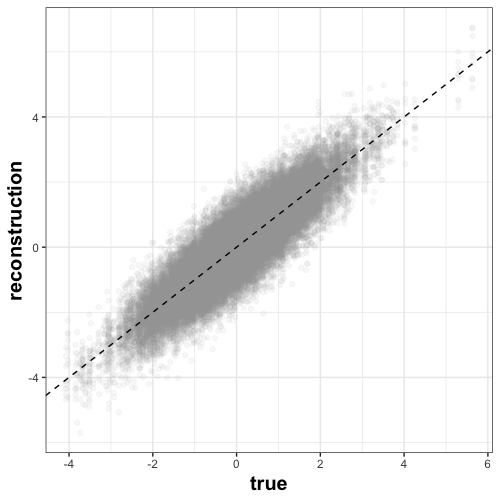}
  \caption*{\tiny Flat Sparse $M$}
\end{subfigure}
\begin{subfigure}{.16\textwidth}
  \centering
  \includegraphics[width=1\linewidth]{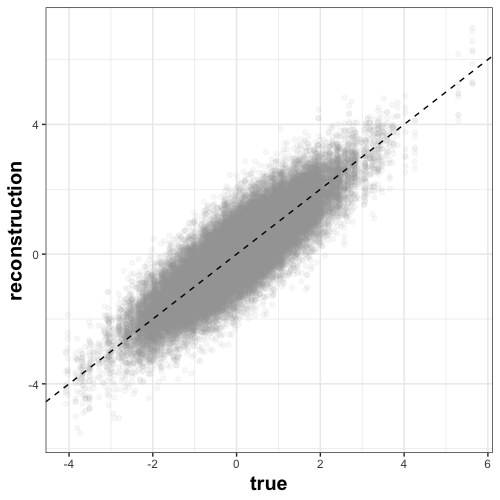}
  \caption*{\tiny Normal-SS Sparse $M$}
\end{subfigure}
\begin{subfigure}{.16\textwidth}
  \centering
  \includegraphics[width=1\linewidth]{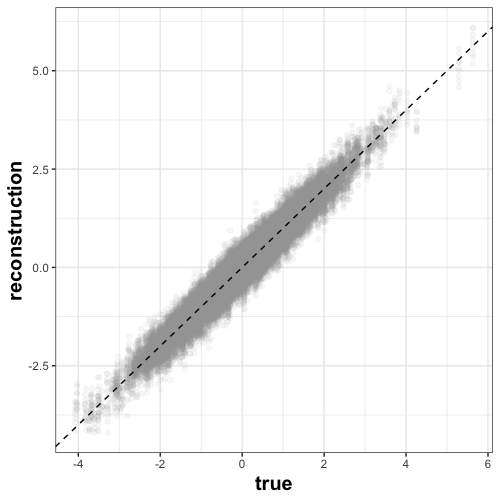}
  \caption*{\tiny MOM-SS Sparse $M$}
\end{subfigure}
\begin{subfigure}{.16\textwidth}
  \centering
  \includegraphics[width=1\linewidth]{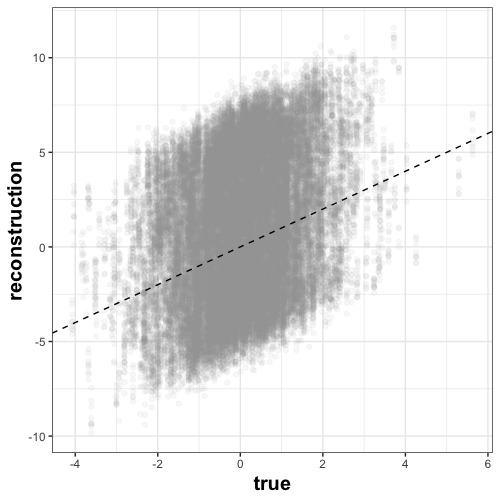}
  \caption*{\tiny ComBat-SS Sparse $M$}
\end{subfigure}
\begin{subfigure}{.16\textwidth}
  \centering
  \includegraphics[width=1\linewidth]{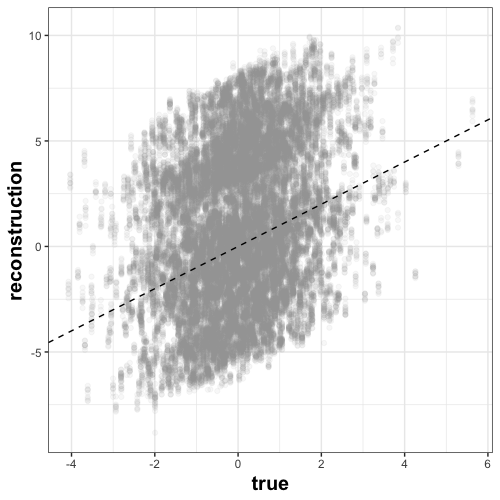}
  \caption*{\tiny FastBFA Sparse $M$}
\end{subfigure}
\begin{subfigure}{.16\textwidth}
  \centering
  \includegraphics[width=1\linewidth]{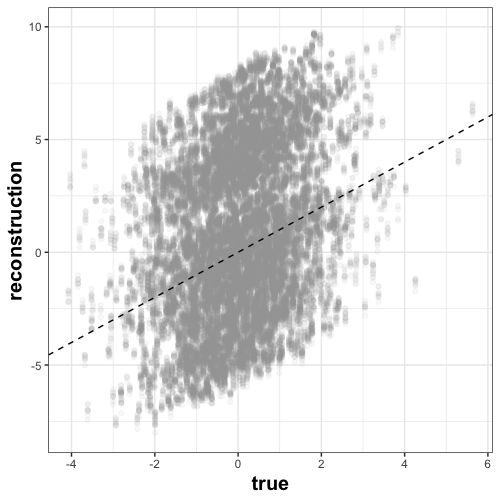}
  \caption*{\tiny LASSO Sparse $M$}
\end{subfigure}
\caption{Scatterplots comparing $ZM^\top$ vs. $\Ex[Z\mid\hat{\Delta},X]\hat{M}^\top$ between the different models  under dense (top) and truly sparse (bottom) loadings $M$ with $q=100$ in simulations with batch effect.}
\label{ZMbe}
\end{figure}

We evaluate our method in our main setting of interest where there are mean and variance batch effects. 
We emphasise that, the competing methods are not designed to account for batch effects; thus, this is not a fair comparison but rather an illustration of how much inference can suffer when not properly accounting for batches.
Also, since Flat-SS, Normal-SS and MOM-SS do incorporate batches, comparing them illustrates the advantages of NLP-based sparsity, e.g.\ see the bottom row in Figure~\ref{ZMbe}.
 
We simulated data with a mean and variance batch effect, $\xv_i=\theta^* \vv_i + M^* \zv_i + \beta^* \bv_i + \ev_i$, sample size $n=200$ and growing $p=250$ or $p=500$. 
We set  $q^*=10$, $p_v=1$ and $p_b = 2$ batches and considered the truly sparse and dense loadings $M^*$ in Figure~\ref{HeatmapMtoy}. 
Factors $\zv_i$ were drawn from $N(0,\I_q)$, errors $\ev_i$ from $N(0,\Sb^{-1})$, where $\tau_{j1}^{-1}=0.5$ and $\tau_{j2}^{-1}=1.5\tau_{j1}^{-1}$ for $j=1,\dots,p$; $\vv_i$ from a continuous Uniform(0,3) and $\bv_i$ from a discrete Uniform\{0,1\}.
We set the first $p/2$ values of $\theta^* \in \R^p$ to -2 and the other $p/2$ to 2 and $\beta_{j1}^* =0$, $\beta_{j2}^* =2$ for $j=1,\dots,p$ we fixed to 2 for the first batch and 0 for the second.
We compared our models with FastBFA and LASSO-BIC without batch effect correction for illustration of the importance of a proper mean and variance batch effect adjustment; and with empirical Bayes batch effect correction, ComBat, followed with an MLE estimation of the parameters ComBat-MLE.  Table~\ref{table:n200BE} shows the results.
The following plots show the comparison between the true $ZM^\top$ against their reconstruction $\Ex[Z\mid\hat{\Delta},X]$ in the scenario with sparsity with factors $q=100$.


Firstly, we considered the scenario when one correctly guesses $q=10$ and loadings are truly dense and sparse solutions could provide poor estimations. 
MOM-SS and Normal-SS achieved similar performance as the case without batch effect and similar results were observed for the $q=100$ case.
MOM-SS estimated correctly the latent cardinality $q^*=10$ and achieved a small estimation error for $\Ex[X]$.

Secondly, we studied the scenario with sparse factors. 
MOM-SS achieved a small estimation error for the mean and was effective in estimating $q^*=10$. 
LASSO-BIC had a small estimation error of the mean, although solutions were generally less sparse in the number of non-zero loadings.

It is important to highlight that even though ComBat-MLE, FastBFA and LASSO-BIC achieved a precise reconstruction of $\Ex[X]$ for purposes of dimensionality reduction the estimates of $ZM^\top$ are less precise as shown in Table~\ref{table:n200BE} and Figure~\ref{ZMbe} (right panels). 
Furthermore, the estimated covariance of the model displayed in the heatmap in Supplementary Sections~\ref{Sup:PlotsS2} and \ref{Sup:PlotsD2}, Supplementary Figures \ref{fig:PlotsS2}  (j)-(l) and \ref{fig:PlotsD2} (j)-(l) are nowhere close to the generating truth.
We remark that for FastBFA and LASSO-BIC these results mainly highlight that one should take into account batch effects. For Combat-MLE they highlight the limitations of using two-step procedures relative to a joint estimation of the factor model and batch effects

\subsection{Applications to cancer datasets}
\label{Sec:Cancer}
We applied our method to two high-dimensional cancer datasets, related to ovarian and lung cancer.
For the ovarian cancer we combined information from two datasets from the package \texttt{curatedOvarianData 1.16.0}. 
The first was the Illumina Human microRNA array expression dataset \texttt{E.MTAB.386}, formed by Angiogenic mRNA and microRNA gene expression signature with $n_1=129$ patients \citep{EMTAB386r}. 
The second was the NCI-60 GEO dataset \texttt{GSE30161} and consisted of multi-gene expression predictors of single drug responses to adjuvant chemotherapy in ovarian carcinoma for $n_2=52$  patients \citep{GSE30161}.
For the lung cancer, we used microarray and mRNA-array, data from two different high-throughput platforms: Affymetrix Human Genome \texttt{U133A 2.0} Array with $n_1=133$ patients and Affymetrix Human \texttt{Exon 1.0} ST Array with $n_2=112$ \citep{TCGAr}.

We considered two main tasks: to give a visual representation of the latent factors of the data, i.e.\ an unsupervised dimension reduction task and a supervised survival analysis using the factors obtained in our method as predictions. 
Prior to our analyses, we selected the 10\% genes with highest total variance across all samples obtaining $p=1,007$ for ovarian and $p=1,198$ for lung. 
All data sets have been normalised to zero mean and unit variance. 
We included the age at initial pathologic diagnosis as a covariate.

\subsubsection{Unsupervised: Data visualisation}

Our first goal was to demonstrate the usefulness of our method as a data visualisation tool. 
We remark that there are no other model-based approaches to jointly adjust for batch effects and estimate latent factors. 
Thus, for comparison we first corrected the data using ComBat and then estimated the latent parameters via MLE and FastBFA akin to Section~\ref{Sec:BE}.
To decide the number of factors for ComBat-MLE, we carried a principal component analysis to the corrected data prior to factor analysis and chose a number of components $\hat{q}$ that explained 90\% or 70\% of the total variance. 
It is important to notice that we are doing an over-optimistic assessment of ComBat-MLE and ComBat-FastBFA as we are doing a cross-validated factor analysis over the ComBat-corrected data, as opposed to also running ComBat in an out-of-sample fashion.

\begin{figure}[h!]
\centering
\begin{subfigure}{.475\textwidth}
  \centering
  \includegraphics[width=.75\linewidth]{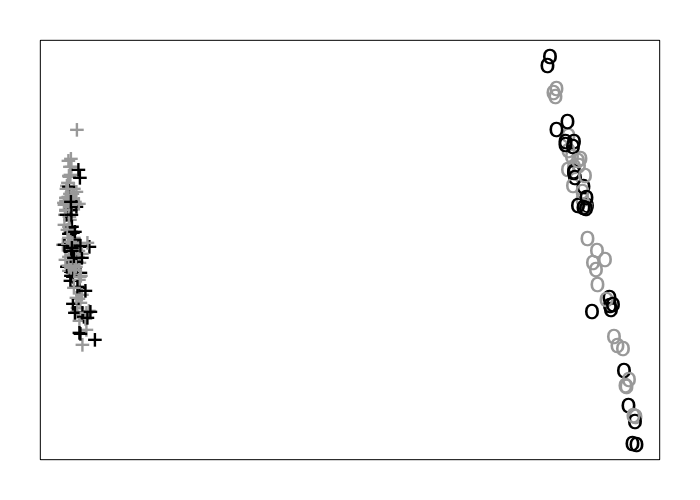}
  \caption{Ovarian no correction}
  \label{fig:OvNB}
\end{subfigure}
\begin{subfigure}{.475\textwidth}
  \centering
  \includegraphics[width=.75\linewidth]{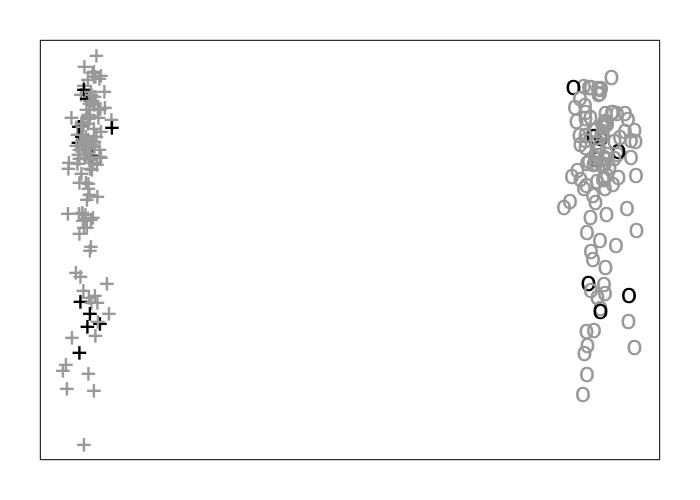}
  \caption{Lung no correction}
  \label{fig:LunNB}
\end{subfigure}\\
\begin{subfigure}{.475\textwidth}
  \centering
  \includegraphics[width=.75\linewidth]{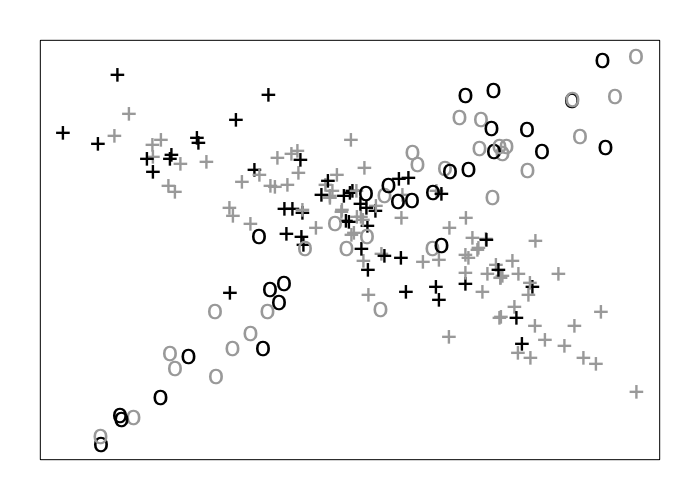}
  \caption{Ovarian ComBat-MLE}
  \label{fig:ovComBat}
\end{subfigure}
\begin{subfigure}{.475\textwidth}
  \centering
  \includegraphics[width=.75\linewidth]{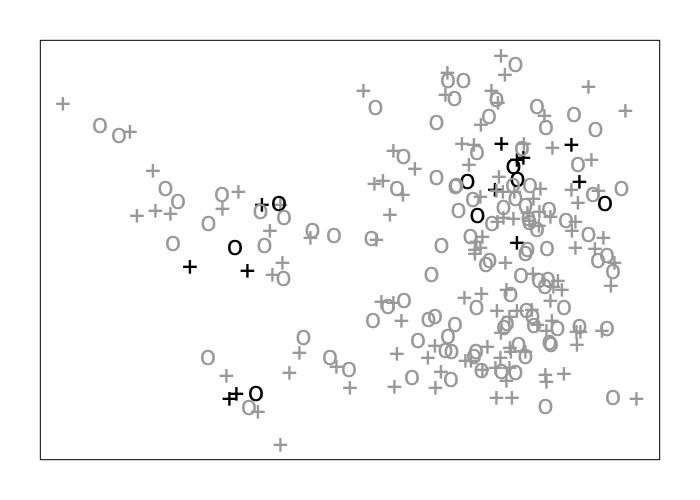}
  \caption{Lung ComBat-MLE}
  \label{fig:LungComBat}
\end{subfigure}\\
\begin{subfigure}{.475\textwidth}
  \centering
  \includegraphics[width=.75\linewidth]{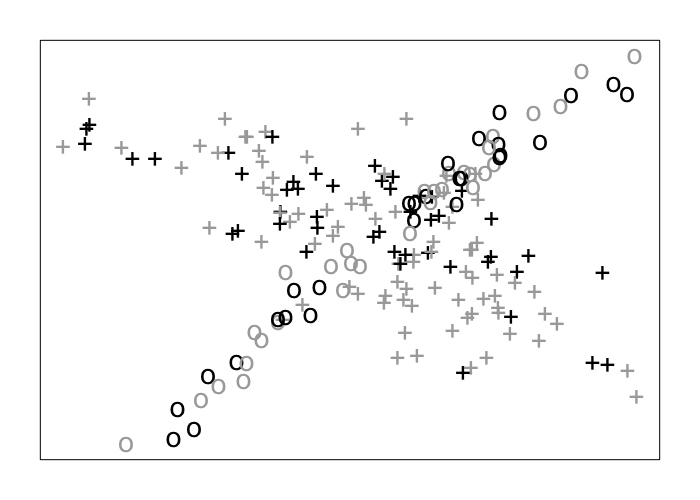}
  \caption{Ovarian ComBat-FastBFA}
  \label{fig:ovComBatFastBFA}
\end{subfigure}
\begin{subfigure}{.475\textwidth}
  \centering
  \includegraphics[width=.75\linewidth]{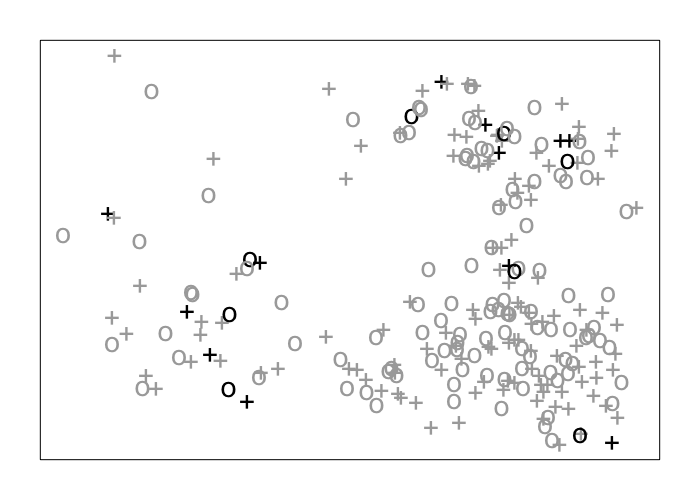}
  \caption{Lung ComBat-FastBFA}
  \label{fig:LungComBatFastBFA}
\end{subfigure}\\
\begin{subfigure}{.475\textwidth}
  \centering
  \includegraphics[width=.778\linewidth]{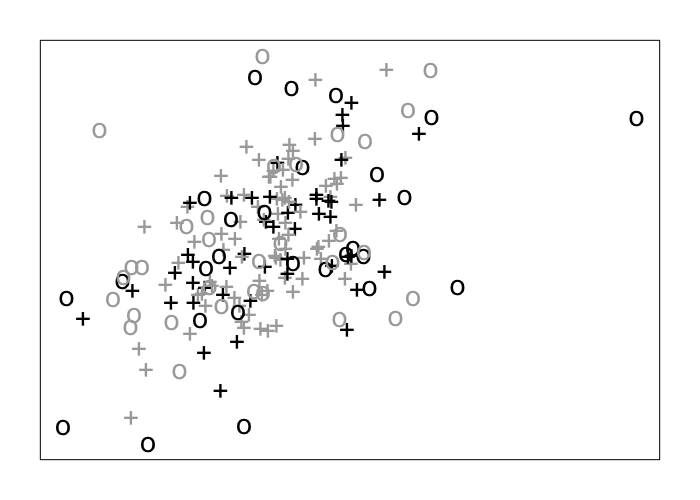}
  \caption{Ovarian MOM-SS unsupervised}
  \label{fig:OvMOMunad}
\end{subfigure}
\begin{subfigure}{.475\textwidth}
  \centering
  \includegraphics[width=.778\linewidth]{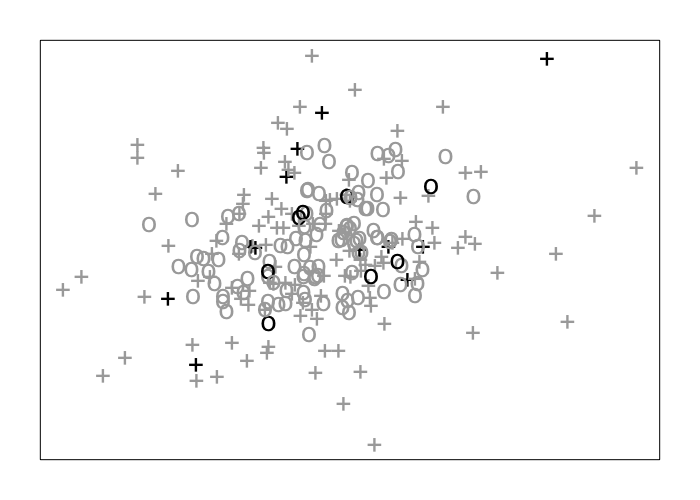}
  \caption{Lung MOM-SS unsupervised}
  \label{fig:LungMOMunad}
\end{subfigure}
\caption{Scatterplot of the first two factors of ovarian (left) and lung (right) datasets for the two different batches (pluses and circles) and displaying in black the patients who died within the first three years. Comparison between models without batch effect adjustment, ComBat-MLE, ComBat-FastBFA and MOM-SS. }
\label{Cancer}
\end{figure}

Figure~\ref{Cancer} illustrates the advantages of our method. 
We can clearly see the usefulness of ComBat correction (middle panels) compared to scenarios without correction (top panels), ComBat removes systematic differences in location and scale across the 2 batches. 
Nonetheless, the latent coordinates displayed distinct covariances for the ovarian cancer dataset. 
Such covariances were not presented in the MOM-SS latent factors (bottom panels).
Figures~\ref{Cancer} (g) and (h) show the two factors that contribute the most to the covariance, i.e.\  the ones with highest $\sum_{j=1}^p \hat{m}^2_{jk}$. 
The latent coordinates were post-processed to standardised their variance $\text{Cov}(\zv_i\mid\hat{\Delta},X)$ as explained in Section \ref{Postprocessing}.

\subsubsection{Supervised: Survival analysis}

We also illustrate the potential of our method as a surpervised tool, performing a survival analysis that aims to predict the time until death. 
To do that, we applied a Cox proportional hazards model \citep{CoxModel} using as covariates the latent coordinates obtained in our models. 
We used the \texttt{coxph} function of the  \textbf{R} package \texttt{survival 2.38} \citep{survival-package}.
We then used the concordance index to asses the quality of our predictions. 
This index is a non-parametric metric to quantify the power of a prediction rule via a pair-wise comparison that measures the probability of concordance between the predicted and the observed survival time \citep{concordance}. 
To obtain the concordance index we used the function \texttt{concordance.index} in the \textbf{R} package \texttt{survcomp} \ \citep{survcomp}. 
The presented results are from 10 independent runs of 10-fold cross-validation.
We initialised MOM-SS with the values obtained for the Flat model along with the other initialisations discussed in Section~\ref{IniEMNS} and chose the one with smallest leave-one-out cross-validated concordance index.


For the cancer data sets,  Table~\ref{table:CancerSup} shows that Flat-SS achieved a high concordance index, even though loadings are not sparse.
Normal-SS gave sparse loading representations but displayed a concordance index lower than Flat-SS; this illustrates a lack of power to detect truly non-zero loadings.
In general, MOM-SS provided sparse loadings and a good concordance index.
In the ovarian cancer data, MOM-SS achieved a concordance index similar to ComBat-MLE 90\% with considerably less factors (4 instead of 101) and a bit higher than Normal-SS. 
In the lung cancer data MOM-SS achieved a high concordance index, particularly relative to Normal-SS and ComBat-MLE 70\%.
The competing methods generally lead to less sparse solutions and their performance fluctuates across scenarios.
In the lung cancer data ComBat-MLE, despite its good performance, had a concordance index that proved to be sensitive to the number of factors (see ComBat-MLE 90\% vs 70\%).
ComBat-FastBFA provided competitive results with a non-sparse reconstruction, recovering values in the latent loadings that were close to zero (even though not exactly zero) and smaller  than the ones of the Flat-SS.
MOM-SS proved to have practical advantages as a supervised tool in comparison with the two-step approaches considered here.
Overall, MOM-SS provided a more stable performance that achieved a good balance between sparsity and prediction accuracy.

\begin{table}[h!]
\centering
\caption{Survival analysis for ovarian ($p=1,007$ genes) and lung ($p=1,198$ genes) cancer data sets.}
\scalebox{0.8}{
\begin{tabular}{r|rrc|rrc}
  \hline
  &\multicolumn{3}{c|}{Ovarian}&\multicolumn{3}{c}{Lung} \\
	\hline
 & $\hat{q}$ & $|\hat{M}|_0$ & \small{Concordance index} & $\hat{q}$ & $|\hat{M}|_0$ & \small{Concordance index} \\ 
  \hline
  Flat & 100.0 & 100700.0 & 0.634 & 100.0 & 119800.0 & 0.669 \\ 
  Normal-SS & 7.8 & 7854.6 & 0.568 & 11.0 & 13178.0 & 0.489 \\ 
  \rowcolor{lightgray} 
  MOM-SS & 4.0 & 4028.0 & 0.588 & 74.0 & 88652.0 & 0.665 \\ 
  ComBat-MLE 90\% & 101.0 & 101707.0 & 0.589 & 79.0 & 94642.0 & 0.688 \\ 
  ComBat-MLE 70\% & 41.0 & 41287.0 & 0.588 & 30.0 & 35940.0 & 0.568 \\ 
  ComBat-FastBFA & 100.0 & 100700.0 & 0.527 & 100.0 & 119800.0 & 0.707 \\ 
   \hline
\end{tabular}
}
\label{table:CancerSup}
\end{table}

\section{Discussion}
\label{sec:Conclusion}
We have presented a novel model to integrate data from multiple sources using joint dimension reduction and batch effect adjustment via high-dimensional latent factor regression.
We outlined three different prior configurations for the loadings and Laplace-tailed extensions whose deeper analysis remain as future work. 
To our knowledge this is the first time NLPs are implemented in the factor analysis context. 
We gave novel EM algorithms to obtain posterior modes. 
We showed that the use of sparse models increases the quality of our estimations even in the absence of batches. 
In our empirical results MOM-SS priors proved to be appealing, improving the estimation of factor cardinality and encouraging parsimony and selective shrinkage.

We illustrated the utility of our method in unsupervised and supervised frameworks. 
MOM-SS provided dimension reduction that corrected distinct covariance patterns present in two-stage methods that adjust variances separately from fitting the factor model. 
Such patterns are highly likely to be technical artefacts, since patients from different batches are believed to be exchangeable.
Our model demonstrated to be useful for downstream analyses, achieving a competitive concordance indexes, in some cases with substantially less factors. 
It is important to notice that although our examples focus on gene expression of cancer datasets, the applications should also be useful in other settings. 

We also remark that our novel MOM-SS and its closed-form EM updates can be extended to frameworks of interest beyond factor models such as: linear regression, generalised linear models as well as graphical models.

Our model assumes common factors across the datasets being integrated. 
An interesting extension for future research is to consider more complex settings where some of the factors differ across data sources or where one wishes to integrate datasets by adding variables (as opposed to adding individuals as we did here), or where potentially same variables were only recorded for a subset of the individuals.

\section*{Supplementary Materials}
The supplementary materials are as follow: 
EM algorithm under a flat, Normal-SS, MOM-SS, Laplace-SS and Laplace-MOM-SS on the loadings, a pseudo-code-algorithm for the weighted 10-fold cross-validation, and heatmaps for $\hat{M}$, $\widehat{\text{Cov}}(\xv_i\mid\cdot)^{-1}$ and $\hat{\gamma}$ for the different simulated scenarios and setting $q=100$.

\section*{Acknowledgements}
We thank Chris Yau for valuable insights and Veronika Ročková for providing the FastBFA package. 
\section*{Funding}
Alejandra Avalos-Pacheco gratefully acknowledges \textit{the Mexican National Council of Science and Technology (CONACYT) grant no.\ CVU5464444}.
David Rossell was partially funded by the \textit{NIH grant R01 CA158113-01, RyC-2015-18544} and \textit{Ayudas Fundación BBVA a equipos de investigación científica 2017}.

\clearpage
\appendix
\bigskip
\begin{center}
{\large\bf APPENDIX}
\end{center}
\section{Proof of Lemma \ref{Lemma:maxMjk}, $m_{jk}$ MOM-SS global mode}
\label{app:maxMjk}
\begin{proof}
 Our goal is to max $f(m_{jk})= a m_{jk}^2 + b m_{jk} + c \log(m_{jk}^2)$. Take derivative with respect to $m_{jk}$
\begin{align}
\frac{d}{dm_{jk}}=2a m_{jk} + b + 2c/m_{jk} = 0 \implies 2a m_{jk}^2 + b m_{jk} + 2c=0. \nonumber
\end{align}
Roots are $\underbar{m}_{jk} := \frac{-b - \sqrt{b^2 - 16ac}}{4a}$ and $\bar{m}_{jk} := \frac{-b + \sqrt{b^2 - 16ac}}{4a}$.\\

If $f(\bar{m}_{jk})-f(\underbar{m}_{jk})>0$ then the global max is $\bar{m}_{jk}$, else the global max is $\underbar{m}_{jk}$. After trivial algebra, $f(\bar{m}_{jk}) - f(\underbar{m}_{jk}) = \frac{b}{4a} \sqrt{b^2 - 16ac} + c \log\left(\left[\frac{-b + \sqrt{b^2 - 16ac} }{ b + \sqrt{b^2 - 16ac}}\right]^2\right)$.\\

For ease of notation let $z= \sqrt{b^2-16ac}$. Note that $z>0$ and that, since $a<0$, $c>0$, that implies that $z-b>0$. 
Then $f(\bar{m}_{jk})-f(\underbar{m}_{jk})>0$ if and only if $\frac{bz}{4a}  > c \log \left(\left[ \frac{z+b}{z-b}\right]^2\right)= 2c \log \left(\left[ \frac{z+b}{z-b}\right]\right)$.
Equivalently, $f(\bar{m}_{jk})-f(\underbar{m}_{jk})>0$ if and only if $\frac{bz}{8ac} > \log(z+b) - \log(z-b)$.

\begin{itemize}
\item Suppose $b>0$. Then left-hand side is $<0$, and right-hand side is $>0$. Hence $f(\bar{m}_{jk})-f(\underbar{m}_{jk})<0 \implies$ global maximum is $\underbar{m}_{jk}$
\item Suppose $b<0$. Then left-hand side is $>0$, and right-hand side is $<0$. Hence $f(\bar{m}_{jk})-f(\underbar{m}_{jk})>0 \implies$ global maximum is $\bar{m}_{jk}$
\qedhere
\end{itemize}
\end{proof}

\section{Proof of Lemma~\ref{Lemma:maxMjkLaplaceSS},$m_{jk}$ Laplace-SS global mode.}
\label{app:maxMjkLSS}
\begin{proof}
Our purpose is to find the maximum of $f(m_{jk})=  a m_{jk}^2 + b m_{jk} + c |m_{jk}|$, where $a<0$, and $c<0$. 
Setting $\frac{\partial Q_1}{\partial m_{jk}}=0$, we obtain
\[
\frac{\partial Q_1}{\partial m_{jk}}=2a m_{jk} + b + c \cdot \sign(m_{jk}) = 0.
\]

\begin{itemize}
\item For $m_{jk}>0$, we look for the solutions of $2a m_{jk} + b+c=0$. Note $a<0$ and $c<0$. Thus
\[
arg \max_{m_{jk} \geq 0} f(m_{jk})=
\begin{cases}
m_{jk}^+ :=\frac{-(b+c) }{2a} & b >-c\\
0 & \text{ otherwise}
\end{cases}
\]
\item For $m_{jk}<0$, we look for the solutions of $2a m_{jk} + b-c=0$. Thus
\[
arg \max_{m_{jk} \leq 0} f(m_{jk})=
\begin{cases}
m_{jk}^- :=\frac{-(b-c) }{2a} & b<c\\
0 & \text{ otherwise}
\qedhere
\end{cases}
\]
\end{itemize}
\end{proof}

\section{Proof of Lemma \ref{Coro:maxMjk}, $m_{jk}$ Laplace-MOM-SS global mode}
\label{app:maxMjkL}
\begin{proof}
We aim to find the maximum of $f(m_{jk})=  a m_{jk}^2 + b m_{jk} + c |m_{jk}|+ d \log(m_{jk}^2)$, where $a<0$, $c<0$ and $d>0$. 
Note that when $m_{jk}=0$, $Q_1(m_{jk}=0)=-\infty$.
Thus, the maximum of $f$ is one of its critical points.
Setting $\frac{\partial Q_1}{\partial m_{jk}}=0$, we obtain
\[
\frac{\partial Q_1}{\partial m_{jk}}=2a m_{jk} + b + c \cdot \sign(m_{jk}) + 2d/m_{jk} = 0 \implies 2a m_{jk}^2 + b m_{jk} + c \cdot \sign(m_{jk}) m_{jk}+ 2d=0.
\]

\begin{itemize}
\item For $m_{jk}>0$, we look for the solutions of $2a m_{jk}^2 + (b+c) m_{jk} + 2d=0$.

The roots of this polynomial are $\frac{-(b+c) \pm \sqrt{(b+c)^2 - 16ad}}{4a}$.
Note that $\sqrt{(b+c)^2 - 16ad}>\lvert b+c \rvert$ since $a<0$ and $d>0$. 
Hence, the only acceptable root is $m_{jk}^+ := \frac{-(b+c) - \sqrt{(b+c)^2 - 16ad}}{4a}>0$, as the other one is negative.
\item For $m_{jk}<0$, we look for the solutions of $2a m_{jk}^2 + (b-c) m_{jk} + 2d=0$.

The roots of this polynomial are $ \frac{-(b-c) \pm \sqrt{(b-c)^2 - 16ad}}{4a}$. 
As before, $\sqrt{(b-c)^2 - 16ad}>\lvert b-c \rvert$. 
Hence, the only acceptable root is $m_{jk}^- := \frac{-(b-c) + \sqrt{(b-c)^2 - 16ad}}{4a}<0$, as the other one is positive.
\end{itemize}

\begin{itemize}
\item Suppose $b=0$. Then clearly $f(m_{jk}) = f(-m_{jk})$ for all $m_{jk}$, i.e.\ the function is even. Therefore, $m_{jk}^+$ and $m_{jk}^-$ are opposite and both arg maxima.
\item Suppose $b>0$. By definition of $f$, $f(m_{jk}) > f(-m_{jk})$ for all $m_{jk} > 0$. In particular, $\max_{m_{jk}>0} f(m_{jk}) \geq \max_{m_{jk}<0} f(m_{jk})$ and $m^+_{jk}= \arg \max_{m_{jk}} f(m_{jk})$.
\item Suppose $b<0$. Then $f(m_{jk}) < f(-m_{jk})$ for all $m_{jk} > 0$. In particular, $\max_{m_{jk}>0} f(m_{jk}) \leq \max_{m_{jk}<0} f(m_{jk})$ and $m^-_{jk}= \arg \max_{m_{jk}} f(m_{jk})$.
\qedhere
\end{itemize}

\end{proof}

\bibliographystyle{chicagoff}
\bibliography{ms} 

\newpage
\setcounter{page}{1}
\appendix
\bigskip
\begin{center}
{\large\bf SUPPLEMENTARY MATERIAL}
\end{center}

\section{EM algorithm under a flat prior on the loadings}
\label{appendix:EMV1}
Here, we outline the derivation of the Expectation-Maximisation (EM) algorithm to fit the latent factor regression model with mean and variance adjustment presented in Section~\ref{sec:NSalgorithm}  via Maximum a posteriori (MAP) estimation.
Our goal is to find values $(\theta,M,\beta,\Tau)$ that maximise the log-posterior 
\begin{equation}
\label{eq:VanillaLogpost}
\log\Prob(M, \theta, \beta, \Tau \mid X) \propto \log\Prob(X \mid M, \theta, \beta, \Tau) +\log\Prob(M, \theta, \beta, \Tau) 
\end{equation}
To maximise \eqref{eq:VanillaLogpost} the EM algorithm make use of complete-data log-posterior associated to $(X,V,B,Z)$
\begin{equation}
\label{Logpostvanilla}
\log\Prob(M, \theta, \beta, \Tau \mid X,Z) \propto \log \Prob (X,Z \mid \theta,M,\beta,\Tau) +\log\Prob(M, \theta, \beta, \Tau)
\end{equation}

For simplicity we will denote by $\Prob(Z\mid \hat{\Delta},X)=\Prob(Z\mid M=M^{(t)}, \theta=\theta^{(t)}, \beta=\beta^{(t)}, \Tau=\Tau^{(t)},X)$ the probability with respect to the latent variables and conditioning upon $\Delta=(\theta, \beta, M, \Tau)$ at time t. 
Similarly, $\Ex[\zv_i \mid\hat{\Delta},X]$ the mean conditional on $X$ and all other model parameters $\Delta$, and likewise for $\Ex[\zv_i \zv_i^\top\mid\hat{\Delta},X]$.

We first outline the E-step, which is based on taking the expectation of Expression~\eqref{Logpostvanilla} with respect to $\Prob(Z\mid \hat{\Delta},X)$, namely:
\begin{equation}
\label{EMvanilla}
\begin{split}
Q(\Delta)=&\Ez\left[ \log \Prob (X,Z \mid \theta,M,\beta,\Tau) +\log\Prob(M, \theta, \beta, \Tau) \right] \\
=& C -\frac{1}{2} \sum_{i=1}^n \left[(\xv_i-\hat{\theta} \vv_i - \hat{\beta} \bv_i)^\top \hat{\Tau}_{\bv_i} (\xv_i-\hat{\theta} \vv_i - \hat{\beta} \bv_i) \right.  \\
&\left. -2 (\xv_i-\hat{\theta} \vv_i - \hat{\beta} \bv_i)^\top \hat{\Tau}_{\bv_i} \hat{M}\Ex[\zv_i \mid\hat{\Delta},X]   + \Tr\left(\hat{M}^{\top} \hat{\Tau}_{\bv_i} \hat{M} \Ex[\zv_i \zv_i^\top \mid\hat{\Delta},X]\right) \right] \\
&+ \sum_{l=1}^{p_b} \left[ \frac{n_l+ \eta  -2 }{2} \log ( \mid\hat{\Tau}_{l}\mid )-\frac{\eta \xi}{2} \Tr( \hat{\Tau}_{l} ) \right] -\frac{1}{2} \sum_{j=1}^p (\hat{\theta}_j,\hat{\beta}_j)^\top \frac{1}{\psi}\I (\hat{\theta}_j,\hat{\beta}_j),
\end{split}
\end{equation}
where $C$ is a constant, and we have defined as usual $\Tau_{l}:=$ diag$(\tau_{1l},\dots,\tau_{pl})$ and $l_i$ to be the unique $l = 1,\dots,p_b$ such that $b_{il}=1$.\\

Expression~\eqref{EMvanilla} only depends on $Z$ through the conditional posterior mean 
\begin{equation}
\label{EZx}
\Ex[\zv_i| \hat{\Delta},X] = (\I_q + \hat{M}^\top \hat{\Sb}\hat{M})^{-1} \hat{M}^\top\hat{\Sb}(\xv_i -\hat{\theta} \vv_i - \hat{\beta} \bv_i)
\end{equation}
and the conditional second moments
\begin{equation}
\label{EZZx}
\Ex[\zv_i \zv_i^\top\mid\hat{\Delta},X] = (\I_q + \hat{M}^{\top} \hat{\Tau}_{\bv_i} \hat{M})^{-1} +\Ex[\zv_i\mid\hat{\Delta},X]\Ex[\zv_i\mid\hat{\Delta},X]^\top,
\end{equation}

The M-step consists in maximising Equation~\eqref{EMvanilla} with respect to $\Delta$. To this end, we set its partial derivatives to 0, as shown below.

\begin{equation}
\label{Mvanila}
\frac{\partial Q}{\partial M}=-\frac{1}{2}\sum_{i=1}^n \left[-2 \hat{\Tau}_{\bv_i}(\xv_i-\hat{\theta} \vv_i - \hat{\beta} \bv_i) \Ex[\zv_i^\top \mid\hat{\Delta},X] + 2 \hat{\Tau}_{\bv_i} \hat{M} \Ex[\zv_i \zv_i^\top \mid\hat{\Delta},X] \right]=0
\end{equation}

The maximum of the $j^{th}$ row of matrix $M$  can be found solving \eqref{Mvanila} as:
\begin{equation}
\label{MVanillaT}
\hat{m}_j=\left[  \sum_{i=1}^n \left[ \hat{\tau}_j^{\top}\bv_i(x_{ij}-\hat{\theta} v_{ij} - \hat{\beta} b_{ij}) \Ex[\zv_i^\top \mid\hat{\Delta},X] \right]  \right] \left[ \sum_{i=1}^n \left[ \hat{\tau}_j^{\top} \bv_i \Ex[\zv_i \zv_i^\top \mid\hat{\Delta},X] \right] \right]^{-1}
\end{equation}
for $ j=1,\dots,p.$

Maximisation of $\Tau$ for a fixed batch $l$ is obtained by taking the derivative with respect to $\Tau_l$ 
\begin{equation}
\label{SigmaVanilla}
\begin{split}
\frac{\partial Q}{\partial \Tau_l}=&-\frac{1}{2}\sum_{i\colon 
	b_{il} = 1}\left[(\xv_i-\hat{\theta} \vv_i - \hat{\beta} \bv_i) (\xv_i-\hat{\theta} \vv_i - \hat{\beta} \bv_i)^\top \right.\\
&\left. -2 (\xv_i-\hat{\theta} \vv_i - \hat{\beta} \bv_i) \Ex[\zv_i \mid\hat{\Delta},X]^\top  \hat{M}^{\top} + \hat{M}  \Ex[\zv_i \zv_i^\top \mid\hat{\Delta},X] \hat{M}^{\top} \right]\\
&+\frac{n_l + \eta  -2 }{2} \hat{\Tau}_l^{-1}-\frac{\eta \xi}{2} \I_p= 0 .
\end{split}
\end{equation}

Solving Equation~\eqref{SigmaVanilla} and using the diagonal constraint we obtain:
\begin{equation}
\label{SigmaVanillaT}
\hat{\Tau}_l^{-1}=\frac{1}{n_l+ \eta  -2}\text{diag}\left\{\sum_{i\colon 
	b_{il} = 1}\left( \tilde{\xv}_i \tilde{\xv}_i^\top  -2\tilde{\xv}_i \Ex[\zv_i \mid\hat{\Delta},X]^\top  \hat{M}^{\top} + \hat{M}  \Ex[\zv_i \zv_i^\top \mid\hat{\Delta},X] \hat{M}^{\top} \right) + \eta \xi \I_p \right\}
\end{equation}
with $\tilde{\xv}_i=\xv_i-\hat{\theta} \vv_i - \hat{\beta} \bv_i$.

To maximise with respect to $(\theta, \beta)$ we set
\begin{equation}
\label{ThetaVanilla}
\frac{\partial Q}{\partial (\theta, \beta)}=-\sum_{i=1}^n \left[ \hat{\Tau}_{\bv_i}(\hat{\theta},\hat{\beta})(\vv_i,\bv_i)(\vv_i,\bv_i)^{\top} -\hat{\Tau}_{\bv_i} (\xv_i - \hat{M} \Ex[\zv_i \mid\hat{\Delta},X] ) (\vv_i,\bv_i)^\top \right] -\frac{1}{\psi}  (\hat{\theta},\hat{\beta}) =0
\end{equation}

Taking the $j^{th}$ row of matrix $(\hat{\theta}, \hat{\beta})$ and solving Equation~\eqref{ThetaVanilla}:
\begin{equation}
\label{ThetaVanillaT}
(\hat{\theta}_j^{\top}, \hat{\beta}_j^{\top}) = \sum_{i=1}^n \left[ \hat{\tau}_{j}^\top \bv_i (x_{ij} - \hat{m}_j^\top \Ex[\zv_{i} \mid\hat{\Delta},X] ) (\vv_i,\bv_i)^\top \right] \left[ \sum_{i=1}^n \left[ \hat{\tau}_{j}^\top \bv_i (\vv_i,\bv_i)(\vv_i,\bv_i)^{\top} \right]  +\frac{1}{\psi} \I\right]^{-1}
\end{equation}

Equation~\eqref{ThetaVanillaT} has the form of a ridge regression estimator with penalty $\psi$, inducing an equal shrinkage to each coefficient of $(\hat{\theta}, \hat{\beta}).$

\section{EM algorithm under Normal-SS}
\label{app:EMSSL}
Akin to the Flat prior, we first take the expectation of the complete-data log-posterior with respect to the latent variables and conditioning upon the current $\Delta=(M,\theta,\beta,\Tau,\zeta)$:
\begin{equation}
\label{ELogpostSSL}
Q(\Delta)\propto\Ezg \left[ \log \Prob (X,Z,\gamma \mid M,\theta,\beta,\Tau,\zeta) + \log \Prob(M,\theta,\beta,\Tau,\zeta) \right]
\end{equation}
Due to the conjugate Normal-SS hierarchical construction, Expression~\eqref{ELogpostSSL} can be split in order to simplify the EM algorithm as $Q(\Delta) =C+ Q_1(\theta, M, \beta, \Tau)+Q_2(\zeta)$,  where:
\begin{align}
\label{Ezx}
Q_1(\theta, M, \beta, \Tau)=
&-\frac{1}{2} \sum_{i=1}^n \left[(\xv_i-\hat{\theta} \vv_i - \hat{\beta} \bv_i)^\top \hat{\Tau}_{\bv_i} (\xv_i-\hat{\theta} \vv_i - \hat{\beta} \bv_i) -2 (\xv_i-\hat{\theta} \vv_i - \hat{\beta} \bv_i)^\top \hat{\Tau}_{\bv_i}\hat{M}\Ex[\zv_i \mid\hat{\Delta},X]   \right. \nonumber \\
&\left.  + \Tr\left(\hat{M}^{\top} \hat{\Tau}_{\bv_i} \hat{M} \Ex[\zv_i \zv_i^\top \mid\hat{\Delta},X]\right) \right]+ \sum_{l=1}^{p_b} \frac{n_l+ \eta  -2}{2} \log \mid\hat{\Tau}_{l}\mid -\sum_{l=1}^{p_b} \frac{\eta \xi}{2} \Tr( \hat{\Tau}_{l} ) \\
& -\frac{1}{2} \sum_{j=1}^p (\hat{\theta}_{j}^\top,\hat{\beta}_{j}^\top) \frac{1}{\psi}\I (\hat{\theta}_{j},\hat{\beta}_{j})  -\frac{1}{2} \sum_{j=1}^p \sum_{k=1}^q  \hat{m}^2_{jk} \Ex \left[ \frac{1}{(1-\gamma_{jk}) \lambda_{0} + \gamma_{jk}\lambda_{1}} \mid \hat{\Delta}  \right] , \nonumber \\
\label{Q2NSS}
Q_2(\zeta)=&\sum_{j=1}^p\sum_{k=1}^q  \log\left( \frac{\hat{\zeta}_k}{1-\hat{\zeta}_k}\right)  \Ex [\gamma_{jk}\mid \hat{\Delta}] +\sum_{k=1}^q \left((\frac{a_\zeta}{k}-1) \log(\hat{\zeta}_k)+(p+b_\zeta-1) \log(1-\hat{\zeta}_k)\right) .
\end{align}

The E-step for $Q_1$ resembles the one for the flat prior model shown in Supplement \ref{appendix:EMV1}, plus an extra conditional expectation:
\[
\Ex \left[ \frac{1}{(1-\gamma_{jk}) \lambda_{0} + \gamma_{jk}\lambda_{1}} \mid \hat{\Delta}  \right] = \frac{1-\hat{p}_{jk}}{\lambda_0} +\frac{\hat{p}_{jk}}{\lambda_1} ,
\]
with $\hat{p}_{jk}=\Prob(\gamma_{jk}=1 \mid \hat{\Delta})$ given by
\[\hat{p}_{jk}=\frac{1}{1+\sqrt{\frac{\llN}{\loN}} \exp\left(-\frac{1}{2} \hat{m}_{jk}^2 \left( \frac{1}{\loN} - \frac{1}{\llN} \right) \right) \frac{1-\Ex[\zeta_j]}{\Ex[\zeta_j]}}\]

The first and second moments $\Ex[\zv_i \mid\hat{\Delta},X]$ and $\Ex[\zv_i \zv_i^\top\mid\hat{\Delta},X]$ respectively are given in Supplement \ref{appendix:EMV1}.

For $Q_2(\zeta)$ corresponds to a beta-binomial prior on $\gamma_{jk}$, with conditional expectations $
\Ex[\gamma_{jk} \mid \hat{\Delta}] = \Prob(\gamma_{jk}=1\mid\hat{\Delta}) =\hat{p}_{jk}$.

In the M-step we proceed by optimising $Q_1$ and $Q_2$ independently, in 2 steps: a maximisation of $Q_1$ with respect to $M$, $\hat{\Tau}_{l}$ and $(\theta, \beta)$, followed by a maximisation of $Q_2$ with respect to $\zeta$.
Setting to 0 the partial derivative with respect to $M$ gives:
\begin{equation}
\label{MSSL}
\begin{split}
\frac{\partial Q}{\partial M}=&-\frac{1}{2}\sum_{i=1}^n \left[-2 \hat{\Tau}_{\bv_i}(\xv_i-\hat{\theta} \vv_i - \hat{\beta} \bv_i) \Ex[\zv_i^\top \mid\hat{\Delta},X] + 2 \hat{\Tau}_{\bv_i} \hat{M} \Ex[\zv_i \zv_i^\top \mid\hat{\Delta},X] \right] \\
&-\hat{M} \circ \Ex[D_{\gamma}\mid \hat{\Delta}] = 0 ,
\end{split}
\end{equation}
with $D_{\gamma}\in \R^{p \times q}$, $d_{jk}=\frac{1}{(1-\gamma_{jk}) \lambda_{0} + \gamma_{jk}\lambda_{1}}$ and $A\circ B$ being the Hadamard (element-wise) product of two matrices $A$ and $B$.
Taking the $j^{th}$ row of matrix $M$ and solving equation~\eqref{MSSL} we obtain:
{\scriptsize
\[
\hat{m}_j=\left[\sum_{i=1}^n \left( \hat{\tau}_j^{\top}\bv_i(x_{ij}-\hat{\theta} v_{ij} - \hat{\beta} b_{ij}) \Ex[\zv_i^\top \mid\hat{\Delta},X] \right)\right]\left[  \text{diag}\{\Ex[d_{j1}\mid \hat{\Delta}], \dots ,\Ex[d_{jq}\mid \hat{\Delta}] \}+\sum_{i=1}^n \left( \hat{\tau}_j^{\top} \bv_i \Ex[\zv_i \zv_i^\top \mid\hat{\Delta},X] \right) \right]^{-1},
\]}
for $j=1,\dots,p.$ 
Updates for $\hat{\Tau}_l$ and $(\hat{\theta}, \hat{\beta})$ are the same ones given in Supplement \ref{appendix:EMV1}.

Finally,
\begin{align}
\label{ZetaSSL}
\frac{\partial Q_2}{\partial \zeta_k}=& \frac{ \sum_{j=1}^p \Ex[\gamma_{jk} \mid \hat{\Delta}]}{\hat{\zeta}_k-\hat{\zeta}_k^{2}}+\frac{\frac{a_\zeta}{k}-1}{\hat{\zeta}_k}-\frac{p+b_\zeta-1}{1-\hat{\zeta}_k}=0.
\end{align}

Solving Equation~\eqref{ZetaSSL} and substituting $\Ex[\gamma_{jk} \mid \hat{\Delta}]$:

\begin{align}
\label{eq:ZetaNormal}
\hat{\zeta}_k=\frac{\sum_{j=1}^p \hat{p}_{jk} + \frac{a_\zeta}{k} -1}{\frac{a_\zeta}{k} + b_\zeta + p -1 } .
\end{align}

\section{EM algorithm under MOM-SS}
\label{app:EMpMOM}

Analogous to Normal-SS, we first take the expected complete-data log-posterior $Q(\Delta) =C+ Q_1(\theta, M, \beta, \Sb)+Q_2(\zeta)$.  
By construction $Q_2$ is of the same form than in Equation~\eqref{Q2NSS} and $Q_1$ is given by
\begin{equation}
\begin{split}
\label{Ap:MOMQ1}
Q_1(\theta, M, \beta, \Tau)=
&-\frac{1}{2} \sum_{i=1}^n \left[(\xv_i-\hat{\theta} \vv_i - \hat{\beta} \bv_i)^\top \hat{\Tau}_{\bv_i} (\xv_i-\hat{\theta} \vv_i - \hat{\beta} \bv_i) -2 (\xv_i-\hat{\theta} \vv_i - \hat{\beta} \bv_i)^\top \hat{\Tau}_{\bv_i}\hat{M}\Ex[\zv_i \mid\hat{\Delta},X]   \right. \\ 
&\left.  + \Tr\left(\hat{M}^{\top} \hat{\Tau}_{\bv_i} \hat{M} \Ex[\zv_i \zv_i^\top \mid\hat{\Delta},X]\right) \right]+ \sum_{l=1}^{p_b} \frac{n_l+ \eta  -2}{2} \log \mid\hat{\Tau}_{l}\mid -\sum_{l=1}^{p_b} \frac{\eta \xi}{2} \Tr( \hat{\Tau}_{l} )\\
& -\frac{1}{2} \sum_{j=1}^p (\hat{\theta}_{j}^\top,\hat{\beta}_{j}^\top) \frac{1}{\psi}\I (\hat{\theta}_{j},\hat{\beta}_{j})  \\
& -\frac{1}{2} \sum_{j=1}^p \sum_{k=1}^q   \hat{m}_{jk}^{2} \Ex \left[ \frac{1}{(1-\gamma_{jk}) \loM + \gamma_{jk}\llM} \mid \hat{\Delta}  \right] + \sum_{j=1}^p \sum_{k=1}^q 2\Ex [\gamma_{jk}\mid \hat{\Delta}] \log(\hat{m}_{jk}),
\end{split}
\end{equation}

For the E-step $\Ex[\zv_i| \hat{\Delta},X]$ and $\Ex[\zv_i \zv_i^\top\mid\hat{\Delta},X]$ are the same as the ones in Supplement \ref{appendix:EMV1} for the flat prior. 
The new conditional expectation for the inclusion probability $\Ex[\gamma_{jk} \mid \hat{\Delta}] = \hat{p}_{jk}$ is \[
\hat{p}_{jk} = \frac{1}{1+ \frac{\llM}{\tilde{m}^2_{jk}}\sqrt{\frac{\llM}{\loM}} \exp\left(-\frac{1}{2} \hat{m}_{jk}^2 \left( \frac{1}{\loM} - \frac{1}{\llM} \right) \right) \frac{1-\Ex[\zeta_j]}{\Ex[\zeta_j]}}\] 
and $\Ex [d_{jk}^{-1} \mid \hat{\Delta}] = \Ex \left[ \frac{1}{(1-\gamma_{jk}) \tilde{\lambda}_{0} + \gamma_{jk}\tilde{\lambda}_{1}} \mid \hat{\Delta}  \right] = \frac{1-\hat{p}_{jk}}{\tilde{\lambda}_0} +\frac{\hat{p}_{jk}}{\tilde{\lambda}_1}$.\par
For the M-step of the loadings, we consider using a coordinate descent algorithm (CDA) that leads to closed-form expressions for $m_{jk}$. 
The partial derivative of \eqref{Ap:MOMQ1} is:
\begin{equation}
\label{MpMoM}
\begin{split}
\frac{\partial Q_1}{\partial M}=&\sum_{i=1}^n \left[\hat{\Tau}_{\bv_i}(\xv_i-\hat{\theta} \vv_i - \hat{\beta} \bv_i) \Ex[\zv_i^\top \mid\hat{\Delta},X] - \hat{\Tau}_{\bv_i} \hat{M} \Ex[\zv_i \zv_i^\top \mid\hat{\Delta},X] \right] \\
&- \hat{M} \circ \Ex[D_{\gamma}\mid \hat{\Delta}] +2 \Ex[\gamma \mid\hat{\Delta}] \circ \hat{M}_{inv} = 0 ,
\end{split}
\end{equation}
with $D_{\gamma} \in \R^{p \times q}$, $d_{jk}=((1-\gamma_{jk})\lambda_0+\gamma_{jk}\lambda_1)^{-1}$, $\hat{M}_{inv}$ a matrix with elements $1/\hat{m}_{jk}$ and $A\circ B$ being the Hadamard (element-wise) product of two matrices $A$ and $B$.\par
Viewing \eqref{MpMoM} with respect to $m_{jk}$:
\begin{equation}
\label{mjkpMoM}
\begin{split}
\frac{\partial Q_1}{\partial m_{jk}}=&- \left( \Ex[d_{jk}] +\sum_{i=1}^n \hat{\tau}_j^{\top} \bv_i \Ex[z_{ik} z_{ik}^\top \mid\hat{\Delta},X] \right) \hat{m}_{jk} + \left( \sum_{i=1}^n \Bigg[ \hat{\tau}_j^{\top}\bv_i(x_{ij}-\hat{\theta} v_{ij} - \hat{\beta} b_{ij}) \Ex[z_{ik} \mid\hat{\Delta},X] \right.  \\
&\left.\left. -\sum_{r\neq k}^q \hat{m}_{jr}  \hat{\tau}_j^{\top} \bv_i \Ex[z_{ir} z_{ik}^\top \mid\hat{\Delta},X] \right] \right) + \frac{2 \Ex[\gamma_{jk}]}{\hat{m}_{jk}} \\
=&- \left( \Ex[d_{jk}] +\sum_{i=1}^n \hat{\tau}_j^{\top} \bv_i \Ex[z_{ik} z_{ik}^\top \mid\hat{\Delta},X] \right) \hat{m}^2_{jk} + \left( \sum_{i=1}^n \Bigg[ \hat{\tau}_j^{\top}\bv_i(x_{ij}-\hat{\theta} v_{ij} - \hat{\beta} b_{ij}) \Ex[z_{ik} \mid\hat{\Delta},X] \right.  \\
&\left.\left. -\sum_{r\neq k}^q \hat{m}_{jr}  \hat{\tau}_j^{\top} \bv_i \Ex[z_{ir} z_{ik}^\top \mid\hat{\Delta},X] \right] \right) \hat{m}_{jk} + 2 \Ex[\gamma_{jk}] \\
=& a \hat{m}^2_{jk} + b \hat{m}_{jk} +c = 0
\end{split}
\end{equation}
for $j=1,\dots,p$ .\par
Define $\underbar{m}_{jk} := \frac{-b - \sqrt{b^2 - 4ac}}{2a}$ and $\bar{m}_{jk} := \frac{-b + \sqrt{b^2 - 4ac}}{2a}$. The global maximum is $\hat{m}_{jk}= \underbar{m}_{jk}$ if $b>0$ or $\hat{m}_{jk}= \bar{m}_{jk}$ if $b<0$. See Appendix~\ref{app:maxMjk} for details.\par
Finally, the updates for $\hat{\Tau}_l$, $(\hat{\theta}, \hat{\beta})_j$ and $\hat{\zeta}_k$ are equivalent to the ones obtained for Normal-SS.

\section{EM algorithm under Laplace-SS}
\label{app:EML}
Now, the part of the expected complete-data log-posterior $Q_1$ for the Laplace-SS is
\begin{equation}
\begin{split}
\label{Q1LaplaceSS}
Q_1(\theta, M, \beta, \Tau)=
&-\frac{1}{2} \sum_{i=1}^n \left[(\xv_i-\hat{\theta} \vv_i - \hat{\beta} \bv_i)^\top \hat{\Tau}_{\bv_i} (\xv_i-\hat{\theta} \vv_i - \hat{\beta} \bv_i) -2 (\xv_i-\hat{\theta} \vv_i - \hat{\beta} \bv_i)^\top \hat{\Tau}_{\bv_i}\hat{M}\Ex[\zv_i \mid\hat{\Delta},X]   \right. \\
&\left.  + \Tr\left(\hat{M}^{\top} \hat{\Tau}_{\bv_i} \hat{M} \Ex[\zv_i \zv_i^\top \mid\hat{\Delta},X]\right) \right]+ \sum_{l=1}^{p_b} \frac{n_l+ \eta  -2}{2} \log \mid\hat{\Tau}_{l}\mid -\sum_{l=1}^{p_b} \frac{\eta \xi}{2} \Tr( \hat{\Tau}_{l} ) \\
& -\frac{1}{2} \sum_{j=1}^p (\hat{\theta}^\top_{j},\hat{\beta}^\top_{j}) \frac{1}{\psi}\I (\hat{\theta}_{j},\hat{\beta}_{j})  - \sum_{j=1}^p \sum_{k=1}^q  |\hat{m}_{jk}| \Ex \left[\frac{1-\gamma_{jk}}{\loL} + \frac{\gamma_{jk}}{\llL}\mid \hat{\Delta}\right],
\end{split}
\end{equation}
In $Q_2$ the conditional expectations $\Ex[\gamma_{jk} \mid \hat{\Delta}]=\hat{p}_{jk}$ is
\[
\hat{p}_{jk} =\frac{1}{1+\frac{\llL}{\loL} \exp\left(- \mid \hat{m}_{jk} \mid \left( \frac{1}{\loL} - \frac{1}{\llL} \right) \right) \frac{1-\Ex[\zeta_j]}{\Ex[\zeta_j]}}
\]
and with $\Ex[d_{jk} \mid \hat{\Delta}]=\Ex \left[\frac{1-\gamma_{jk}}{\loL} + \frac{\gamma_{jk}}{\llL}\mid \hat{\Delta}\right] = \frac{1-\hat{p}_{jk}}{\lambda_0} +\frac{\hat{p}_{jk}}{\lambda_1}$ for $Q_1$.

The M-step update for $M$ is obtain by setting to 0 the partial derivative with respect to $M$, for $m_{jk} \neq 0$
\begin{equation}
\begin{split}
\frac{\partial Q}{\partial M}=&-\frac{1}{2}\sum_{i=1}^n \left[-2 \hat{\Tau}_{\bv_i}(\xv_i-\hat{\theta} \vv_i - \hat{\beta} \bv_i) \Ex[\zv_i^\top \mid\hat{\Delta},X] + 2 \hat{\Tau}_{\bv_i} \hat{M} \Ex[\zv_i \zv_i^\top \mid\hat{\Delta},X] \right] -D^{\gamma,M}= 0 ,
\end{split}
\end{equation}
with $D^{\gamma,M} \in \R^{p \times q}$ with element $jk$: $d_{jk}^{\gamma,M}=\sign(\hat{m}_{jk})\Ex[d_{jk} \mid \hat{\Delta}]$.

Taking the partial derivative of \eqref{Q1LaplaceSS} with respect to $m_{jk}$, when $m_{jk} \neq 0$ and setting it to 0 we obtain:
\begin{equation}
\label{mjkpLSS}
\begin{split}
\frac{\partial Q_1}{\partial m_{jk}}=&- \left(\sum_{i=1}^n \hat{\tau}_j^{\top} \bv_i \Ex[z_{ik} z_{ik}^\top \mid\hat{\Delta},X] \right) \hat{m}_{jk} + \left( \sum_{i=1}^n \Bigg[ \hat{\tau}_j^{\top}\bv_i(x_{ij}-\hat{\theta} v_{ij} - \hat{\beta} b_{ij}) \Ex[z_{ik} \mid\hat{\Delta},X] \right.  \\
&\left.\left. -\sum_{r\neq k}^q \hat{m}_{jr}  \hat{\tau}_j^{\top} \bv_i \Ex[z_{ir} z_{ik}^\top \mid\hat{\Delta},X] \right] - \sign(\hat{m}_{jk})\left[ \frac{1-\hat{p}_{jk}}{\lambda_0} +\frac{\hat{p}_{jk}}{\lambda_1}\right] \right) \\
=& a \hat{m}_{jk} + b + c\cdot \sign{(\hat{m}_{jk})} = 0
\end{split}
\end{equation}
for $j=1,\dots,p$ .

Define $m_{jk}^+ := \frac{-(b+c)}{a}$ and $m^-_{jk} := \frac{-(b-c)}{a}$.

If $b>-c$, then $m_{jk}^+= \arg \max_{m_{jk}} f(m_{jk})$. If $b<c$, then $m_{jk}^-= \arg \max_{m_{jk}} f(m_{jk})$. If $c \leq b \leq -c$, then $0= \arg \max_{m_{jk}} f(m_{jk})$.
See Appendix~\ref{app:maxMjkLSS} for details.

\section{EM algorithm under Laplace-MOM-SS}
\label{app:EMMOML}
Finally, $Q_1$ for Laplace-MOM-SS is given by:
\begin{equation}
\begin{split}
\label{EzxpMoMQL}
Q_1(\theta, M, \beta, \Tau)=
&-\frac{1}{2} \sum_{i=1}^n \left[(\xv_i-\hat{\theta} \vv_i - \hat{\beta} \bv_i)^\top \hat{\Tau}_{\bv_i} (\xv_i-\hat{\theta} \vv_i - \hat{\beta} \bv_i) -2 (\xv_i-\hat{\theta} \vv_i - \hat{\beta} \bv_i)^\top \hat{\Tau}_{\bv_i}\hat{M}\Ex[\zv_i \mid\hat{\Delta},X]   \right. \\
&\left.  + \Tr\left(\hat{M}^{\top} \hat{\Tau}_{\bv_i} \hat{M} \Ex[\zv_i \zv_i^\top \mid\hat{\Delta},X]\right) \right]+ \sum_{l=1}^{p_b} \frac{n_l+ \eta  -2}{2} \log \mid\hat{\Tau}_{l}\mid -\sum_{l=1}^{p_b} \frac{\eta \xi}{2} \Tr( \hat{\Tau}_{l} )   \\
& -\frac{1}{2} \sum_{j=1}^p (\hat{\theta},\hat{\beta})^\top_{j} \frac{1}{\psi}\I (\hat{\theta},\hat{\beta})_{j}  - \sum_{j=1}^p \sum_{k=1}^q  \mid \hat{m}_{jk} \mid \Ex \left[ \frac{(1-\gamma_{jk})}{\loLM} + \frac{(\gamma_{jk})}{\llLM} \mid \hat{\Delta}  \right]  \\
&+\sum_{j=1}^p \sum_{k=1}^q  2 \Ex [\gamma_{jk}\mid \hat{\Delta}] \log(\hat{m}_{jk}), \\
\end{split}
\end{equation}

The new conditional expectation for the inclusion probability $\Ex[\gamma_{jk} \mid \hat{\Delta}]= \hat{p}_{jk}$ is 
\[
\hat{p}_{jk} =\frac{1}{1+\frac{2 \llLM^2}{\hat{m}^2_{jk}}\frac{\llLM}{\loLM} \exp\left(- \mid \hat{m}_{jk} \mid \left( \frac{1}{\loLM} - \frac{1}{\llLM} \right) \right) \frac{1-\Ex[\zeta_j]}{\Ex[\zeta_j]}}
\]
and $ \Ex \left[ \frac{1}{(1-\gamma_{jk}) \loLM + \gamma_{jk}\llLM} \mid \hat{\Delta}  \right] = \frac{1-\hat{p}_{jk}}{\loLM} +\frac{\hat{p}_{jk}}{\llLM}$.\par
For the M-step of the loadings, we consider using a coordinate descent algorithm (CDA) that performs successive univariate optimisation with respect to each $m_{jk}$.

Notice that when $m_{jk}=0$, the value of $Q(m_{jk}=0)=-\infty$, thus the solution for the optimisation is given by setting $\frac{\partial Q_1}{\partial m_{jk}}=0$.

The partial derivative of \eqref{EzxpMoMQL} w.r.t.\ $M$ is
\begin{equation}
\label{MpMoML}
\begin{split}
\frac{\partial Q}{\partial M}=&\sum_{i=1}^n \left[\hat{\Tau}_{\bv_i}(\xv_i-\hat{\theta} \vv_i - \hat{\beta} \bv_i) \Ex[\zv_i^\top \mid\hat{\Delta},X] - \hat{\Tau}_{\bv_i} \hat{M} \Ex[\zv_i \zv_i^\top \mid\hat{\Delta},X] \right] \\
&- D^{\gamma,M}+2 \Ex[\gamma \mid\hat{\Delta}] \circ \hat{M}_{inv} = 0 ,
\end{split}
\end{equation}
with $D^{\gamma,M} \in \R^{p \times q}$ with element $jk$: $d_{jk}^{\gamma,M}=\sign(\hat{m}_{jk})\Ex \left[ \frac{1}{(1-\gamma_{jk}) \loLM + \gamma_{jk}\llLM} \mid \hat{\Delta}  \right]$, $\hat{M}_{inv}$ a matrix with elements $1/\hat{m}_{jk}$ and $A\circ B$ being the Hadamard (element-wise) product of two matrices $A$ and $B$.\par
Taking the partial derivative of \eqref{EzxpMoMQL} with respect to $m_{jk}$, when $m_{jk} \neq 0$ and setting it to 0 we obtain:
\begin{equation}
\label{mjkpMoML}
\begin{split}
\frac{\partial Q_1}{\partial m_{jk}}=&- \left(\sum_{i=1}^n \hat{\tau}_j^{\top} \bv_i \Ex[z_{ik} z_{ik}^\top \mid\hat{\Delta},X] \right) \hat{m}_{jk} + \left( \sum_{i=1}^n \Bigg[ \hat{\tau}_j^{\top}\bv_i(x_{ij}-\hat{\theta} v_{ij} - \hat{\beta} b_{ij}) \Ex[z_{ik} \mid\hat{\Delta},X] \right.  \\
&\left.\left. -\sum_{r\neq k}^q \hat{m}_{jr}  \hat{\tau}_j^{\top} \bv_i \Ex[z_{ir} z_{ik}^\top \mid\hat{\Delta},X] \right] - \sign(\hat{m}_{jk}) \Ex \left[ \frac{1}{(1-\gamma_{jk}) \loLM + \gamma_{jk}\llLM} \mid \hat{\Delta}  \right] \right) + \frac{2 \Ex[\gamma_{jk}]}{\hat{m}_{jk}} \\
=&- \left(\sum_{i=1}^n \hat{\tau}_j^{\top} \bv_i \Ex[z_{ik} z_{ik}^\top \mid\hat{\Delta},X] \right) \hat{m}^2_{jk} + \left( \sum_{i=1}^n \Bigg[ \hat{\tau}_j^{\top}\bv_i(x_{ij}-\hat{\theta} v_{ij} - \hat{\beta} b_{ij}) \Ex[z_{ik} \mid\hat{\Delta},X] \right.  \\
&\left.\left. -\sum_{r\neq k}^q \hat{m}_{jr}  \hat{\tau}_j^{\top} \bv_i \Ex[z_{ir} z_{ik}^\top \mid\hat{\Delta},X] \right] - \sign(\hat{m}_{jk}) \left[ \frac{1-\hat{p}_{jk}}{\loLM} +\frac{\hat{p}_{jk}}{\llLM}\right] \right) \hat{m}_{jk} + 2 \Ex[\gamma_{jk}] \\
=& a \hat{m}^2_{jk} + b \hat{m}_{jk} + c\cdot \sign{(\hat{m}_{jk})} \hat{m}_{jk}+ d = 0
\end{split}
\end{equation}
for $j=1,\dots,p$ .

Define $m_{jk}^+ := \frac{-(b+c) - \sqrt{(b+c)^2 - 4ad}}{2a}$ and $m^-_{jk} := \frac{-(b-c) + \sqrt{(b-c)^2 - 4ad}}{2a}$.

If $b>0$, then $m_{jk}^+= \arg \max_{m_{jk}} f(m_{jk})$. If $b<0$, then $m_{jk}^-= \arg \max_{m_{jk}} f(m_{jk})$. If $b=0$, then $m_{jk}^+= \arg \max_{m_{jk}} f(m_{jk})$ or $m_{jk}^-=\arg \max_{m_{jk}} f(m_{jk})$.
See Appendix~\ref{app:maxMjkL} for details.

\newpage
\section{Weighted 10-fold cross-validation}
\label{app:WCV}

We aim to a pseudo-code-algorithm for the weighted 10-fold cross-validation used through this paper.

\RestyleAlgo{boxruled}
\begin{algorithm}[h!]
 \textbf{initialise} $\epsilon_X =0$\\
 \textbf{set} 10 random cross-validation subsets of\\\begin{tabular}{rl}
 Observations: & $\{\xv^{[1]},\dots,\xv^{[10]}\} \in \R^{\frac{n}{10}\times p}$\\
 Covariates: & $\{\vv^{[1]}\dots,\vv^{[10]}\}\in \R^{\frac{n}{10}\times p_v}$\\
 Batches: & $\{\bv^{[1]},\dots,\bv^{[10]}\}\in \R^{\frac{n}{10}\times p_b}$\\
 \end{tabular}\\
 \For{$r \leftarrow 1,\dots,10$}{
  \textbf{set}: Cross-validation subsets\\
  $\tilde{\xv} := (\xv^{[1]},\dots,\xv^{[r-1]},\xv^{[r+1]},\dots,\xv^{[10]}),\tilde{\vv} := (\vv^{[1]},\dots,\vv^{[r-1]},\vv^{[r+1]},\dots,\vv^{[10]})$\\
 and $\tilde{\bv} := (\bv^{[1]},\dots,\bv^{[r-1]},\bv^{[r+1]},\dots,\bv^{[10]}),$\\

  \textbf{compute}: EM algorithm\\
  \begin{tabular}{rl}
  input: &$\tilde{\xv},\tilde{\vv},\tilde{\bv}$\\
  output: &$\hat{M},\hat{\Sb},\hat{\theta},\hat{\beta},\hat{\zeta}$  
  \end{tabular}\\
   \textbf{set}: Test factors $\hat{\zv}_i=(\I_q + \hat{M}^\top \hat{\Sb}\hat{M})^{-1} \hat{M}^\top\hat{\Sb}(\xv^{[r]}-\hat{\theta} \vv^{[r]} - \hat{\beta} \bv^{[r]})$\\ 
\textbf{compute} $\epsilon_X = \epsilon_X  + \sum_{i} || \left[ \xv_i^{[r]} - (\hat{\theta} \vv_i^{[r]} + \hat{M} \hat{\zv}_i +\hat{\beta} \bv_i^{[r]}) \right] \hat{\Sb} ||_F$
 }
 \textbf{set} $\epsilon_X = \frac{\epsilon_X}{10}$
 \caption{Weighted 10-fold cross-validation}
 \label{alg:EMvanilla}
\end{algorithm}

\newpage
\section{Plots simulations of truly sparse loadings $M$ scenario and no batch effect}
\label{Sup:PlotsS}
The following plots present the heatmaps of the reconstruction of $\hat{M}, \widehat{\text{Cov}}(\xv_i\mid\cdot)^{-1}$  and $\hat{\gamma}$ for the scenario with truly sparse loadings $M$, setting $q=100$ and without batch effects.

\begin{figure}[h!]
	\begin{subfigure}{.19\textwidth}
		\centering
		\includegraphics[width=.9\linewidth]{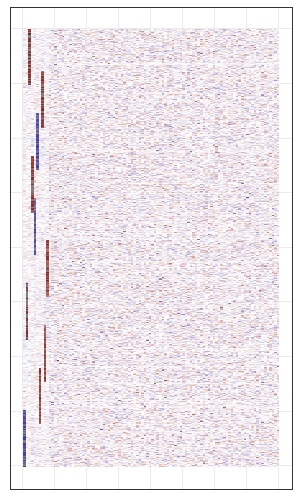}
		\caption{$\hat{M}$ Flat}
	\end{subfigure}
	\begin{subfigure}{.19\textwidth}
		\centering
		\includegraphics[width=.9\linewidth]{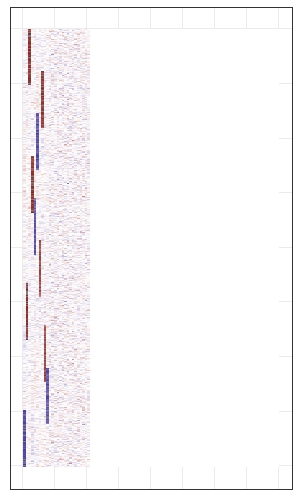}
		\caption{$\hat{M}$ Normal}
	\end{subfigure}
	\begin{subfigure}{.19\textwidth}
		\centering
		\includegraphics[width=.9\linewidth]{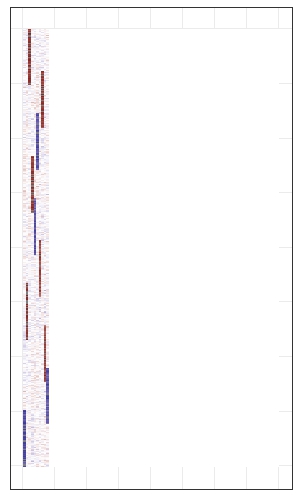}
		\caption{$\hat{M}$ MOM}
	\end{subfigure}
\begin{subfigure}{.19\textwidth}
	\centering
	\includegraphics[width=.9\linewidth]{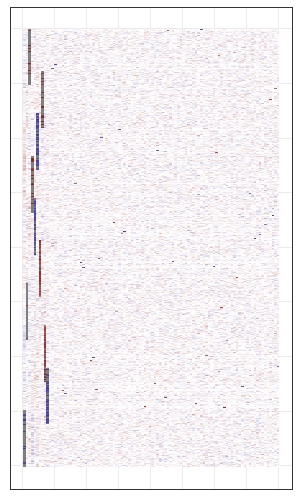}
	\caption{$\hat{M}$ FastBFA}
\end{subfigure}
	\begin{subfigure}{.19\textwidth}
		\centering
		\includegraphics[width=.9\linewidth]{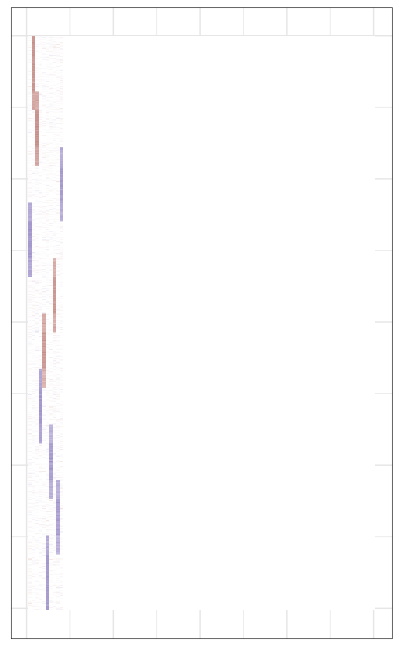}
		\caption{$\hat{M}$ LASSO}
	\end{subfigure}
	\begin{subfigure}{.19\textwidth}
		\centering
		\includegraphics[width=.9\linewidth]{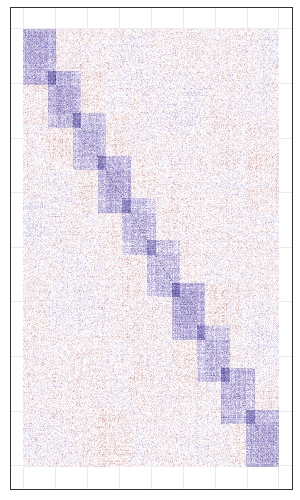}
		\caption{\tiny$\widehat{\text{Cov}}(\xv_i\mid\cdot)^{-1}$ Flat}
	\end{subfigure}
	\begin{subfigure}{.19\textwidth}
		\centering
		\includegraphics[width=.9\linewidth]{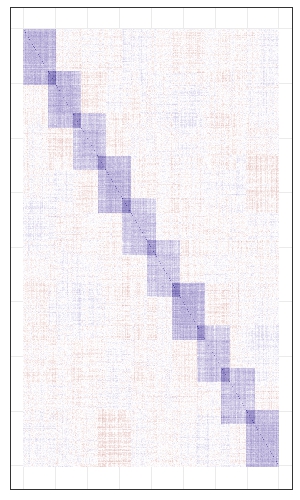}
		\caption{\tiny$\widehat{\text{Cov}}(\xv_i\mid\cdot)^{-1}$ Normal}
	\end{subfigure}
	\begin{subfigure}{.19\textwidth}
		\centering
		\includegraphics[width=.9\linewidth]{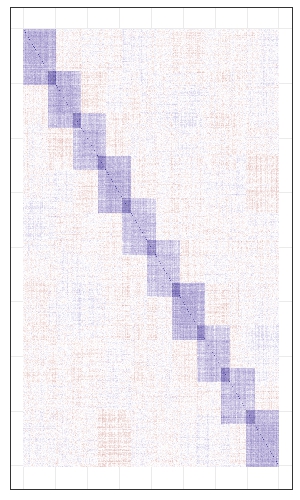}
		\caption{\tiny $\widehat{\text{Cov}}(\xv_i\mid\cdot)^{-1}$ MOM}
	\end{subfigure}
	\begin{subfigure}{.19\textwidth}
	\centering
	\includegraphics[width=.9\linewidth]{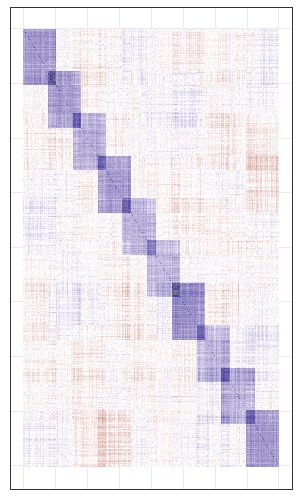}
	\caption{\tiny $\widehat{\text{Cov}}(\xv_i\mid\cdot)^{-1}$ FastBFA}
\end{subfigure}
	\begin{subfigure}{.19\textwidth}
		\centering
		\includegraphics[width=.9\linewidth]{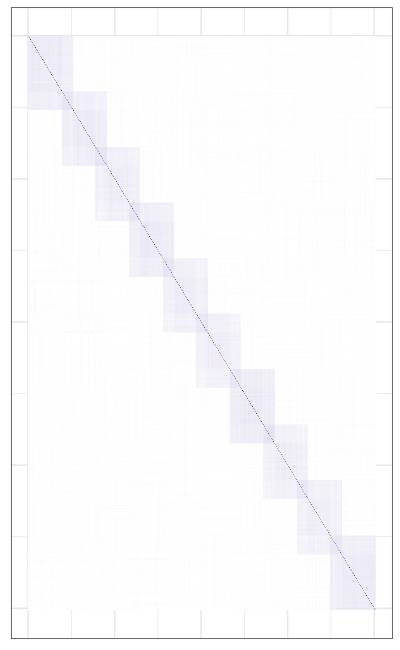}
		\caption{\tiny $\widehat{\text{Cov}}(\xv_i\mid\cdot)^{-1}$ LASSO}
	\end{subfigure}
	\caption{Heatmaps of loadings and covariance (red denotes large negative values, blue large positive values, white denotes zero).}
\end{figure}

\begin{figure}[h!]
\centering
\begin{subfigure}{.2\textwidth}
  \centering
  \includegraphics[width=.9\linewidth]{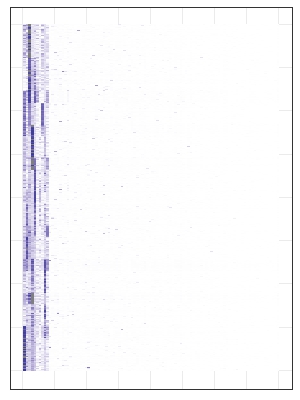}
  \caption{$\hat{\gamma}$ (Normal)}
  \label{fig:GammaSSLNB} 
\end{subfigure}
\begin{subfigure}{.2\textwidth}
  \centering
  \includegraphics[width=.9\linewidth]{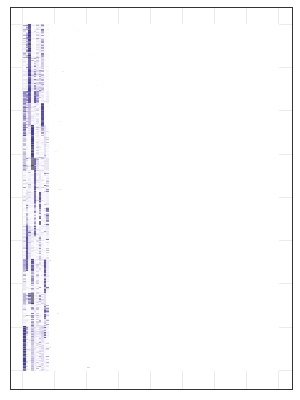}
  \caption{$\hat{\gamma}$ (MOM)}
  \label{fig:GammapMOMNB}
\end{subfigure}
\begin{subfigure}{.2\textwidth}
  \centering
  \includegraphics[width=.9\linewidth]{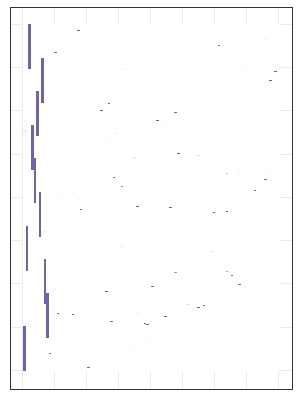}
  \caption{$\hat{\gamma}$ (FastBFA)}
  \label{fig:GammaLaplaceNB}
\end{subfigure}
\caption{Heatmaps of inclusion probability (white denotes 0, dark blue denotes 1).}
\label{fig:HeatGammapMOMvsSSLNB}
\end{figure}

\newpage
\section{Plots simulations dense loadings $M$ and no batch effect}
The following plots show the heatmaps of the reconstruction of $\hat{M}, \widehat{\text{Cov}}(\xv_i\mid\cdot)^{-1}$  and $\hat{\gamma}$ for the scenario with dense loadings $M$, setting $q=100$ and without batch effects.
\label{Sup:PlotsD}
\begin{figure}[h!]
	\begin{subfigure}{.19\textwidth}
		\centering
		\includegraphics[width=.9\linewidth]{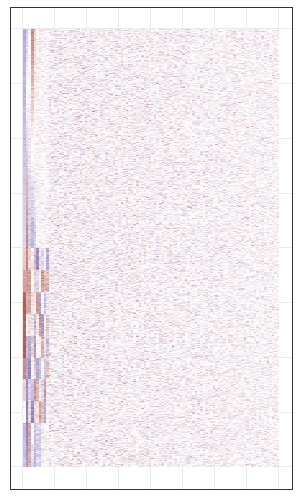}
		\caption{$\hat{M}$ Flat}
	\end{subfigure}
	\begin{subfigure}{.19\textwidth}
		\centering
		\includegraphics[width=.9\linewidth]{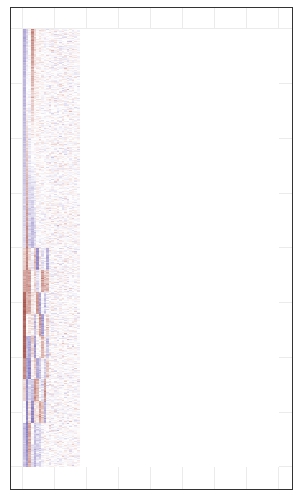}
		\caption{$\hat{M}$ Normal}
	\end{subfigure}
	\begin{subfigure}{.19\textwidth}
		\centering
		\includegraphics[width=.9\linewidth]{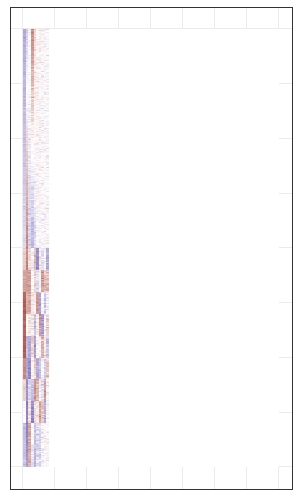}
		\caption{$\hat{M}$ MOM}
	\end{subfigure}
	\begin{subfigure}{.19\textwidth}
	\centering
	\includegraphics[width=.9\linewidth]{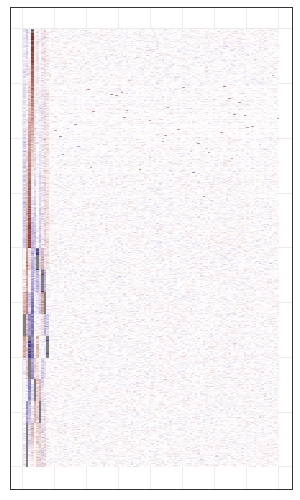}
	\caption{$\hat{M}$ FastBFA}
\end{subfigure}
	\begin{subfigure}{.19\textwidth}
		\centering
		\includegraphics[width=.9\linewidth]{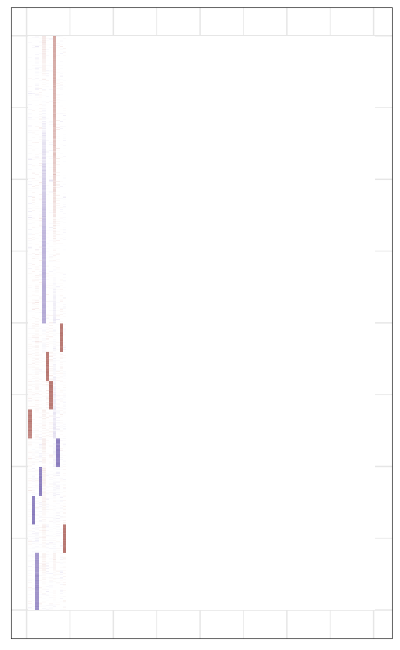}
		\caption{$\hat{M}$ LASSO}
	\end{subfigure}
	\begin{subfigure}{.19\textwidth}
		\centering
		\includegraphics[width=.9\linewidth]{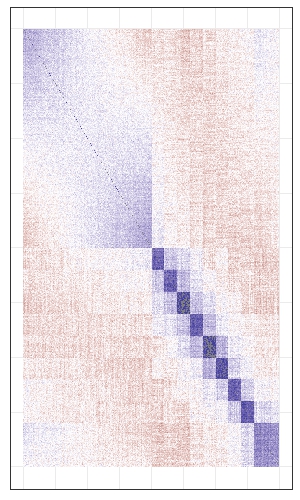}
		\caption{\tiny $\widehat{\text{Cov}}(\xv_i\mid\cdot)^{-1}$ Flat}
	\end{subfigure}
	\begin{subfigure}{.19\textwidth}
		\centering
		\includegraphics[width=.9\linewidth]{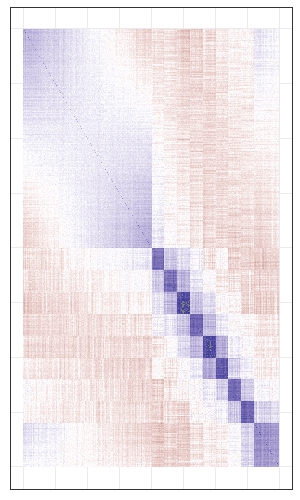}
		\caption{\tiny $\widehat{\text{Cov}}(\xv_i\mid\cdot)^{-1}$ Normal}
	\end{subfigure}
	\begin{subfigure}{.19\textwidth}
		\centering
		\includegraphics[width=.9\linewidth]{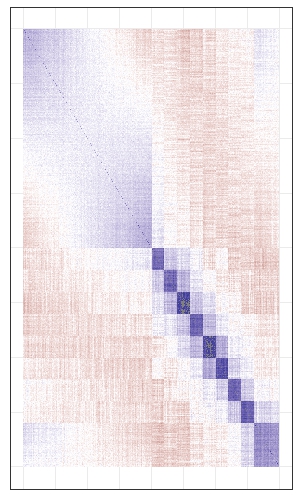}
		\caption{\tiny $\widehat{\text{Cov}}(\xv_i\mid\cdot)^{-1}$ MOM}
	\end{subfigure}
	\begin{subfigure}{.19\textwidth}
	\centering
	\includegraphics[width=.9\linewidth]{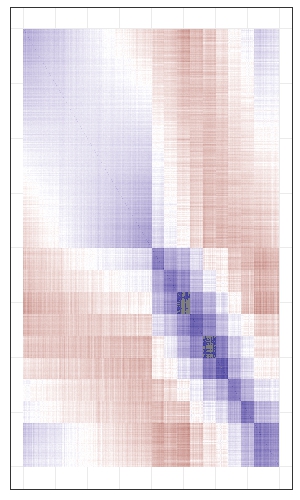}
	\caption{\tiny $\widehat{\text{Cov}}(\xv_i\mid\cdot)^{-1}$ FastBFA}
\end{subfigure}
	\begin{subfigure}{.19\textwidth}
		\centering
		\includegraphics[width=.9\linewidth]{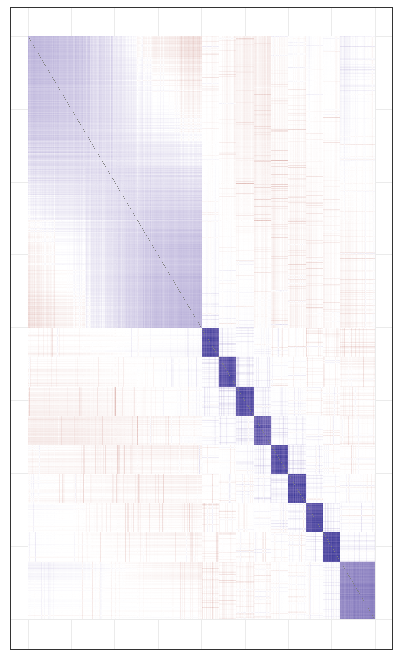}
		\caption{\tiny $\widehat{\text{Cov}}(\xv_i\mid\cdot)^{-1}$ LASSO}
	\end{subfigure}
	\caption{Heatmaps of loadings and covariance (red denotes large negative values, blue large positive values, white denotes zero).}
\end{figure}

\begin{figure}[h!]
\centering
\begin{subfigure}{.2\textwidth}
  \centering
  \includegraphics[width=.9\linewidth]{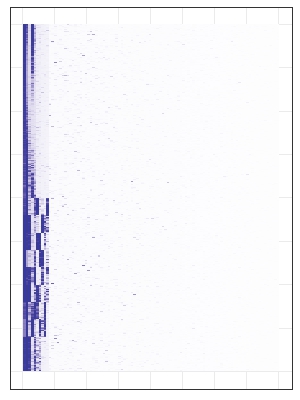}
  \caption{$\hat{\gamma}$ (Normal)}
\end{subfigure}
\begin{subfigure}{.2\textwidth}
  \centering
  \includegraphics[width=.9\linewidth]{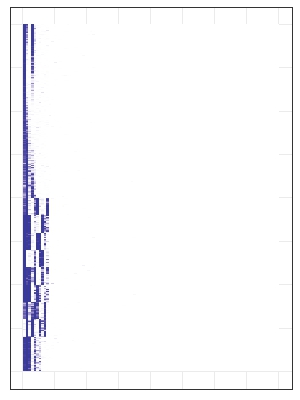}
  \caption{$\hat{\gamma}$ (MOM)}
  \label{fig:GammapMOMNB}
\end{subfigure}
\begin{subfigure}{.2\textwidth}
  \centering
  \includegraphics[width=.9\linewidth]{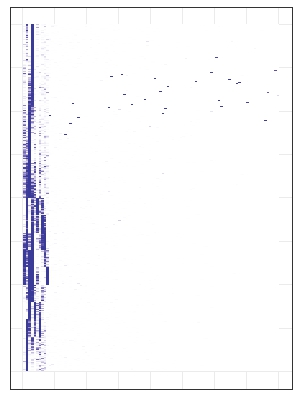}
  \caption{$\hat{\gamma}$ (FastBFA)}
\end{subfigure}
\caption{Heatmaps of inclusion probability (white denotes 0, dark blue denotes 1).}
\label{fig:HeatGammapMOMvsSSLNB02}
\end{figure}

\newpage
\section{Plots simulations truly sparse loadings $M$ and batch effect}
Visual representation of the reconstruction of $\hat{M}, \widehat{\text{Cov}}(\xv_i\mid\cdot)^{-1}$  and $\hat{\gamma}$ for the scenario with truly sparse loadings $M$, setting $q=100$ and with mean and variance batch effects.
\label{Sup:PlotsS2}
\begin{figure}[h!]
	\begin{subfigure}{.16\textwidth}
		\centering
		\includegraphics[width=.9\linewidth]{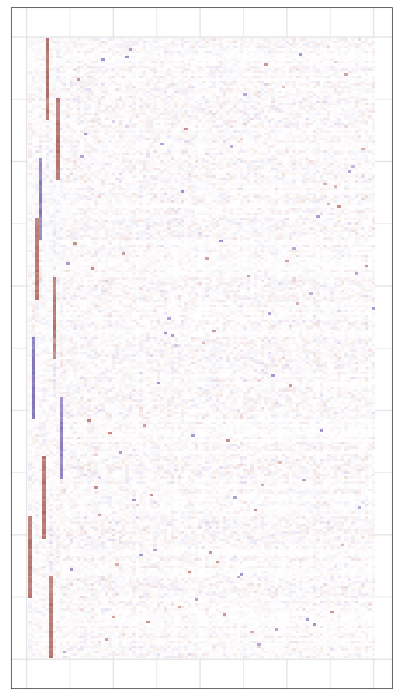}
		\caption{$\hat{M}$ Flat}
	\end{subfigure}
	\begin{subfigure}{.16\textwidth}
		\centering
		\includegraphics[width=.9\linewidth]{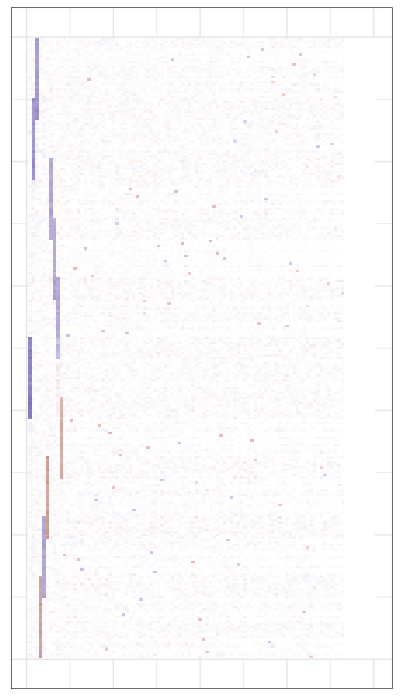}
		\caption{$\hat{M}$ Normal}
	\end{subfigure}
	\begin{subfigure}{.16\textwidth}
		\centering
		\includegraphics[width=.9\linewidth]{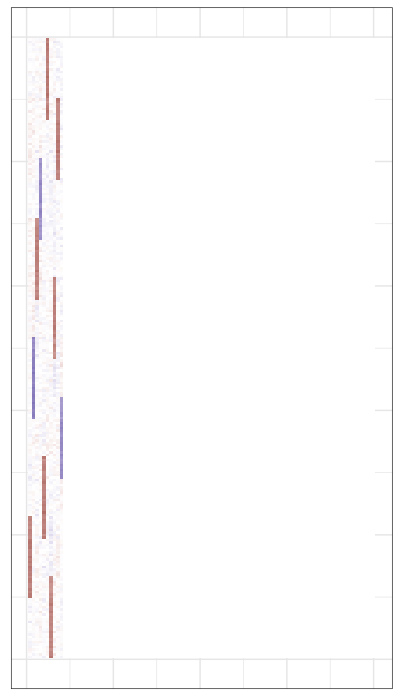}
		\caption{$\hat{M}$ MOM}
	\end{subfigure}
	\begin{subfigure}{.16\textwidth}
		\centering
		\includegraphics[width=.9\linewidth]{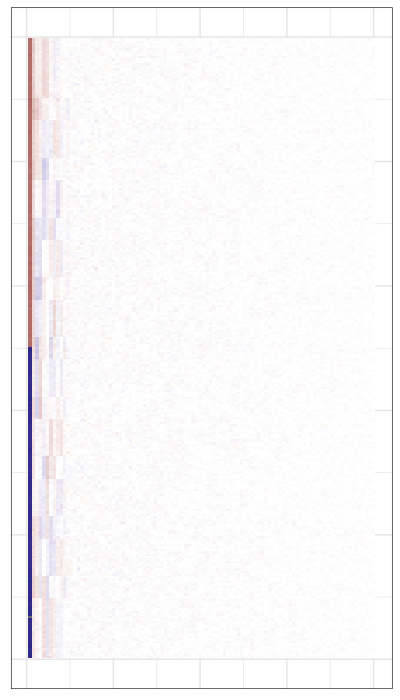}
		\caption{$\hat{M}$ ComBat}
	\end{subfigure}
\begin{subfigure}{.16\textwidth}
	\centering
	\includegraphics[width=.9\linewidth]{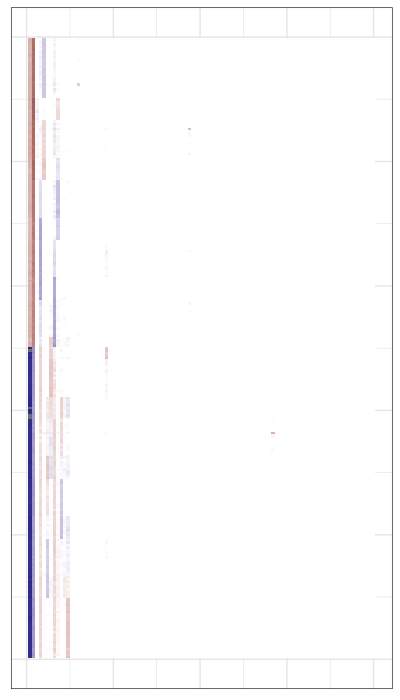}
	\caption{$\hat{M}$ FastBFA}
\end{subfigure}
	\begin{subfigure}{.16\textwidth}
		\centering
		\includegraphics[width=.9\linewidth]{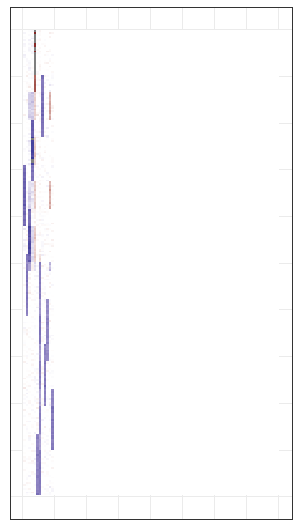}
		\caption{$\hat{M}$ LASSO}
	\end{subfigure}
	\begin{subfigure}{.16\textwidth}
		\centering
		\includegraphics[width=.9\linewidth]{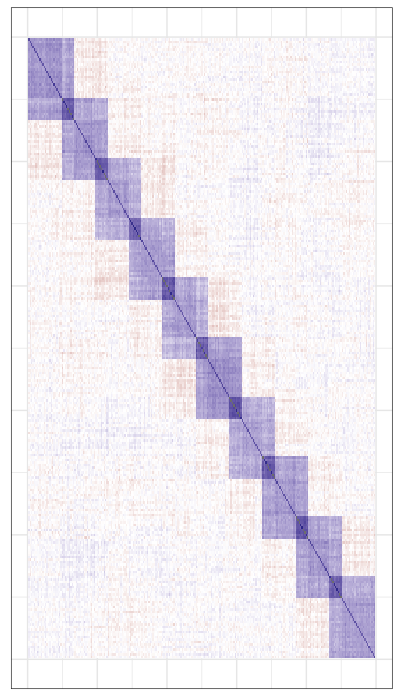}
		\caption{\tiny $\widehat{\text{Cov}}(\xv_i\mid\cdot)^{-1}$ Flat}
	\end{subfigure}
	\begin{subfigure}{.16\textwidth}
		\centering
		\includegraphics[width=.9\linewidth]{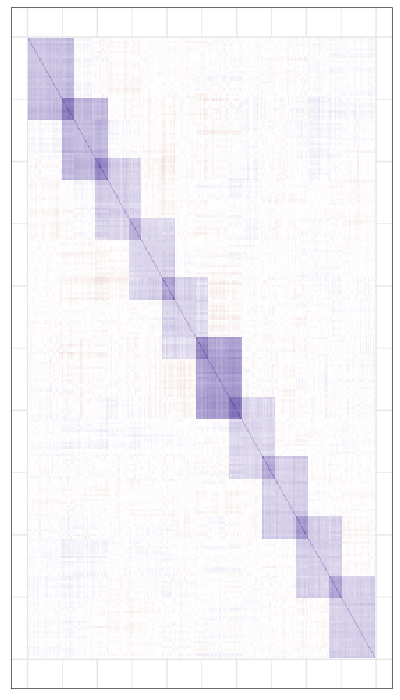}
		\caption{\scalebox{.6}{$\widehat{\text{Cov}}(\xv_i\mid\cdot)^{-1}$ Normal}}
	\end{subfigure}
	\begin{subfigure}{.16\textwidth}
		\centering
		\includegraphics[width=.9\linewidth]{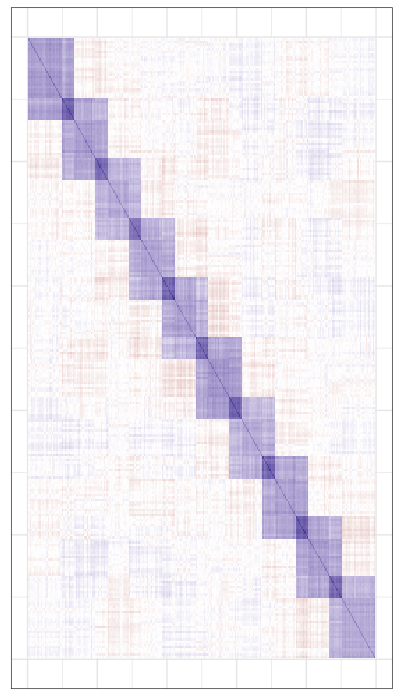}
		\caption{\tiny $\widehat{\text{Cov}}(\xv_i\mid\cdot)^{-1}$ MOM}
	\end{subfigure}
	\begin{subfigure}{.16\textwidth}
		\centering
		\includegraphics[width=.9\linewidth]{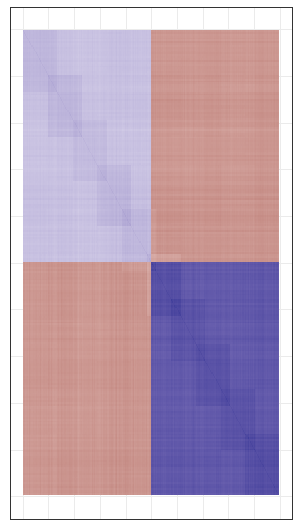}
		\caption{\scalebox{.6}{$\widehat{\text{Cov}}(\xv_i\mid\cdot)^{-1}$ ComBat}}
	\end{subfigure}
\begin{subfigure}{.16\textwidth}
	\centering
	\includegraphics[width=.9\linewidth]{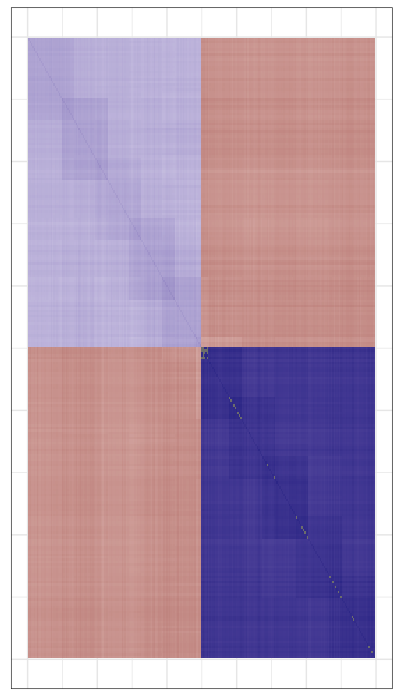}
	\caption{\scalebox{.55}{$\widehat{\text{Cov}}(\xv_i\mid\cdot)^{-1}$ FastBFA}}
\end{subfigure}
	\begin{subfigure}{.16\textwidth}
		\centering
		\includegraphics[width=.9\linewidth]{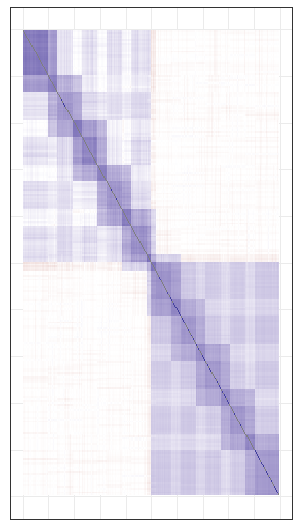}
		\caption{\scalebox{.55}{$\widehat{\text{Cov}}(\xv_i\mid\cdot)^{-1}$ LASSO}}
	\end{subfigure}
	\caption{Heatmaps of loadings and covariance (red denotes large negative values, blue large positive values, white denotes zero).}
	\label{fig:PlotsS2}
\end{figure}

\begin{figure}[h!]
\centering
\begin{subfigure}{.2\textwidth}
  \centering
  \includegraphics[width=.9\linewidth]{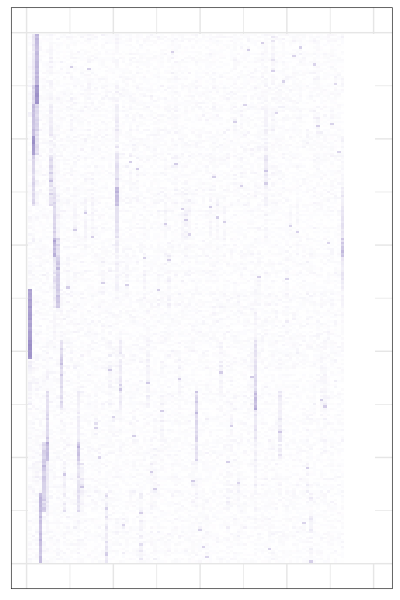}
  \caption{$\hat{\gamma}$ (Normal)}
\end{subfigure}
\begin{subfigure}{.2\textwidth}
  \centering
  \includegraphics[width=.9\linewidth]{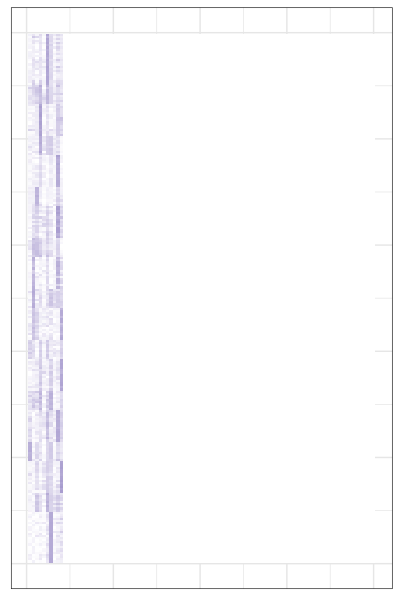}
  \caption{$\hat{\gamma}$ (MOM)}
\end{subfigure}
\begin{subfigure}{.2\textwidth}
  \centering
  \includegraphics[width=.9\linewidth]{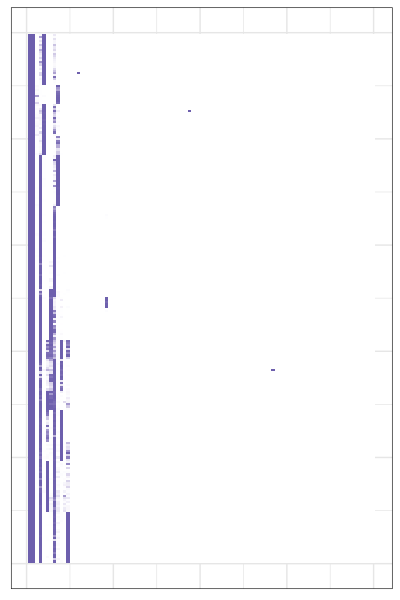}
  \caption{$\hat{\gamma}$ (FastBFA)}
\end{subfigure}
\caption{Heatmaps of inclusion probability (white denotes 0, dark blue denotes 1).}
\end{figure}

\newpage
\section{Plots simulations dense loadings $M$ and  batch effect}
\label{Sup:PlotsD2}
Graphical representation of the reconstruction of $\hat{M}, \widehat{\text{Cov}}(\xv_i\mid\cdot)^{-1}$  and $\hat{\gamma}$ for the scenario with dense loadings $M$, setting $q=100$ and with mean and variance batch effects.

\begin{figure}[h!]
	\begin{subfigure}{.16\textwidth}
		\centering
		\includegraphics[width=.9\linewidth]{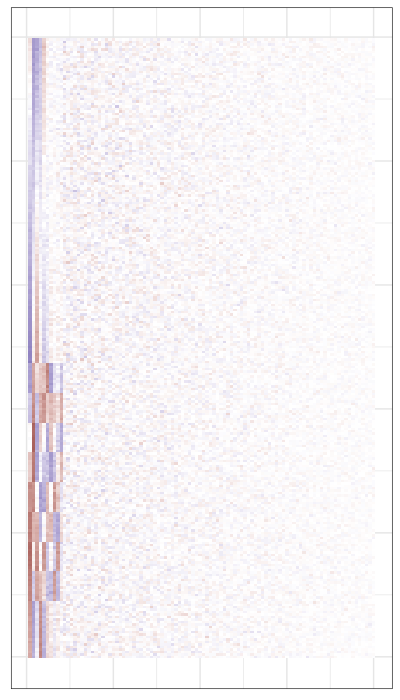}
		\caption{$\hat{M}$ Flat}
	\end{subfigure}
	\begin{subfigure}{.16\textwidth}
		\centering
		\includegraphics[width=.9\linewidth]{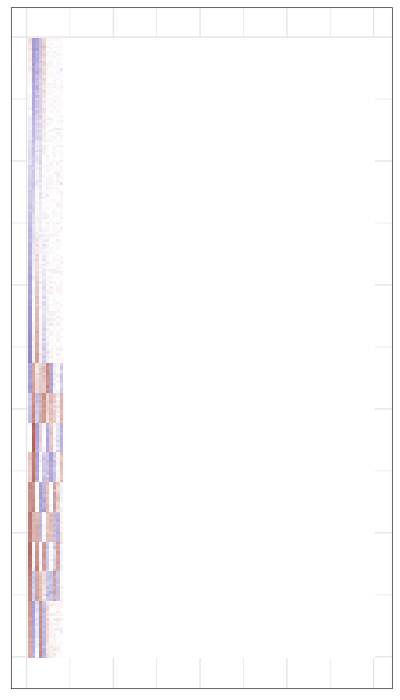}
		\caption{$\hat{M}$ Normal}
	\end{subfigure}
	\begin{subfigure}{.16\textwidth}
		\centering
		\includegraphics[width=.9\linewidth]{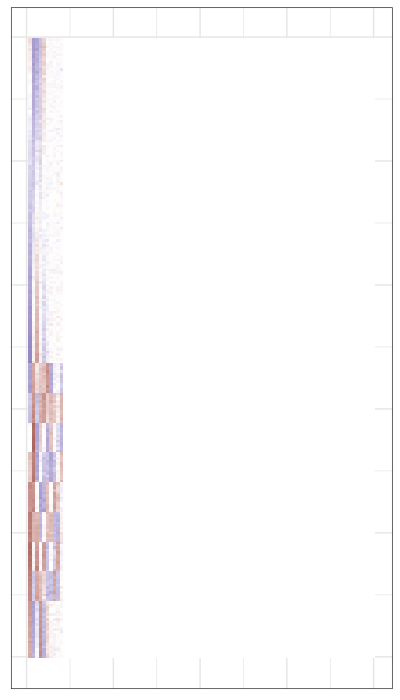}
		\caption{$\hat{M}$ MOM}
	\end{subfigure}
	\begin{subfigure}{.16\textwidth}
		\centering
		\includegraphics[width=.9\linewidth]{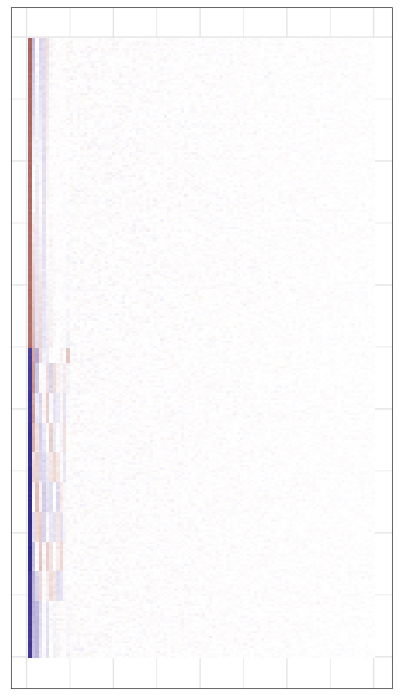}
		\caption{$\hat{M}$ ComBat}
	\end{subfigure}
\begin{subfigure}{.16\textwidth}
	\centering
	\includegraphics[width=.9\linewidth]{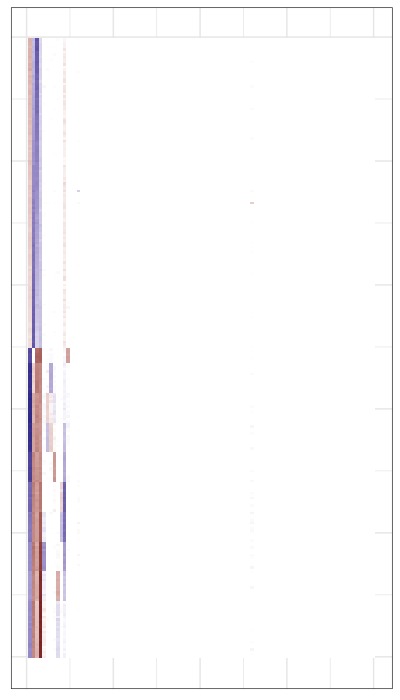}
	\caption{$\hat{M}$ FastBFA}
\end{subfigure}
	\begin{subfigure}{.16\textwidth}
		\centering
		\includegraphics[width=.9\linewidth]{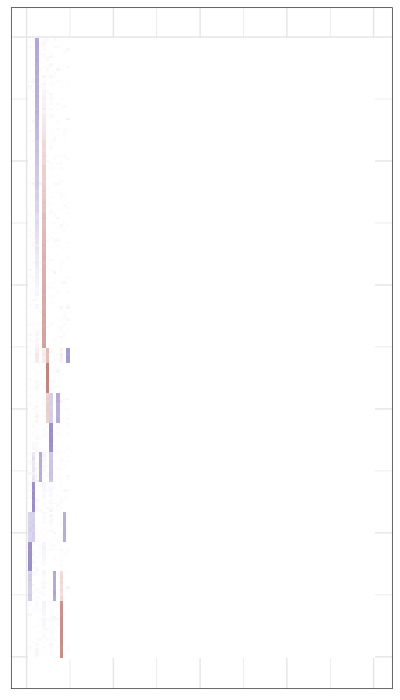}
		\caption{$\hat{M}$ LASSO}
	\end{subfigure}
	\begin{subfigure}{.16\textwidth}
		\centering
		\includegraphics[width=.9\linewidth]{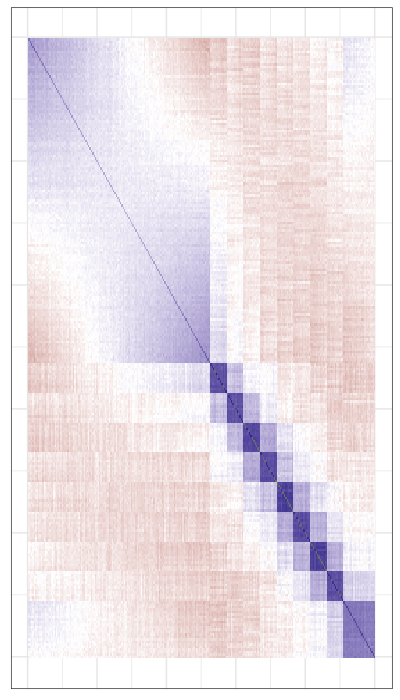}
		\caption{\tiny $\widehat{\text{Cov}}(\xv_i\mid\cdot)^{-1}$ Flat}
	\end{subfigure}
	\begin{subfigure}{.16\textwidth}
		\centering
		\includegraphics[width=.9\linewidth]{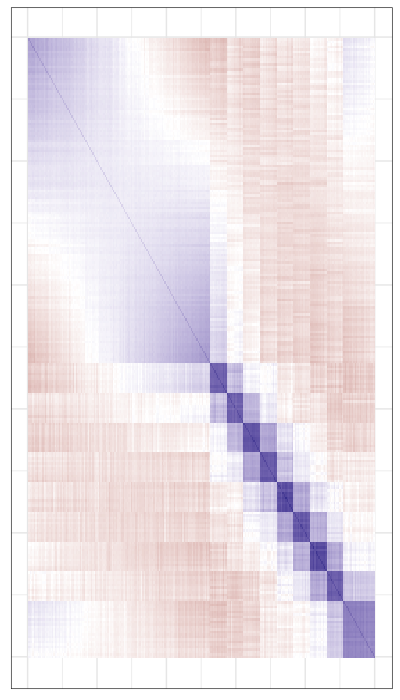}
		\caption{\scalebox{.6}{$\widehat{\text{Cov}}(\xv_i\mid\cdot)^{-1}$ Normal}}
	\end{subfigure}
	\begin{subfigure}{.16\textwidth}
		\centering
		\includegraphics[width=.9\linewidth]{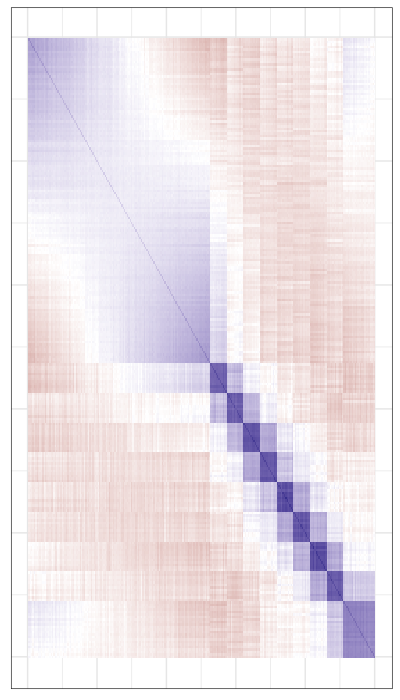}
		\caption{\tiny $\widehat{\text{Cov}}(\xv_i\mid\cdot)^{-1}$ MOM}
	\end{subfigure}
	\begin{subfigure}{.16\textwidth}
		\centering
		\includegraphics[width=.9\linewidth]{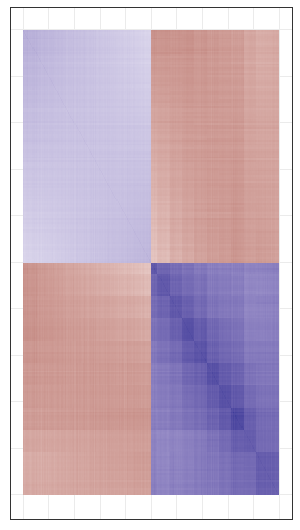}
		\caption{\scalebox{.6}{$\widehat{\text{Cov}}(\xv_i\mid\cdot)^{-1}$ ComBat}}
	\end{subfigure}
\begin{subfigure}{.16\textwidth}
	\centering
	\includegraphics[width=.9\linewidth]{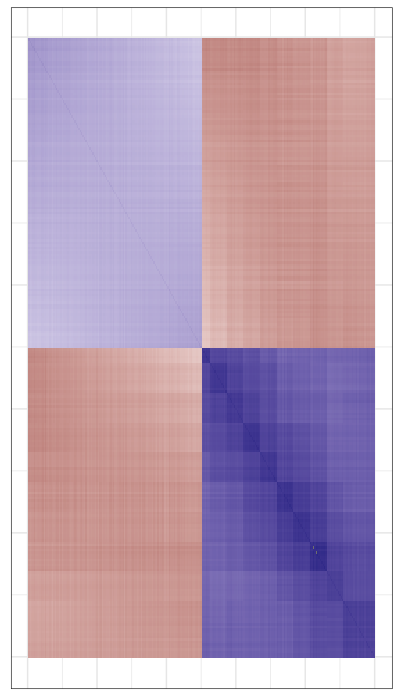}
	\caption{\scalebox{.55}{$\widehat{\text{Cov}}(\xv_i\mid\cdot)^{-1}$ FastBFA}}
\end{subfigure}
	\begin{subfigure}{.16\textwidth}
		\centering
		\includegraphics[width=.9\linewidth]{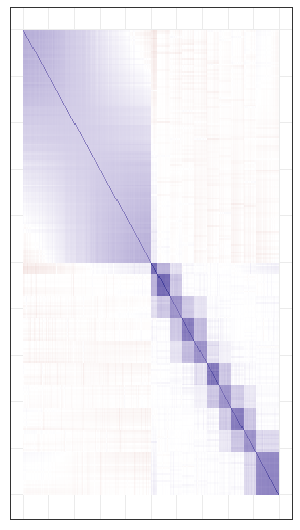}
		\caption{\scalebox{.55}{$\widehat{\text{Cov}}(\xv_i\mid\cdot)^{-1}$ LASSO}}
	\end{subfigure}
	\caption{Heatmaps of loadings and covariance (red denotes large negative values, blue large positive values, white denotes zero).}
	\label{fig:PlotsD2}
\end{figure}

\begin{figure}[h!]
\centering
\begin{subfigure}{.2\textwidth}
  \centering
  \includegraphics[width=.9\linewidth]{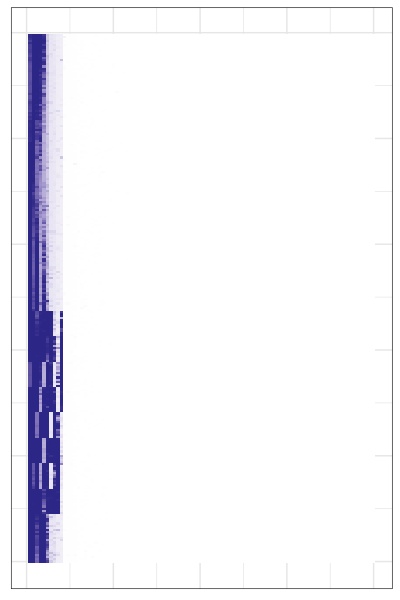}
  \caption{$\hat{\gamma}$ (Normal)}
\end{subfigure}
\begin{subfigure}{.2\textwidth}
  \centering
  \includegraphics[width=.9\linewidth]{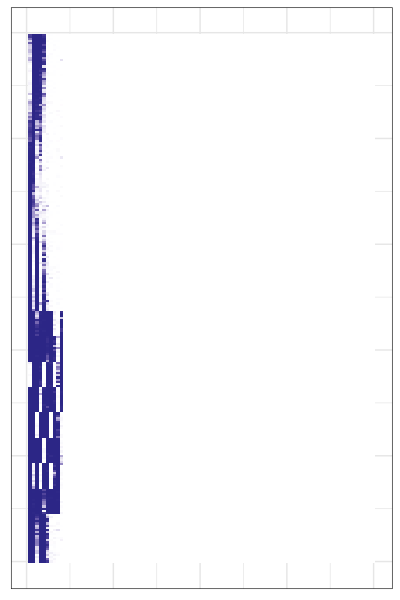}
  \caption{$\hat{\gamma}$ (MOM)}
\end{subfigure}
\begin{subfigure}{.2\textwidth}
  \centering
  \includegraphics[width=.9\linewidth]{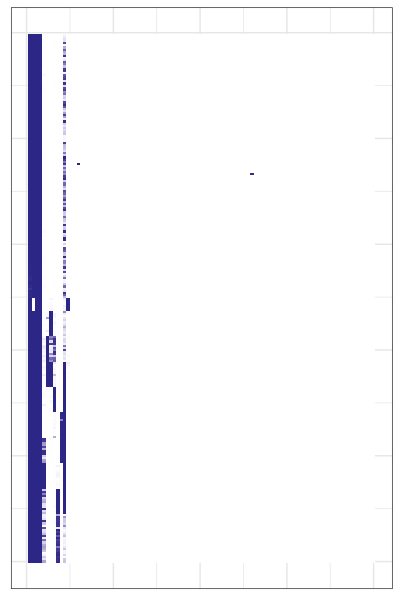}
  \caption{$\hat{\gamma}$ (FastBFA)}
\end{subfigure}
\caption{Heatmaps of inclusion probability (white denotes 0, dark blue denotes 1).}
\label{fig:HeatGammapMOMvsSSLBEsup}
\end{figure}

\end{document}